\documentclass[aps,prd,preprint,superscriptaddress]{revtex4}
\usepackage{graphics}
\usepackage{epstopdf}
\usepackage{graphicx}
\usepackage{dcolumn}
\usepackage{bm}
\usepackage{amsmath} 
\usepackage{multirow}

\newcommand{\be}{\begin{equation}}
\newcommand{\ee}{\end{equation}}
\newcommand{\beq}{\begin{eqnarray}}
\newcommand{\eeq}{\end{eqnarray}}

\def\nue{\mathrel{{\nu_e}}}
\def\numu{\mathrel{{\nu_\mu}}}
\def\nutau{\mathrel{{\nu_\tau}}}
\def\nux{\mathrel{{\nu_x}}}

\def\barnue{\mathrel{{\bar \nu}_e}}
\def\barnumu{\mathrel{{\bar \nu}_\mu}}
\def\barnutau{\mathrel{{\bar \nu}_\tau}}

\def \lta {\mathrel{\vcenter{\hbox{$<$}\nointerlineskip\hbox{$\sim$}}}}
\def \gta {\mathrel{\vcenter{\hbox{$>$}\nointerlineskip\hbox{$\sim$}}}}

\def\t13{\mathrel{{\theta_{13}}}}
\def\y12{\mathrel{{\tan^2 \theta_{12}}}}
\def\c2{\mathrel{{\chi^2 }}}

\def\msun{\mathrel{{M_\odot}}}

\newcommand{\n}{neutrino}
\newcommand{\ns}{neutrinos}

\newcommand{\sn}{supernova}
\newcommand{\sne}{supernovae}

\newcommand{\bh}{black hole-forming}
\newcommand{\nts}{neutron star-forming}

\newcommand{\lar}{LAr}

\newcommand{\ck}{Cherenkov}

\newcommand{\sk}{SuperKamiokande}

\newcommand{\df}{DSNB}

\newcommand{\nsf}{NSFCs}
\newcommand{\bhf}{DBHFCs}

\begin{document}


\title{Neutrinos from failed supernovae at future water and liquid argon detectors}

\author{James G. Keehn}
\affiliation{Arizona State University, Tempe, AZ 85287-1504, USA}%
\affiliation{Missouri University of Science and Technology, Rolla, MO 65409, USA}

\author{Cecilia Lunardini}
\affiliation{Arizona State University, Tempe, AZ 85287-1504, USA}%
\affiliation{RIKEN BNL Research Center, Brookhaven National Laboratory, Upton, NY 11973, USA}

 
\begin{abstract}

We discuss the diffuse flux of electron \ns\ and antineutrinos from cosmological  {\it failed \sne}, stars that collapse directly into a black hole with no explosion. 
This flux has a hotter energy spectrum compared to the flux from regular, neutron-star forming collapses and therefore it dominates the total diffuse flux from core collapses above 20-45 MeV of \n\ energy. 
 Reflecting the features of the originally emitted \ns, the flux of $\nue$ and $\barnue$ at Earth is larger when the survival probability of these species is larger, and also when the equations of state of nuclear matter are stiffer.   
In the 19-29 MeV energy window, the flux from failed \sne\ is susbtantial, ranging from $\sim$7\% to a dominant fraction of the total flux from all core collapses. It can be as large as $\phi^{BH}_{\bar e} = 0.38~{\rm  ~s^{-1}cm^{-2}}$ for $\barnue$ and as large as $\phi^{BH}_{ e} =  0.28 ~ {\rm ~ s^{-1}cm^{-2}}$ for $\nue$,  normalized to a local rate of core collapses of $R_{cc}(0)=10^{-4}~{\rm yr^{-1}Mpc^{-3}}$. 
In 5 years, a 0.45 Mt water \ck\ detector should see $\sim 5-65 $ events from failed \sne, while up to $\sim 160$ events are expected for the same mass with Gadolinium added.  A 0.1 Mt liquid argon experiment should record $\sim 1-11$ events.  Signatures of \ns\ from failed \sne\ are the enhancement of the total rates of events from core collapses (up to a factor of $\sim 2$) and the appearance of high energy tails in the event spectra. 

\end{abstract}                            
 
\maketitle

\section{Introduction}
\label{intro}

Neutrinos are unique probes of the physics of collapsing stars (supernovae).  Diffusing from the dense region surrounding the collapsed stellar core, they can deliver first hand information on the collapse of their stars, on the physics of matter near nuclear density and on the propagation of \ns\ from such high densities to the interstellar space and to detectors on Earth. 

The physics potential of  \ns\ from \sne\ has been studied only minimally due to the scarcity of data. These are limited to handful of events from SN1987A \cite{Hirata:1987hu,Bionta:1987qt}, which was the only recent \sn\ close enough for its neutrino flux to be detectable.  
While the rarity of nearby \sne\ seems an insurmountable problem, a new phase of data taking is expected to begin with  the detection of the {\it  diffuse \sn\ \n\ flux} (or background, \df), on which only upper limits exist \cite{Malek:2002ns,Eguchi:2003gg,Aharmim:2006wq,Lunardini:2008xd}.
Tiny but continuous in time, the  diffuse flux  will give tens to hundreds of events in a few years at future Mt scale detectors, ensuring constant progress for decades.  

Besides the practical advantages, the diffuse flux has a theoretical value of its own: indeed, it has the unique potential to probe the entire \sn\ population of the universe in its diversity.
An important advancement in this direction is the study of the \n\ flux from {\it failed \sne}, stars that collapse directly into a black hole with no explosion and no significant emissions other than \ns\ and gravitational waves.  These direct \bh\ collapses are rare; they are estimated to account for less than $\sim 22\%$ of all collapses \cite{Sumiyoshi:2006id,Lunardini:2009ya}. 
The physics of failed \sne\ were modeled numerically in a number of works, including \cite{Liebendoerfer:2002xn,Sumiyoshi:2006id,Sumiyoshi:2007pp,Fischer:2008rh,Sumiyoshi:2008zw,Nakazato:2008vj,Nakazato:2010ue}, which predicted the emission of a \n\ flux with a higher luminosity and average energy compared to the flux from regular, \nts\ collapses.  

In \cite{Lunardini:2009ya} this result was used to make the first calculation of the diffuse \n\ flux from failed \sne.  The main result was the possibility that, due to their higher energetics,  failed \sne\ might contribute substantially to the \df, with an enhancement of the total flux and event rate in water detectors of up to $\sim 100\%$.

The possibility to detect \ns\ from failed \sne\ in the form of a diffuse flux has several interesting implications. Experimentally, the enhancement of the total flux is attractive because it means that a detection might be closer in time and within the reach of the next phase of  \sk, especially in the configuration with Gadolinium \cite{Beacom:2003nk}.  Theoretically, detecting the diffuse flux would make it possible to learn about direct \bh\ collapses, specifically by constraining their energetics and cosmological rate.   This opportunity is especially precious, considering that failed \sne\ are virtually invisible to telescopes \footnote{An interesting possibility has been suggested recently \cite{Kochanek:2008mp}: observing the disappearance of stars rather than the appearance of \sne. In principle, a ``before vs after"  comparison, together with the non-observation of a \sn\ explosion, could reveal a direct \bh\ collapse.}.  The published \sk\ \n\ data already constrain  the rate of failed \sne\  \cite{Lien:2010yb}.  A new, preliminary, analysis from the \sk\ collaboration \cite{iida,iidathesis} considers the \n\ flux from failed \sne, and limits it to about a factor of two from the most optimistic predictions.  
Neutrinos from failed \sne\ can also increase the amount of Technetium 97 ($^{97}$Tc) that accumulates in  Molybdenum ores over millions of years due to solar and galactic supernova \n\ irradiation \cite{Lazauskas:2009yh}. It was observed that they  also enhance the proposed neutrino-based mechanisms to create amino acid enantiomerism \cite{Boyd:2010ak}.

In this paper we elaborate further on the theme of the diffuse \n\ flux from failed \sne, with a focus on its dependence on the relevant parameters and on its signatures at the next generation of \n\ detectors with 0.1 - 1 Mt masses.  Specifically, we consider a Mt water \ck\ detector and a 0.1 Mt liquid argon (\lar) experiment. 
Our results for water \ck\ detectors elaborate on those of ref. \cite{Lunardini:2009ya}, while the discussion of the potential of liquid argon detectors is presented here for the first time.  The advent of liquid argon technology will be a revolution for the study of \sn\ \ns due to its strong sensitivity to electron \ns, which complements the sensitivity of water detectors to antineutrinos.  A mass of 0.1 Mt is considered to be the minimum mass required to have any sensitivity to diffuse \sn\ \ns.  We will show that this configuration might be particularly suited for \ns\ from failed supernovae: their higher energies imply a larger detection cross section compared to \ns\ from neutron-star forming collapses, and their event energy spectrum might peak above the background of solar \ns. The enhancement of the event rate due to failed \sne\ increases the potential of discovery of the \df\ during the earliest phase of the liquid argon technology development. 

The paper opens with generalities on \ns\ from failed \sne, their expected flux at Earth and basics of their detection and relevant backgrounds (sec. \ref{general}).  We then give results for fluxes and event rates in the antineutrino channel (sec. \ref{antinu}) and the \n\ channel (sec. \ref{nu}).  In sec. \ref{disc} the results are discussed and summarized.

\section{Generalities}
\label{general}

\subsection{Failed supernovae and their \ns}

Core collapse occurs for stars with mass $M\gta 8 M_{\odot}$
(with $M_{\odot}=1.99\cdot 10^{30}$ Kg, the mass of the Sun) at an average rate of $R_{cc}(0)
\sim 10^{-4} ~{\rm Mpc^{-3} yr^{-1}}$ today \cite{Hopkins:2006bw} and
of
\be
R_{cc}(z)\propto
\Bigg\{ \begin{array}{lc}{(1+z)^{\beta}} & z<1 \\
{(1+z)^{\alpha}} & 1<z<4.5\\
{(1+z)^{\gamma}} & 4.5<z\\
\end{array}
\label{rcc}
\ee
at redshift $z$, with $\beta \simeq 3$, $\alpha \simeq 0$ and $\gamma \simeq -8$  \cite{Hopkins:2006bw}). 

\begin{widetext}

\begin{figure}[htbp]
  \centering
 \includegraphics[width=0.45\textwidth]{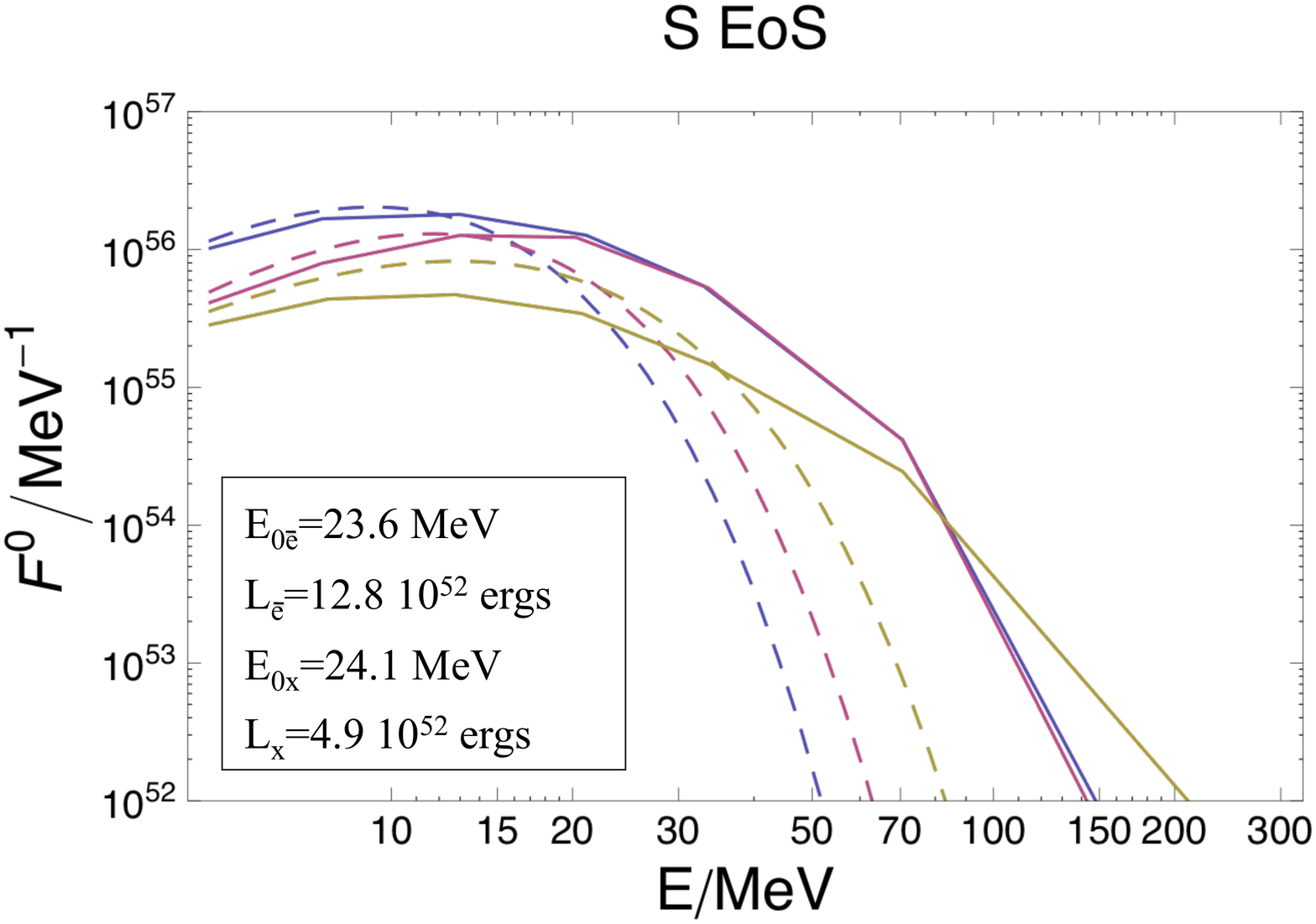}
 \includegraphics[width=0.45\textwidth]{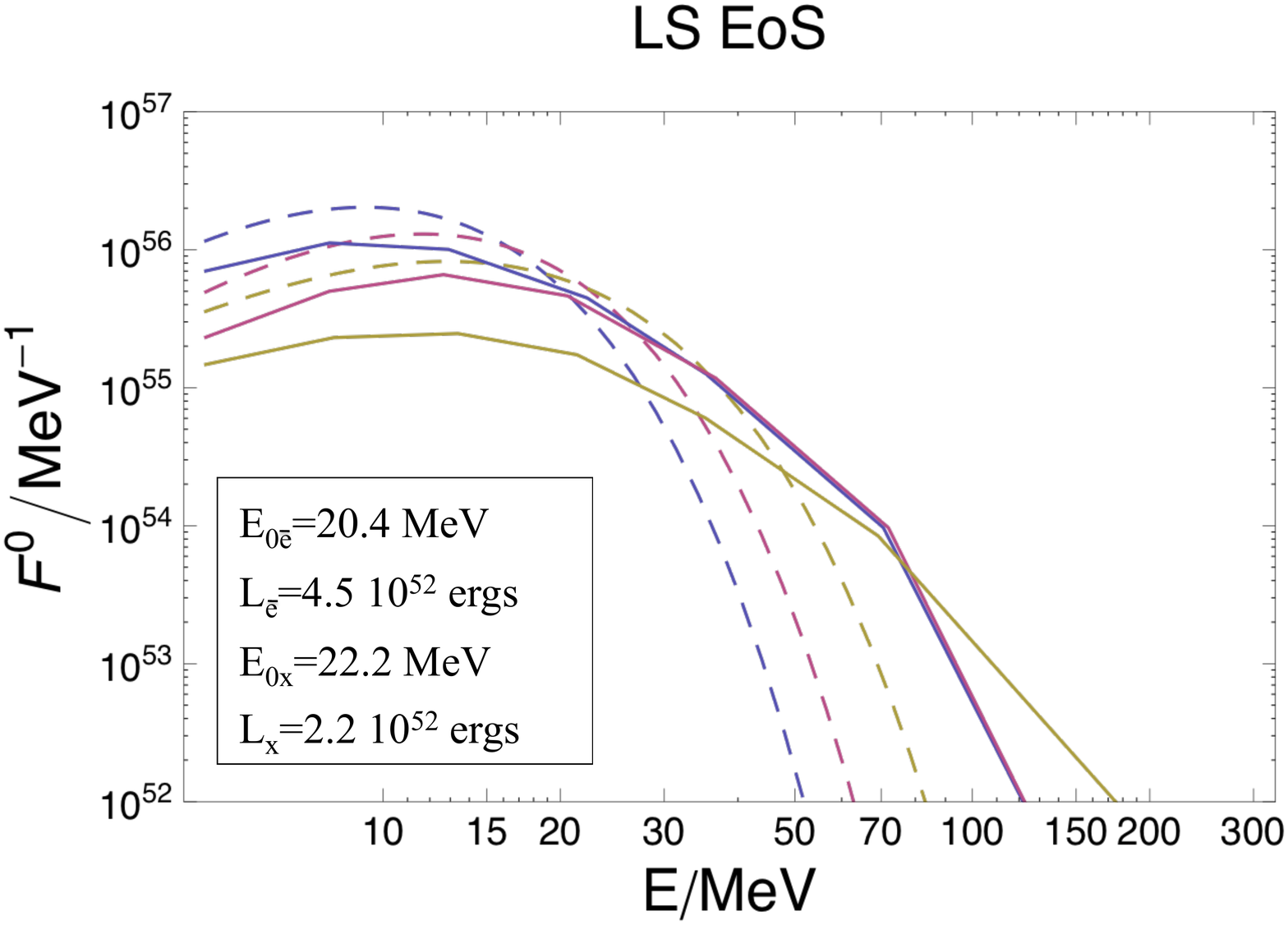}
   \caption{Neutrino fluxes at production inside the star for direct \bh\ collapses (solid, from \cite{Nakazato:2008vj}) ,
   and \nts\  collapses (dashed, Eq. (\ref{dnde})).  In both cases, the curves from upper to lower at 10 MeV correspond to $\nue$, $\barnue$, $\nux$.  
   For direct \bh\ collapses the \n\ spectra are shown for the Shen et al. (left panel) and Lattimer-Swesty (right) equations of state.  For each, the \n\ luminosities and average energies are given (insets).  See text for details.}
\label{spectra}
\end{figure}

\end{widetext}

For $M = 8 - 25 M_{\odot}$  the collapse leads to an explosion, followed by the formation of a
neutron star \cite{Woosley:2002zz}.
Considering that 
 stars are distributed in mass as $\phi(M) \propto
M^{-2.35}$ \cite{Salpeter:1955it}, one gets that these Neutron Star Forming Collapses (\nsf)  are a
fraction $f_{NS}\simeq 0.78$ of the total.
They emit \ns\  in comparable amounts  in the
 six 
 species: $\nue,\barnue,\numu,\barnumu,\nutau,\barnutau$ ($\numu,\barnumu,\nutau,\barnutau=\nux$ from here on).  At the production site, the flux in
 each species $w$, differential in energy, can be described as \cite{Keil:2002in}:
\be
F^0_w\simeq \frac{(1+\alpha_w)^{1+\alpha_w}L_w}
  {\Gamma (1+\alpha_w){E_{0w}}^2}
  \left(\frac{E}{{E_{0w}}}\right)^{\alpha_w}
  e^{-(1+\alpha_w)E/{E_{0w}}},
\label{dnde}
\ee
 where $\Gamma(x)$ stands for the Gamma function. Here $\alpha_w $ controls the spectral shape, $L_w $  is the time integrated luminosity and $E_{0 w}$ is
the average energy.  Here we will use
typical values \cite{Keil:2002in}: $E_{0 e} = 9$ MeV, $E_{0 \bar e} = 15$ MeV, $E_{0
x} = 18$ MeV, $L_{ e}=L_{\bar e}=L_x=5 \cdot 10^{52}$ ergs, $\alpha_{
e}=\alpha_{\bar
e}=3.5$ and $\alpha_x=2.5$.   For these, the fluxes $F^0_e,F^0_{\bar e}, F^0_x$ are illustrated in fig. \ref{spectra} (dashed lines). 

While \nts\ collapses  have been studied in detail, the evolution of higher
mass stars is more uncertain.  For $M \sim 25 - 40 \msun $ (13\% of
the total) a weaker explosion should occur, with a black hole formed
by fallback \cite{Heger:2002by,Woosley:2002zz}. Stars with $M \gta 40 \msun$ (a 9\%
fraction), would instead collapse into a black hole directly.
Simulations of such Direct Black Hole Forming Collapses (\bhf)
\cite{Liebendoerfer:2002xn,Sumiyoshi:2006id,Sumiyoshi:2007pp,Fischer:2008rh,Sumiyoshi:2008zw}
show an emitted \n\ flux that is  more energetic and more luminous than
the NSFC case as a result of the rapid contraction of the protoneutron star (see e.g., \cite{Sumiyoshi:2006id}).  Furthermore, the 
$\nue$ and $\barnue$ fluxes are especially luminous due to the capture of
electrons and positrons on nucleons.

A ``stiffer'' equation of state (EoS) of nuclear matter
\cite{Sumiyoshi:2006id} and/or a smaller accretion rate of matter on the protoneutron star  \cite{Fischer:2008rh} correspond to more
luminous and hotter neutrino fluxes. 
Here we take the fluxes from DBHFCs
from fig. 5 of
Nakazato et al. \cite{Nakazato:2008vj}, using the same linear
interpolation of numerically calculated points (which underestimates the \df\  in the \sk\ energy window by about 10-20\% \cite{Lunardini:2009ya}). 
They are  are shown in fig. \ref{spectra}. These fluxes were obtained for the $40 M_{\odot}$ progenitor in  \cite{woosley} with the stiffer  Shen et al. (S) EoS 
\cite{shenetal} (incompressibility $K=281$ MeV) and the softer  Lattimer-Swesty (LS) EoS  \cite{lattimer91generalized} (with $K=180$ MeV  \cite{Nakazato:2008vj}).  For the different progenitors considered in \cite{Nakazato:2008vj} 
results appear unchanged for the S EoS, while for the LS one 
the luminosity and average energy  may be lower by a factor of two and by 10-20\% respectively.

Flavor oscillations modify the flavor composition of the \n\ flux in a way that is similar, in its generalities, for \nsf\ and \bhf, as was shown in an initial study \cite{Nakazato:2008vj}.   After oscillations the fluxes of the $\nue$ and $\barnue$ species are admixtures of the original fluxes of the different flavors: 
\beq
&&F_{ e} = { p} F_{ e}^0 + (1-{ p}) F_{ x}^0  ~~,
\label{fluxes1}\\
&&F_{\bar e} = {\bar p} F_{\bar e}^0 + (1-{\bar p}) F_{ x}^0 ~,
\label{fluxes2}
\eeq
where the  probabilities $p$ and $\bar p$ 
 depend on the structure of the \n\ mass spectrum and mixing, 
 and on refraction effects due to  \n-\n\ and to \n-electron interaction, with the latter producing two MSW resonances inside the star \cite{Dighe:1999bi}. 
A very general result -- valid if turbulence effects are negligible -- is that $p$ and $\bar p$ vary in the intervals \cite{Dasgupta:2007ws}:
\beq 
&&p= 0 - \sin^2\theta_{12}\simeq 0 -  0.32~ , \nonumber \\
&&\bar p= 0 - \cos^2\theta_{12}\simeq 0 -  0.68~.
\label{rangesprob}
\eeq
Their energy dependence is generally smooth due to the energy dependence of the transition probability in the highest density MSW resonance (see e.g., \cite{Lunardini:2003eh}). An exception is the recently studied ``spectral swap", a step-like structure in the probabilities as functions of energy caused by \n-\n\ interaction (see e.g., \cite{Duan:2005cp,Hannestad:2006nj,Raffelt:2007cb,EstebanPretel:2007ec,Fogli:2007bk} and the review \cite{Duan:2009cd} and references therein).
For the inverted mass hierarchy a sharp, single swap should appear in the $\nue$ spectrum  at the critical energy $E_c$ defined as \cite{Raffelt:2007cb}:
\be
\int_{E_c}^{\infty} (F^0_{e}-F^0_{x})=\int_{0}^{\infty} (F^0_{\bar e}-F^0_{ x})~,
\label{ecrit}
\ee
and the probability $p$ is then given by:
\be
p \simeq \Bigg\{ \begin{array}{lc}
\sin^2 \theta_{12} \simeq 0.32 & (E<E_c) \\
0  & (E>E_c)\\ 
\end{array}
\label{step}
\ee
Typical values of $E_c$ are $E_c \simeq 3-10 $ MeV  \cite{Duan:2007bt}; for our set of parameters we find $E_c \simeq 8$ MeV for \nsf\ and  $E_c \simeq 12$ MeV for \bhf\, using both equations of state. Multiple other swaps could also appear in both the $\nue$ and $\barnue$ channels in a way that is highly dependent  on the original \n\ fluxes and on the mass hierarchy \cite{Dasgupta:2009mg,Fogli:2009rd,Friedland:2010sc}.  

A study of the MSW resonances only (no \n-\n\ effects)  for \bhf\ \cite{Nakazato:2008vj} (see also \cite{Nakazato:2010qy}) indicates that the oscillation pattern is the same for both collapse types in the intervals $\sin^2 \theta_{13}\gta 3 \cdot 10^{-4}$ or $\sin^2 \theta_{13}\lta 3 \cdot 10^{-6}$,  where the higher density MSW resonance is completely adiabatic or completely non-adiabatic. In this case the probabilities are energy-independent and take their extreme values in eq. (\ref{rangesprob}). While swap effects have not been studied for \bhf, we expect that the picture with a  single swap, eq. (\ref{step}), should be valid since it is a typical occurence when $F^0_x \lta F^0_{e}, F^0_{\bar e}$ \cite{Fogli:2009rd}.   

For generality, here we follow ref. \cite{Lunardini:2009ya} and limit our discussion to energy-independent permutation parameters that are equal for both collapse types.  
We give results only for  the extremes of the intervals of the permutation parameters, Eq. (\ref{rangesprob}): from these one can easily derive fluxes and event rates  for intermediate values.  
The approximation of energy-independent permutation should be adequate in the energy windows relevant for the \df\ detection ($E \gta 11-20$ MeV, see sec. \ref{fluxbckg}): indeed,  we have checked that the largest energy modulations due to the MSW resonance cause an effect at the level of 20\% or less on the \df\ spectrum, which is negligible compared to other uncertainties in the problem.   Moreover, our calculated values of $E_c$ produce effects on the \df\ that fall below typical  windows of detection. Still, in sec. \ref{fluxbckg} we briefly discuss the effect of spectral swaps.

\subsection{Fluxes at Earth: signal and backgrounds}
\label{fluxbckg}

Following ref. \cite{Lunardini:2009ya}, we model the neutrino fluxes from \nsf\ and \bhf, and the total diffuse flux for a schematic
two-population scenario, with a fraction $f_{NS}$ ($f_{BH}=1-f_{NS}$)
of identical \n\ emitters of the NSFC (DBHFC) type.  The
total diffuse $\barnue$ flux at Earth, differential in energy and area, is:
\begin{widetext}
\beq
&&\Phi_{\bar e}(E)=\Phi^{BH}_{\bar e}+\Phi^{NS}_{\bar e}~,   \nonumber\\
&&\Phi^{BH}_{\bar e}=\frac{c}{H_0}(1-f_{NS})\int_0^{z_{ max}} R_{cc}(z)  F^{BH}_{\bar e}(E(1+z))\frac{{d}
z}{\sqrt{\Omega_{ m}(1+z)^3+\Omega_\Lambda}}~, \nonumber\\
&&\Phi^{NS}_{\bar e}=\frac{c}{H_0} f_{NS} \int_0^{z_{ max}} R_{cc}(z)  F^{NS}_{\bar e}(E(1+z))\frac{{d}
z}{\sqrt{\Omega_{ m}(1+z)^3+\Omega_\Lambda}}~, 
\label{difflux}
\eeq
\end{widetext}
where $\Omega_{ m}=0.3$ and $\Omega_\Lambda=0.7$ are the fractions of
the cosmic energy density in matter and dark energy, $c$ is the speed
of light and $H_0$ is the Hubble constant. An analogous expression holds for the $\nue$ diffuse flux $\Phi_{e}$. 
 In what follows the values
$R_{cc}(0) = 10^{-4} ~{\rm Mpc^{-3} yr^{-1}}$, $\beta=3.28$, $\alpha=0$
and $z_{max}=4.5$ \cite{Hopkins:2006bw} will be used (results depend weakly on $z_{max}$, at the level of $\sim 7\%$ or less for $z_{max}\gta 3$
\cite{Ando:2004hc}). 
 We
take the interval $f_{NS}= 0.78 - 0.91$,  corresponding to a mass
of $25-40 \msun$ as the upper limit for neutron star-forming collapses, as a way to parametrize the uncertainty in the \n\ fluxes in the region of  transition between robust, neutron-star forming explosions and direct black hole formation. Most of this uncertainty is due to the poorly studied \n\ emission for black hole formation by fallback (see sec. \ref{directions}). There are also uncertainties in the minimum mass required for direct black hole formation:  a value as low as $\sim 17- 20 M_\odot$ is compatible with observations of progenitors of type IIP \sne\ \cite{Smartt:2008zd}, while numerical studies indicate larger minimum masses, with wide variations associated with the star's metallicity and rotation (see e.g., \cite{O'Connor:2010tk}). 

\begin{figure}[htbp]
  \centering
 \includegraphics[width=.45\textwidth]{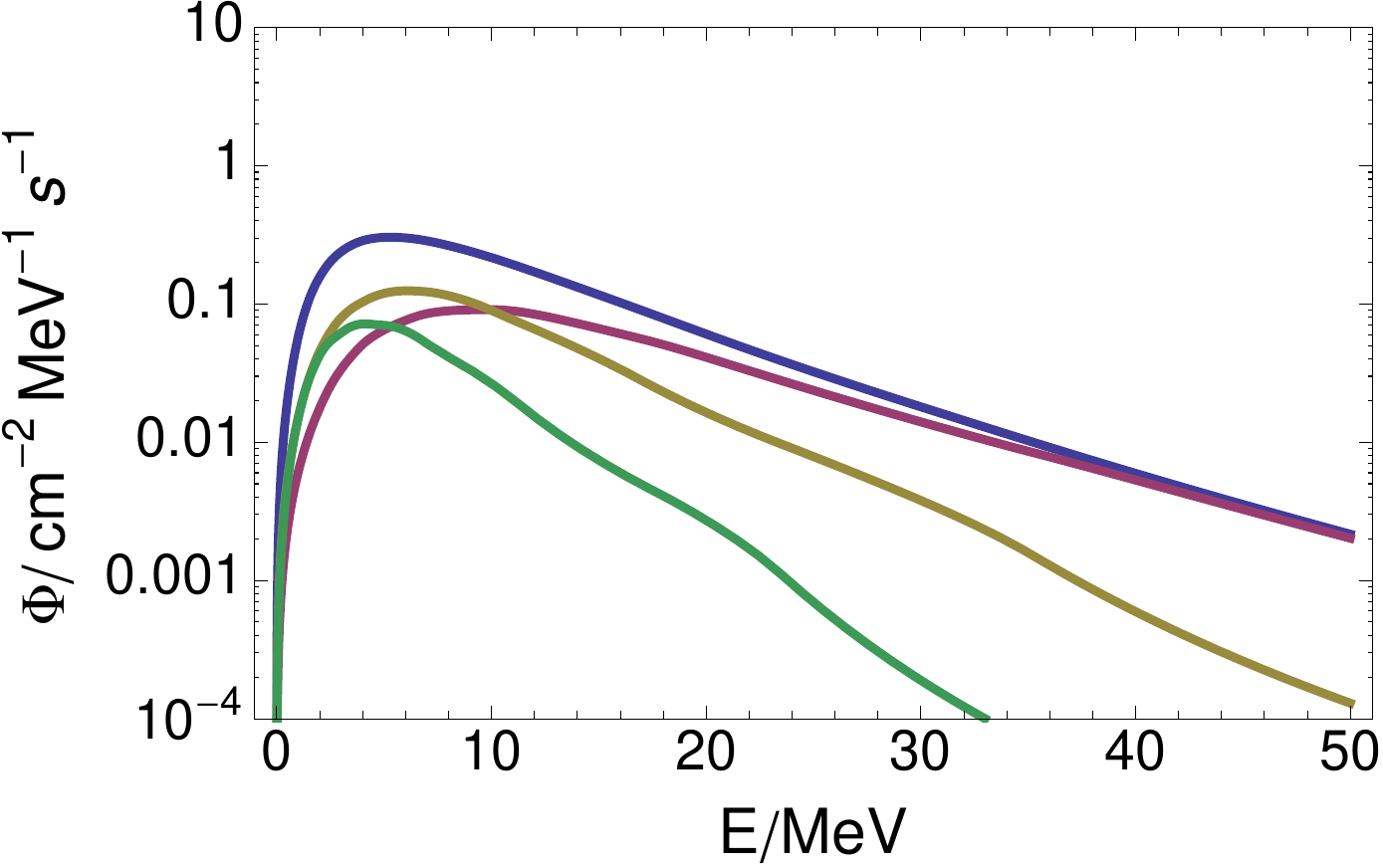}
  \includegraphics[width=.45\textwidth]{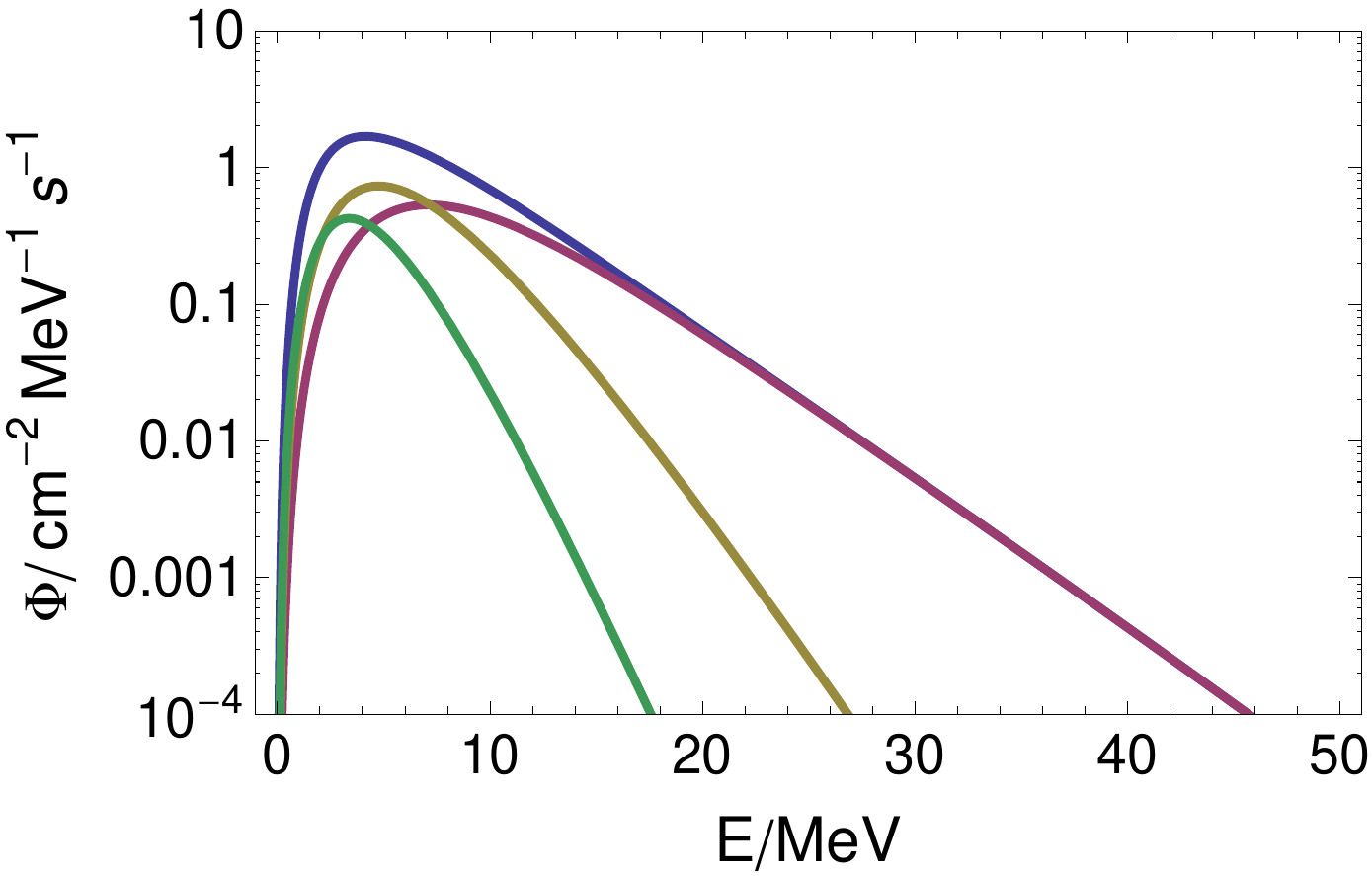}
 \caption{The diffuse flux of $\barnue$ from \bh\  collapses (left panel) and from \nts\ collapses (right panel) from different bins of redshift, $z$. From lower to upper curves at 20 MeV: $2 \leq z<3$,  $1 \leq z<2$,  $0\leq z<1$,  and the total from all redshifts.  We used the S EoS, $\bar p=0.68$ and $f_{NS}=0.78$. }
\label{bhzplot}
\end{figure}

An example of the resulting diffuse \n\ fluxes  is given in fig. \ref{bhzplot}, which shows $\Phi^{BH}_{\bar e}$ and  $\Phi^{NS}_{\bar e}$ as well as the contributions to each from different redshift bins  \footnote{The bumps at 20-30 MeV in the lower curves of fig. \ref{bhzplot} are a numerical artifact due to interpolating a sparse set of points in the highest energy part of the original spectra (fig. \ref{spectra}). Therefore, these curves have indicative value only. }. The parameters that maximize $\Phi^{BH}_{\bar e}$ above 20 MeV have been chosen  (see caption). 
While a detailed discussion is deferred to secs. \ref{antinu} and \ref{nu}, here we observe the main differences between the two fluxes: reflecting the features of the original spectra at production, $\Phi^{BH}_{\bar e}$ has a more energetic spectrum so that, in spite of the rarity of failed \sne, it dominates above 20 MeV or so.  

From the contributions of each redshift bin we see that, as expected, the flux from more distant collapses accumulates at lower energies due to the redshift of energy, and so generally at energies relevant for detection ($E\gta 10-20$ MeV) the flux from sources with $z< 1$ dominates. Still, for \bhf\ the flux from higher redshifts ($z \gta 1$)  is substantial:  at 10 MeV (20 MeV) it  is about 58\% (32\%) of the total from failed \sne.  For other combinations of parameters the fraction varies in the interval 52-58\%  (30-40\%). 

In contrast, the analogous calculation for \nts\ collapses gives 40\% (9\%) (of the total from \nts\ collapses) at 10 MeV (20 MeV).

 The larger cosmological component of the failed \sn\ flux  is explained by the more energetic original \n\ spectra.   In principle, its implications are profound: if seen, this flux can probe the rate of failed \sne\ beyond $z \simeq 1$. This is the limit of current \sn\ surveys, which are not sensitive to direct \bh\ collapses in any case. 

\begin{figure}[htbp]
  \centering
 \includegraphics[width=.55\textwidth]{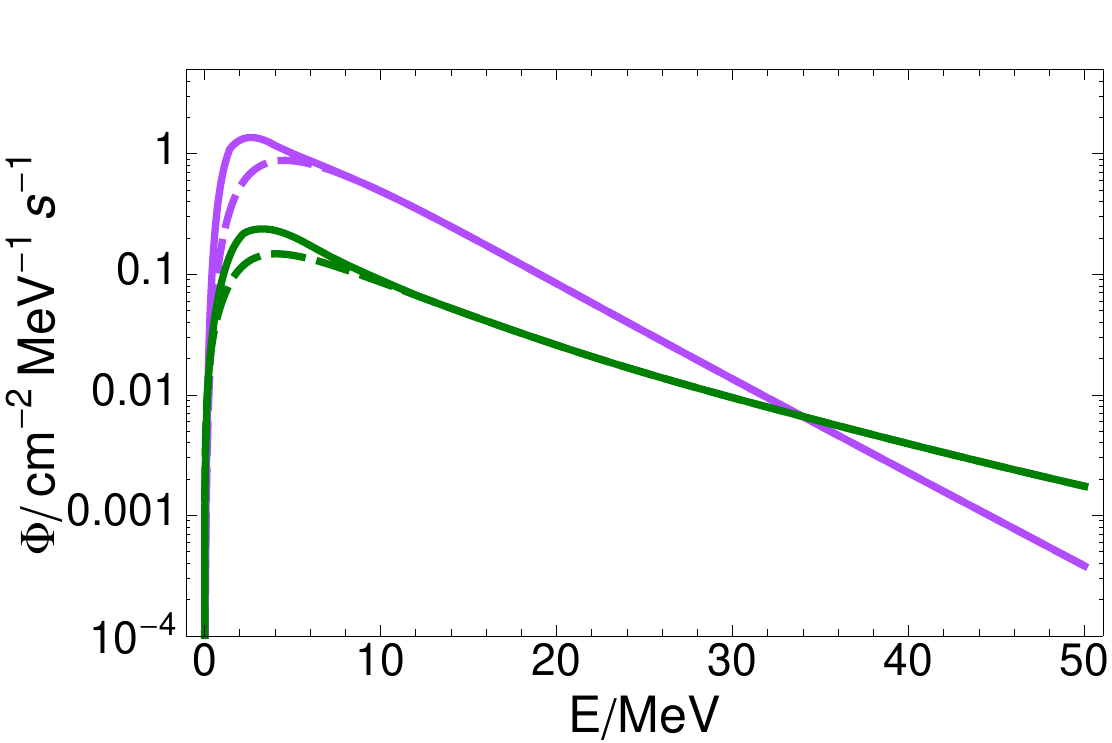}
 \caption{$\nue$ fluxes from direct \bh\ collapses (lower curves at 10 MeV) and from \nts\  collapses (upper curves) for $p=0$ (dashed lines) and for the step-like probability in eq. (\ref{step}) (solid).  The swap energies are $E_c=12.1,8.0$ MeV for the the two fluxes respectively.  We used the S EoS and $f_{NS}=0.78$. }
\label{nueswap}
\end{figure}
Summing over many sources at different redshifts smears out potentially interesting oscillation effects, so that only dramatic features affecting the original fluxes might appear in the \df.  In fig. \ref{nueswap} we give an example of how $\Phi^{BH}_{e}$ and $\Phi^{NS}_e$ are modified by the spectral swap, eq. (\ref{step}), assuming that a single swap is realized according to eq. (\ref{ecrit}).  Due to the smearing, the effect of the swap is a smooth spectral distortion at $E < E_c$. 
In this interval the flux is larger compared to the case of constant $p=0$. This is due to  the larger survival of the original $\nue$ flux, which is the dominant component at these energies.  When compared with the case where $p=0.32$, instead, the step causes a suppression of the flux in the same interval. We do not discuss the swap effects in detail because the spectral distortion falls below detectable energies. The question should be reexamined, however, when a detailed picture of \n-\n\ refraction in failed \sne\ becomes  available. 

\begin{figure}[htbp]
  \centering
 \includegraphics[width=0.55\textwidth]{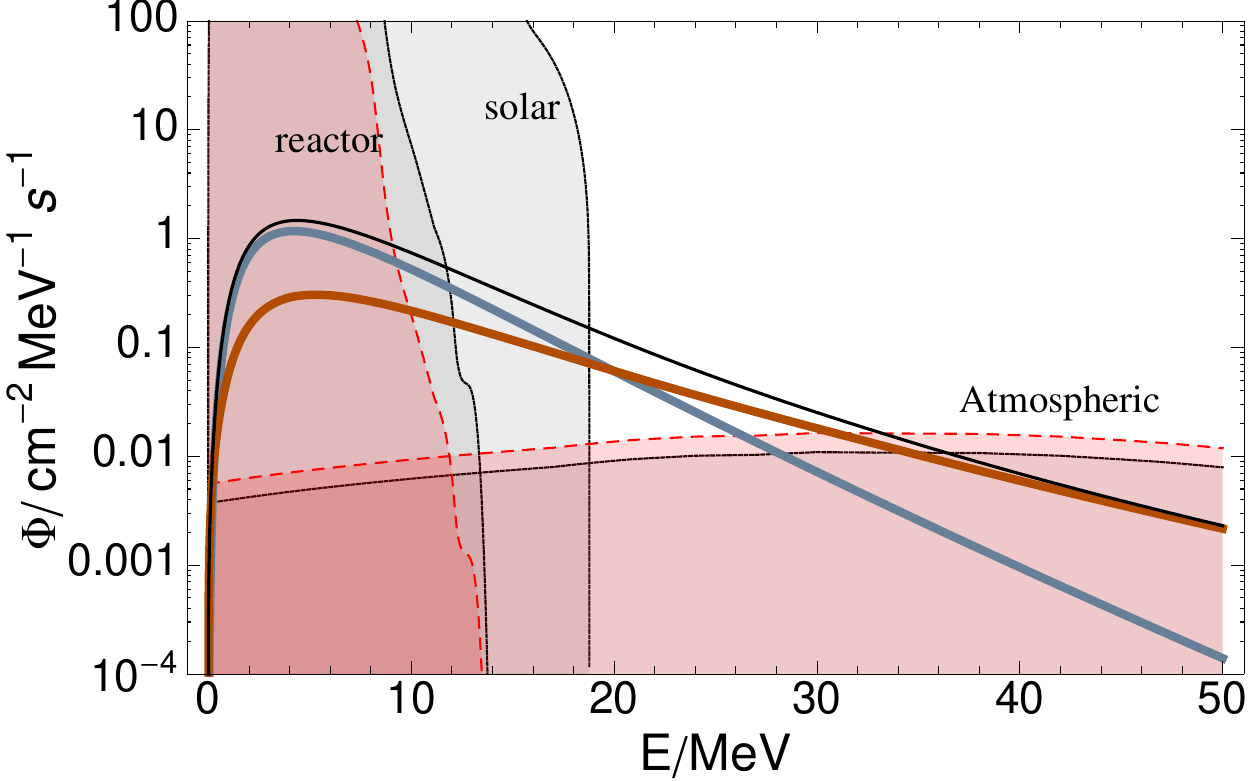}
   \caption{Signal and background  fluxes.  The atmospheric and 
reactor fluxes are shown for the Kamioka (solid, gray) and Homestake (dashed, red) sites. 
The atmospheric fluxes of $\nue$ and  $\barnue$  (from \cite{Battistoni:2005pd}) are very similar, so only one of them is plotted. The 
calculations of the background fluxes include oscillation effects, which are responsible for the 
visible modulation of the reactor spectrum.  The signal flux plotted refers to the $\barnue$ channel, with the S EoS, $p=0.68$ and $f_{NS}=0.78$.  From the lower to upper curves at 40 MeV: the flux from \nts\ collapses, the flux from direct \bh\ collapses and the total flux. }
\label{backgroundsites}
\end{figure}
The potential to detect diffuse \ns\ from core collapses is strongly influenced by backgrounds, which determine the {\it energy window} of sensitivity  (defined as the interval where the \df\ exceeds other \n\ fluxes) and the statistical significance of a signal.  Several neutrino fluxes of other origin constitute ineliminable backgrounds for water and \lar\ experiments; they are shown in fig. \ref{backgroundsites}.   Since water detectors are mostly sensitive to $\barnue$s, their backgrounds are $\barnue$ fluxes from the atmosphere and from reactors.  The atmospheric $\barnue$s are truly indistinguishable from diffuse \sn\ \ns\ because they have the same energy range and the same isotropic distribution in space.  Their flux exceeds the \df\ typically above 30-40 MeV, thus restricting the experimental  sensitivity  to this range.  The restriction is stronger for detectors closer to the Earth's poles, where the atmospheric flux is larger.   Fig. \ref{backgroundsites} illustrates this, showing the atmospheric $\barnue$ flux, taken from the FLUKA group \cite{Battistoni:2005pd}, at Kamioka and Homestake; the second location has a higher flux by a factor $\sim 1.5$, which results in a restriction of the energy window by 2-4 MeV. 

For the same two locations the reactor flux is shown in fig. \ref{backgroundsites} (from \cite{Wurm:2007cy}). It is stronger at Kamioka, reflecting the high concentration of nuclear reactors in Japan, and weaker by a factor $\sim 24$  at Homestake \cite{duselwhite,vaginsprivate}.  It restricts the experimental sensitivity to the \df\ to \n\ energies higher than $\sim 10$ MeV ($\sim 12$ MeV) at Homestake (Kamioka).   Reactor \n\ events could be distinguished in principle from their direction; conceptual studies exist on this in the context of geoneutrino detection \cite{Hochmuth:2005nh,Terashima:2008zz}, but their effectiveness for the \df\ is unclear at this time. 

LAr detectors are mostly sensitive to electron \ns, and therefore solar and atmospheric $\nue$ fluxes are the main backgrounds for them.  The atmospheric $\nue$ flux is very similar to the $\barnue$ one at these energies, so the considerations above apply.   The flux from the $hep$ process in the Sun  prevents studying the \df\ below $\sim 18$ MeV, unless a method is devised to distinguish it using directional information \footnote{Water detectors are sensitive to solar \ns\  via neutrino-electron scattering. However, these events can be effectively subtracted because they are highly directional. This is not the case for $\nue$ charged current scattering on $^{40}Ar$, therefore  solar \n\ events will have to be included in the data analysis of \lar\ detectors. }.  The solar \n\ flux is plotted in fig. \ref{backgroundsites}; it is from the BPS05 model \cite{Bahcall:2004pz} with the inclusion of oscillation effects as in  \cite{Lazauskas:2009yh}.

Clearly, the energy window widens for a larger \df, and therefore a strong flux from failed \sne\ is advantageous for  signal to background discrimination.  For our parameters of choice  and the largest \bhf\ flux, we find that the window is $\sim 12- 36$ MeV for a water detector at Kamioka and $\sim 9 - 32$ MeV for one at Homestake.   Instead, if the flux from \bhf\ is negligible, the window is  $\sim 12-29$ MeV and $\sim 9-27$ MeV in the two cases respectively.     These windows refer to the ideal case in which other, non-\n, backgrounds can be neglected. In practice, however, water detectors are limited by other backgrounds that will be summarized next.

\subsection{Neutrino detection}
\label{detection}

We consider a representative scenario with a water \ck\ detector and a \lar\ detector of 0.45 Mt and 0.1 Mt fiducial masses respectively, as envisioned for the next generation of underground laboratories.   We discuss each briefly and refer to recent reviews \cite{Lunardini:2010ab} for more details. 
 
For a water \ck\ detector the main detection channel is inverse beta decay:
 \be
 \barnue + p \rightarrow n +  e^+ ~,
 \label{beta}
 \ee
which dominates the event rate due to to the larger cross section compared to other relevant processes. Here we consider only inverse beta decay, with the cross section from \cite{Strumia:2003zx}, and present results in terms of the positron energy $E_e \simeq E - 1.3~{\rm MeV}$.  
We consider a representative detection efficiency of 90\%.
Besides atmospheric and reactor neutrinos, a water detector suffers large backgrounds from spallation and invisible muons, which we model following Fogli et al. \cite{Fogli:2004ff}. Spallation products motivate limiting the search to the window $E_e \geq 18$ MeV ($E \geq 19.3$ MeV) at \sk, while invisible muon events are included in the analysis and exceed the signal \cite{Malek:2002ns}.   
Current \sk\ data give a stringent upper limit on the $\barnue$ component of the \df\ \cite{Malek:2002ns,Lunardini:2008xd} 
\footnote{We quote the limit as given in  \cite{Lunardini:2008xd} for a large variety of \n\ spectra and with an updated inverse beta decay cross section. The  limit  given by the \sk\ collaboration in 2002 is  $\phi_{\bar e}(E)(E>19.3~ {\rm MeV})<1.2~{\rm cm^{-2} s^{-1}}$  at  90\% C.L.  \cite{Malek:2002ns}. It remains to be determined what the limit is for the more energetic spectrum of \ns\ from failed \sne\ (see \cite{iida,iidathesis} for preliminary results). }
:
\be
\phi_{\bar e}(E)(E>19.3~ {\rm MeV})<1.4-1.9~{\rm cm^{-2} s^{-1}~~~~at~ 90\% C.L.}~,
\label{sklim}
\ee
where the interval of values accounts for varying \n\ spectra.
 Although our main focus is on Mt class detectors, we will discuss how this  limit constrains the flux of \ns\ from failed \sne\ (sec. \ref{antinu}).

If Gadolinium is dissolved in the water, as in the proposed GADZOOKS design \cite{Beacom:2003nk}, neutron tagging becomes possible, so that spallation can be almost  completely subtracted and the invisible muon background effectively reduces by a 
factor of $\sim 5$ \cite{Beacom:2003nk}.  This allows to search for the \df\ in the whole energy window determined by reactor and atmospheric \ns. 
In sec. \ref{antinu} we will give results for this larger window, as well as for the one relevant to pure water. 

In \lar\ the largely dominant detection channel is charged current $\nue$ scattering:
 \be
 \nue + ^{40}Ar \rightarrow X +  e^-  ~,
 \label{cclar}
 \ee
where $X$ stands for any of the possible products.  The emitted electron is detected by the ionization track it produces in the liquid Argon.
We model the process (\ref{cclar}) following ref. \cite{Cocco:2004ac} and use ref. \cite{Kolbe:2003ys} for the cross section.  The energy of the emitted electron differs from that of the incoming neutrino by $\sim$ 3-4 MeV depending on the nuclear transition taking place \cite{Cocco:2004ac}.  Since detailed information on the spectrum of these transitions is not available, however, here event rates will be discussed in terms of neutrino energy.  For generality, we use a 100\% detection efficiency.  We consider only the solar and atmospheric backgrounds, under the assumption that events of other nature can be effectively identified and subtracted; this is being investigated in the intense R\&D work that is ongoing at this time. 

In the reminder of the paper we consider all backgrounds for the Kamioka site; results for the Homestake location can be inferred using the rescaling factors in sec. \ref{fluxbckg} as well as fig. \ref{backgroundsites} for estimating the energy window.

\section{Antineutrinos: flux and event rates in water}
\label{antinu}

\subsection{$\barnue$ flux}

\begin{figure}[htbp]
  \centering
 \includegraphics[width=0.39\textwidth]{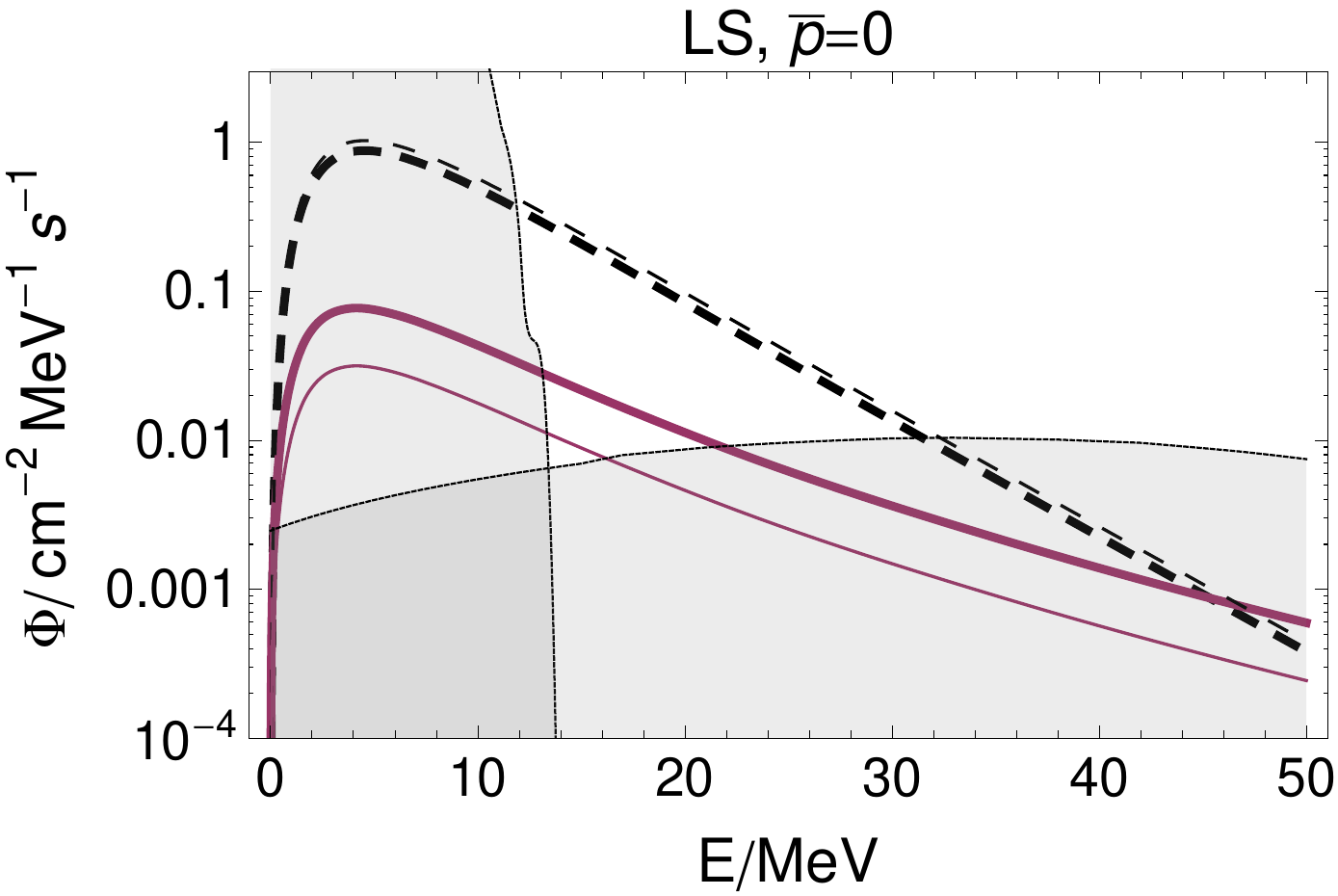}
  \vspace{0.3 truecm}
 \includegraphics[width=0.25\textwidth,height=4 truecm]{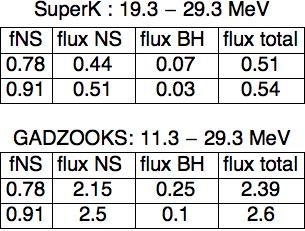}
 \vspace{0.3 truecm}
 \includegraphics[width=0.39\textwidth]{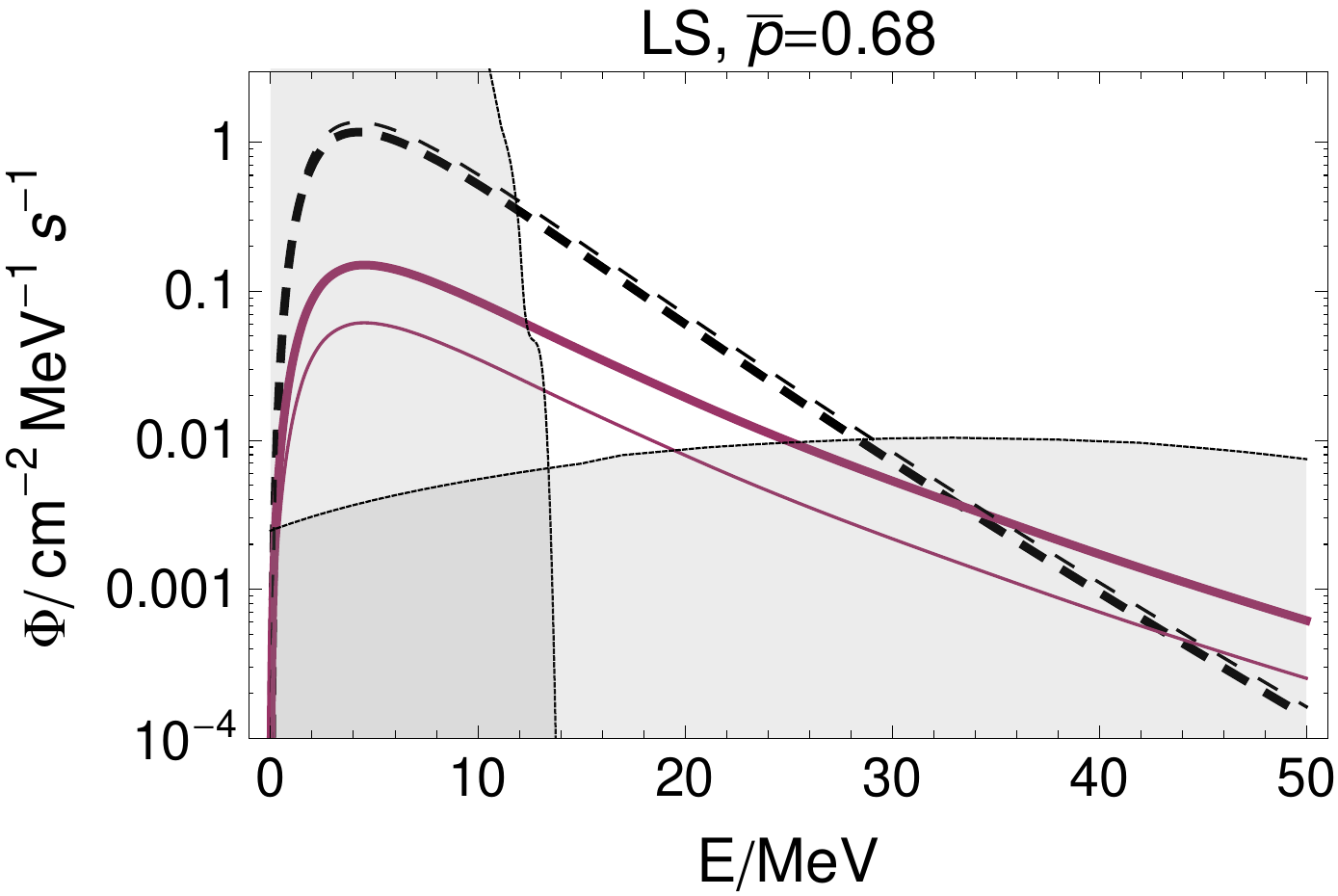}
  \vspace{0.3 truecm}
 \includegraphics[width=0.25\textwidth,height=4 truecm]{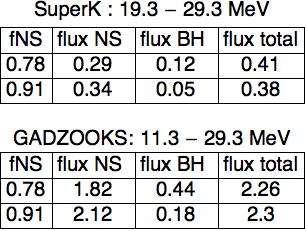}
  \vspace{0.3 truecm}
 \includegraphics[width=0.39\textwidth]{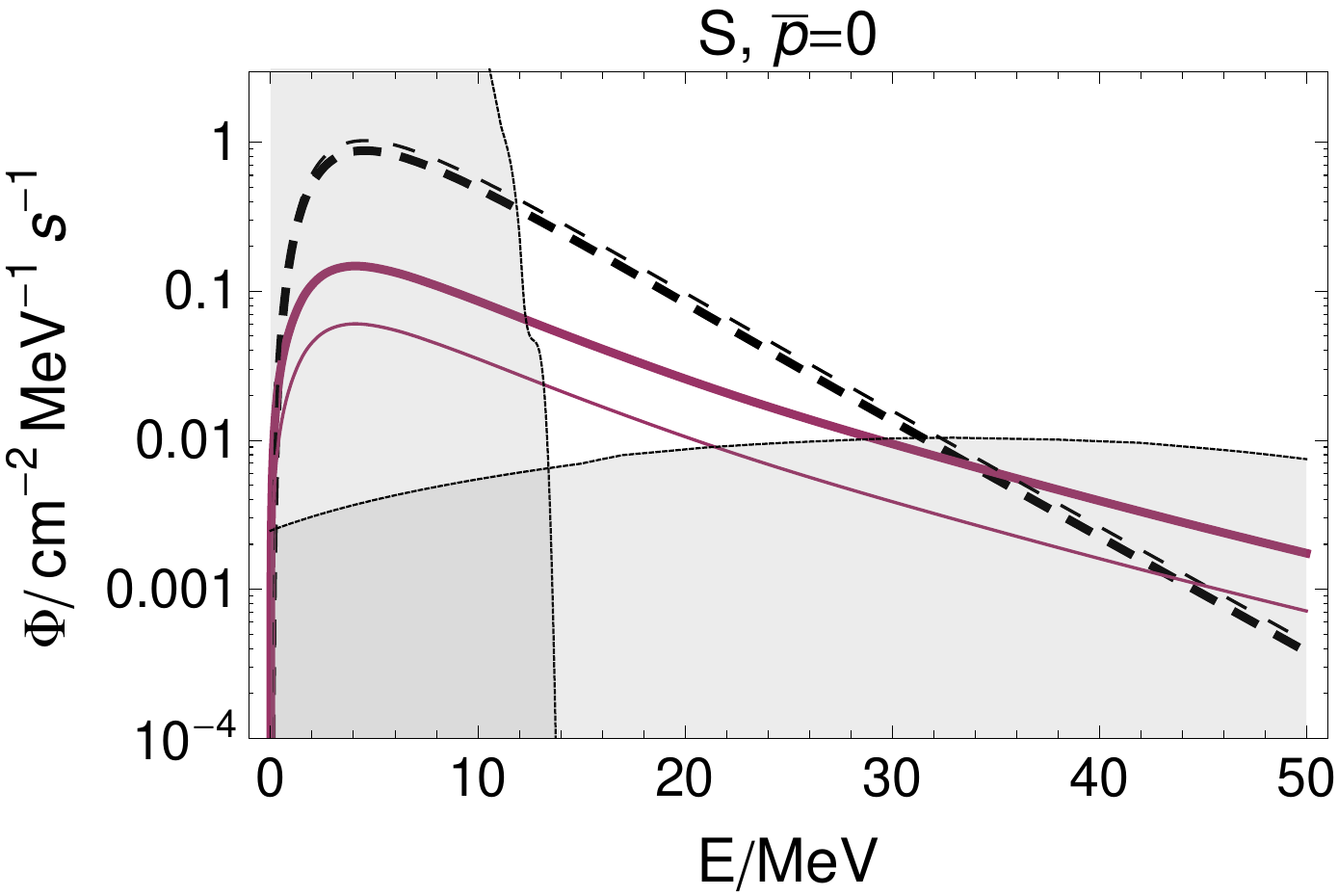}
  \vspace{0.3 truecm}
 \includegraphics[width=0.25\textwidth,height=4 truecm]{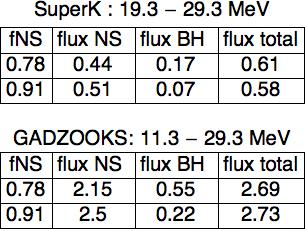}
  \vspace{0.3 truecm}
 \includegraphics[width=0.39\textwidth]{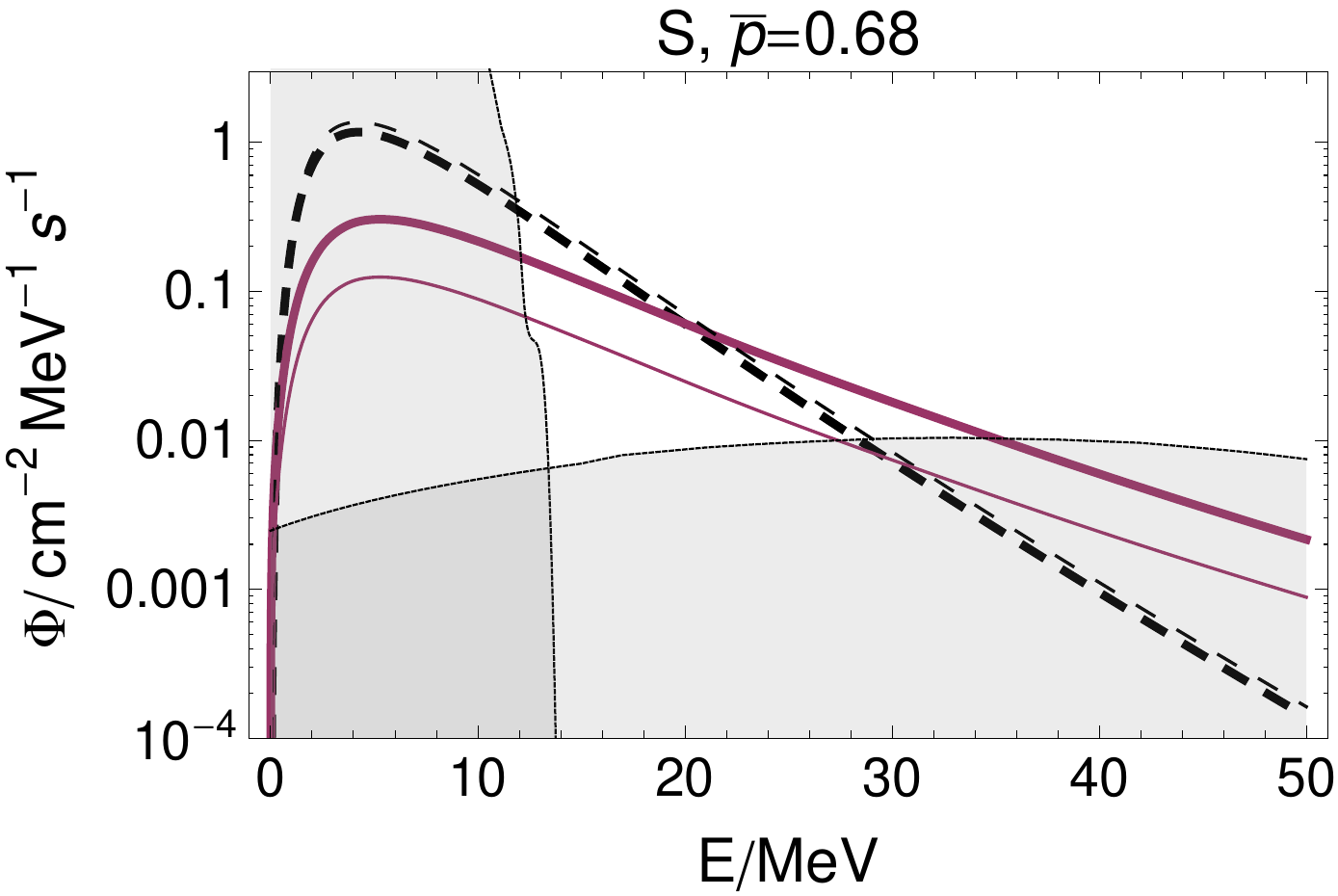}
  \vspace{0.3 truecm}
 \includegraphics[width=0.25\textwidth,height=4 truecm]{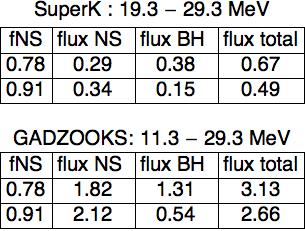}
 \caption{ The diffuse flux of $\barnue$ from neutron star-forming collapses (dashed lines) and from failed supernovae (direct black-hole forming , solid lines).    We use two different equations of state (the S EoS and the LS EoS), the extreme values for the survival probability ($\bar p=0, 0.68$) and two values of the fraction of neutron star-forming collapses: $f_{NS}=0.78$ (thick curves) and $f_{NS}=0.91$ (thin lines).  For each case we give the integrated flux in energy intervals of interest, in units of ${\rm cm^{-2} s^{-1}}$.  }
\label{DiffNueBarFlux}
\end{figure}

Fig. \ref{DiffNueBarFlux} shows the  $\barnue$ fluxes from \nsf\ and \bhf, as well as the corresponding integrated fluxes in the energy windows relevant to pure water and water with Gadolinium. 

As expected, the diffuse fluxes reflect the features of the originally produced fluxes of each \sn\ type.  The diffuse flux from failed \sne\  has a harder spectrum compared to the flux from \nts\ collapses, and is larger for the  S EoS.  Above 20 MeV, $\Phi^{BH}_{\bar e}$ is highest for the S EoS, $f_{NS}=0.78$ and $\bar p=0.68$.  For this ``best case scenario", the contribution from failed \sne\ is dominant in the interval of sensitivity of pure water, reaching $\sim 0.07~{\rm  cm^{-2}s^{-1}MeV^{-1}}$ at 20 MeV and falling almost exponentially at higher energy.
This is the result already highlighted in \cite{Lunardini:2009ya}: the diffuse flux from failed \sne\ could be large and therefore detectable, with interesting implications for the study of direct black hole formation.

Results vary substantially with the parameters, however. An uncertainty of a factor of $\sim 2$ is associated  with the fraction of \bh\ collapses, $1-f_{NS}= 0.09 - 0.22$, and a lower $\Phi^{BH}_{\bar e}$ is expected for the LS EoS and for the total flavor permutation $\bar p=0$.  This latter feature is peculiar of failed \sne\ (the opposite is realized for \nts\ collapses), and is due to the especially luminous and energetic original $\barnue$ flux. When all parameters conspire to suppress it, $\Phi^{BH}_{\bar e}$ is small, exceeding the flux from \nsf\ only above $\sim$45 MeV where the atmospheric background dominates by one order of magnitude. 

Let us now discuss the fluxes integrated over the water detector energy window, $E \sim 19.3 - 29.3$ MeV.
In this interval the failed \sn\ flux varies by about one order of magnitude, from $0.03~{\rm cm^{-2} s^{-1}}$ to $0.38~{\rm cm^{-2} s^{-1}}$, corresponding to $\sim 6 - 57\%$ of the total flux.  The latter is enhanced by up to  a factor $\sim 2.3$ compared to \nts\ collapses only, and can be as high as $0.67~{\rm cm^{-2} s^{-1}}$. 
The flux in the open interval $E \geq 19.3 $ MeV can reach $0.89~{\rm cm^{-2} s^{-1}}$, about a factor of two away from the most conservative  \sk\ limit, eq. (\ref{sklim}).  This indicates the exciting possibility of detection in the near future at \sk\ or GADZOOKS, before the advent of more massive detectors.  

The \sk\ limit can be used to constrain the space of parameters \cite{Lien:2010yb}. 
 For example, if we compare this limit to the total predicted flux, $\Phi^{BH}_{\bar e}+\Phi^{NS}_{\bar e}$, we get a constraint  on the normalization of the \sn\ rate, $R_{cc}(0)$. 
If the most conservative limit is used and all  parameters are fixed to the most optimistic scenario (the largest  $\Phi^{BH}_{\bar e}$), one gets $R_{cc}(0) < 2.1 \cdot 10^{-4}~{\rm yr^{-1}Mpc^{-3}}$.  Alternatively, one can fix $R_{cc}(0) =  10^{-4}~{\rm yr^{-1}Mpc^{-3}}$, and obtain a  constraint on the fraction of  failed \sne: $1-f_{NS} \lta 0.7  $, for the same set of the remaining parameters.   
In general, the \sk\ limit excludes a small portion of the parameter space, the one where $\Phi^{BH}_{\bar e}$ is largest. Constraints on single quantities are loose due to degeneracies.  In the future the neutrino constraints will become more powerful, when degeneracies and uncertainties are reduced by independent measurements.  These could be precision tests of the \sn\ rate from astronomy \cite{snap,snls} or more precise \nsf\ fluxes from a galactic \sn. 

For the larger energy window of water with Gadolinium, $E\sim 11.3 - 29.3$ MeV, results are similar, as shown in fig. \ref{DiffNueBarFlux}.  We note that for the purpose of studying \ns\ from failed \sne, the lowering of the energy threshold is not as crucial as it is for  the \nsf\ flux, since $\Phi^{BH}_{\bar e}$ emerges above $\Phi^{NS}_{\bar e}$ only above 20 MeV or higher.   
Still, the lower threshold would be important to be able to reconstruct \emph{ both} $\Phi^{BH}_{\bar e}$ and $\Phi^{NS}_{\bar e}$ individually: a fit to the lower energy data could be sufficient to reconstruct $\Phi^{NS}_{\bar e}$, so that the higher energy data could be used to distinguish $\Phi^{BH}_{\bar e}$. 
Besides the lower threshold, water with Gadolinium has the advantage of reducing the background in the energy window, so it can establish the \df\ with much better significance, as will be seen in the next section. 

\subsection{Events in water}

\begin{figure}[htbp]
\centering
 \includegraphics[width=0.39\textwidth]{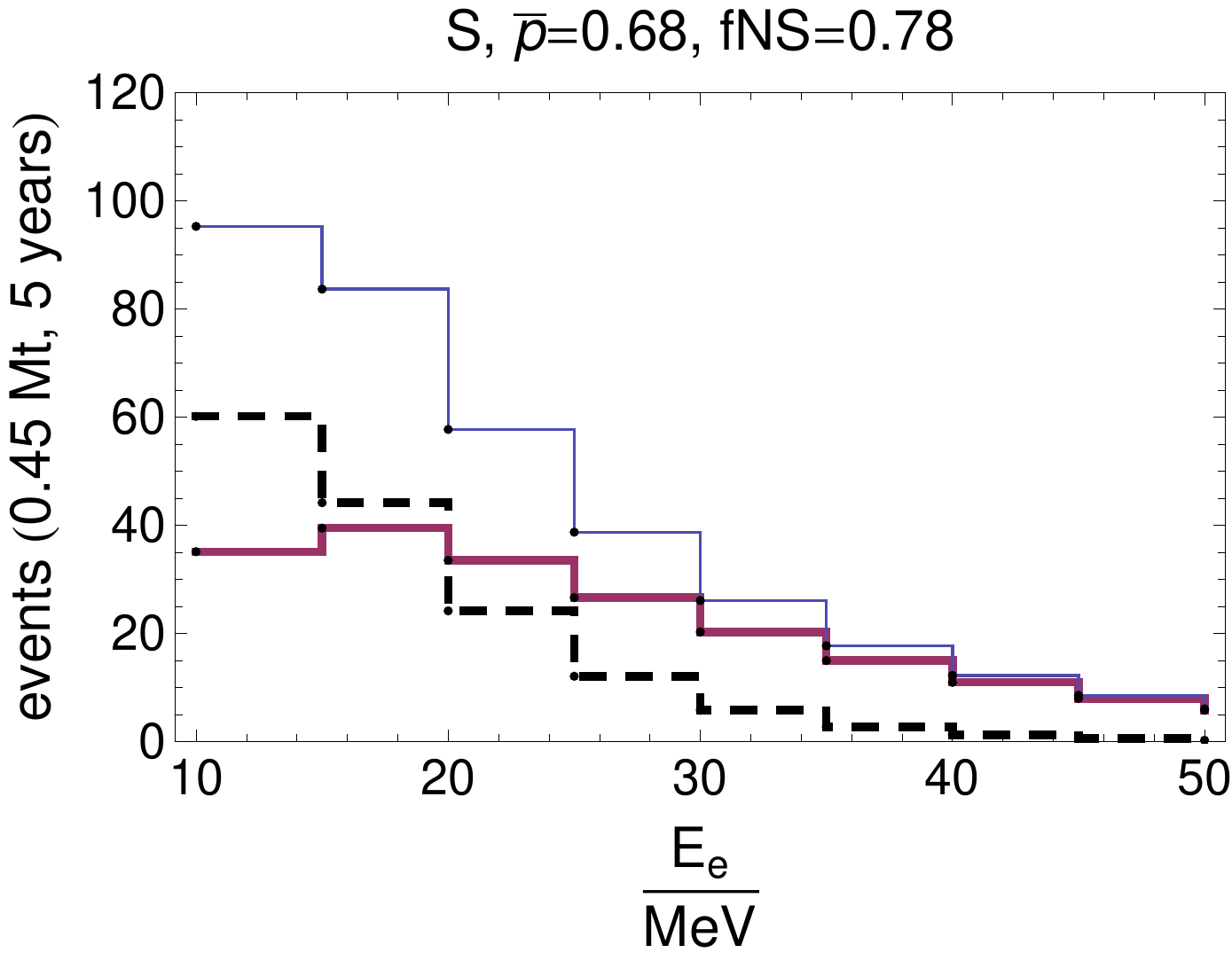}
 \includegraphics[width=0.39\textwidth]{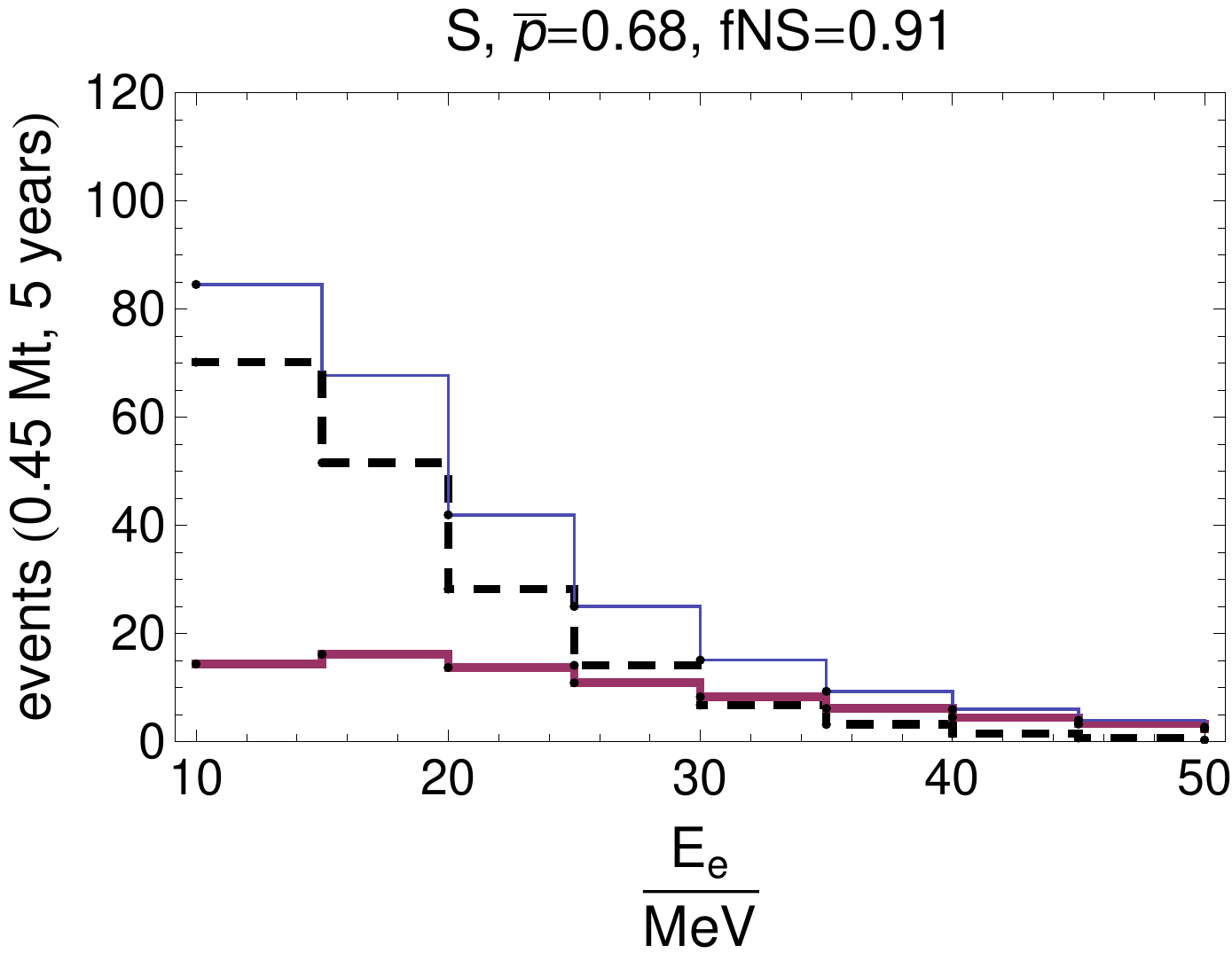}
 \includegraphics[width=0.39\textwidth]{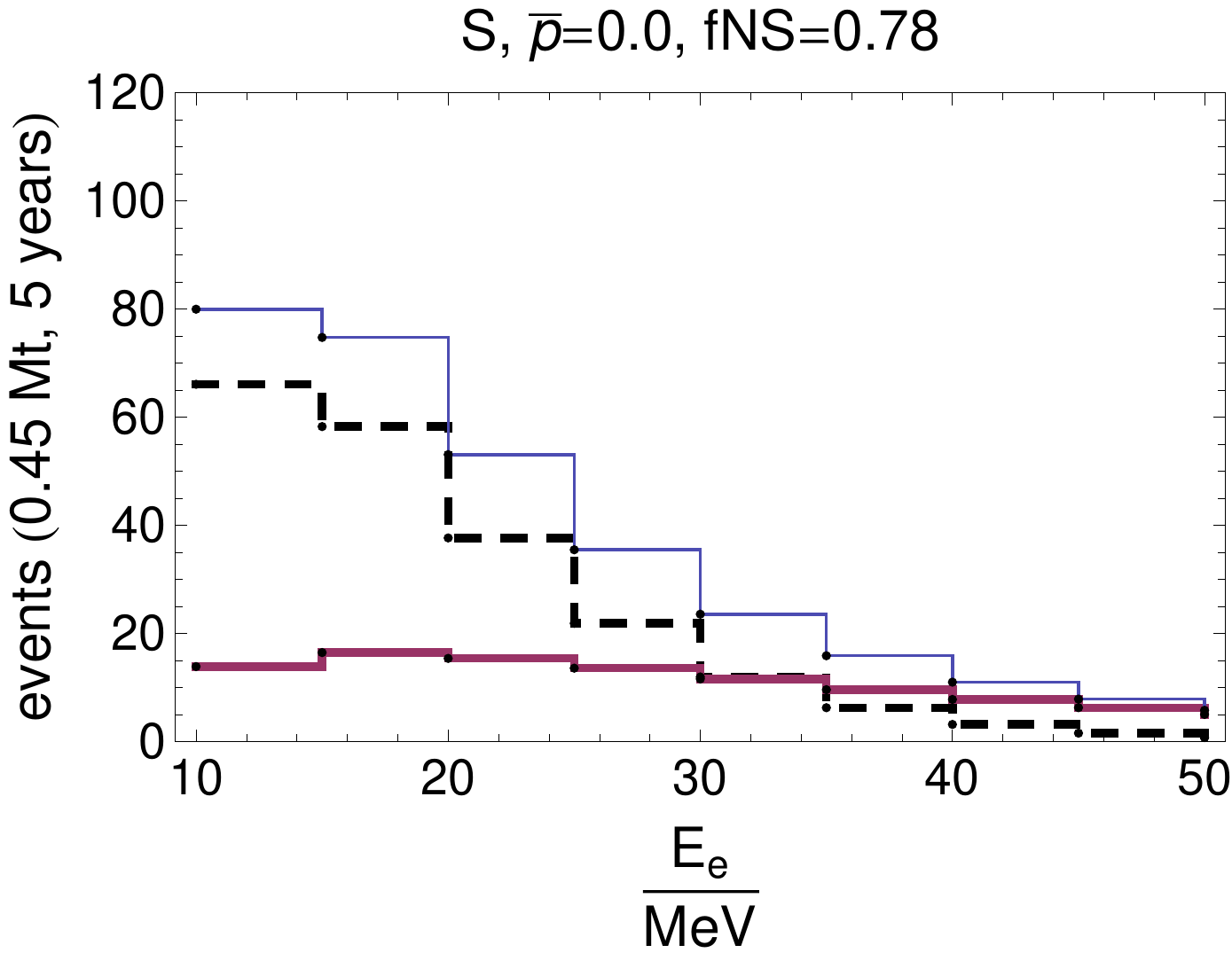}
 \includegraphics[width=0.39\textwidth]{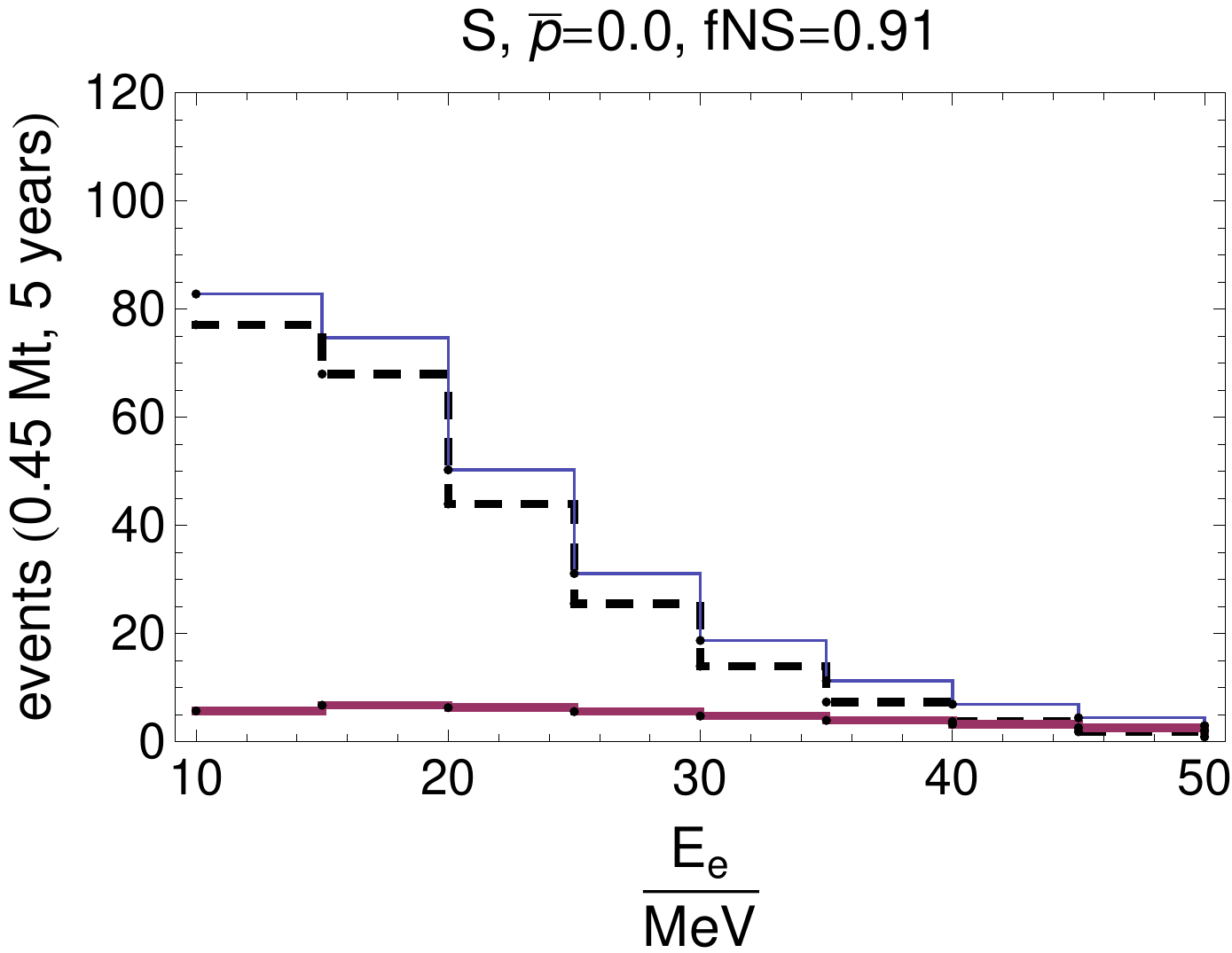}
 \caption{The number of inverse beta decay events in a water \ck\ detector with exposure 2.25 Mt$\cdot$yr. Histograms are shown, with the S EoS, for  direct \bh\ collapses only (solid thick, purple), \nts\  collapses only (dashed, black), and total (solid, thin, blue).
 }
\label{nuebareventplotS}
\end{figure}

\begin{figure}[htbp]
\centering
 \includegraphics[width=0.39\textwidth]{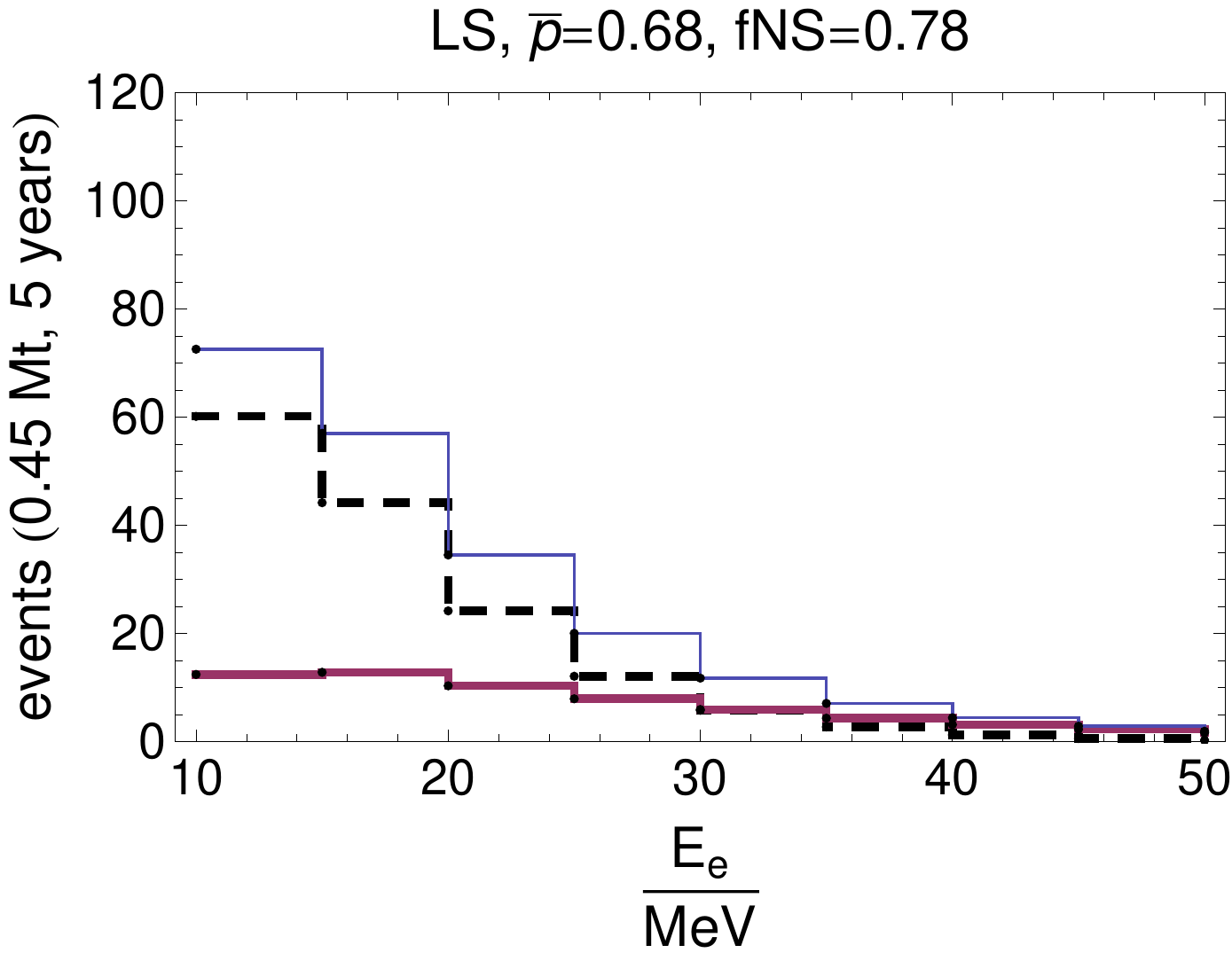}
 \includegraphics[width=0.39\textwidth]{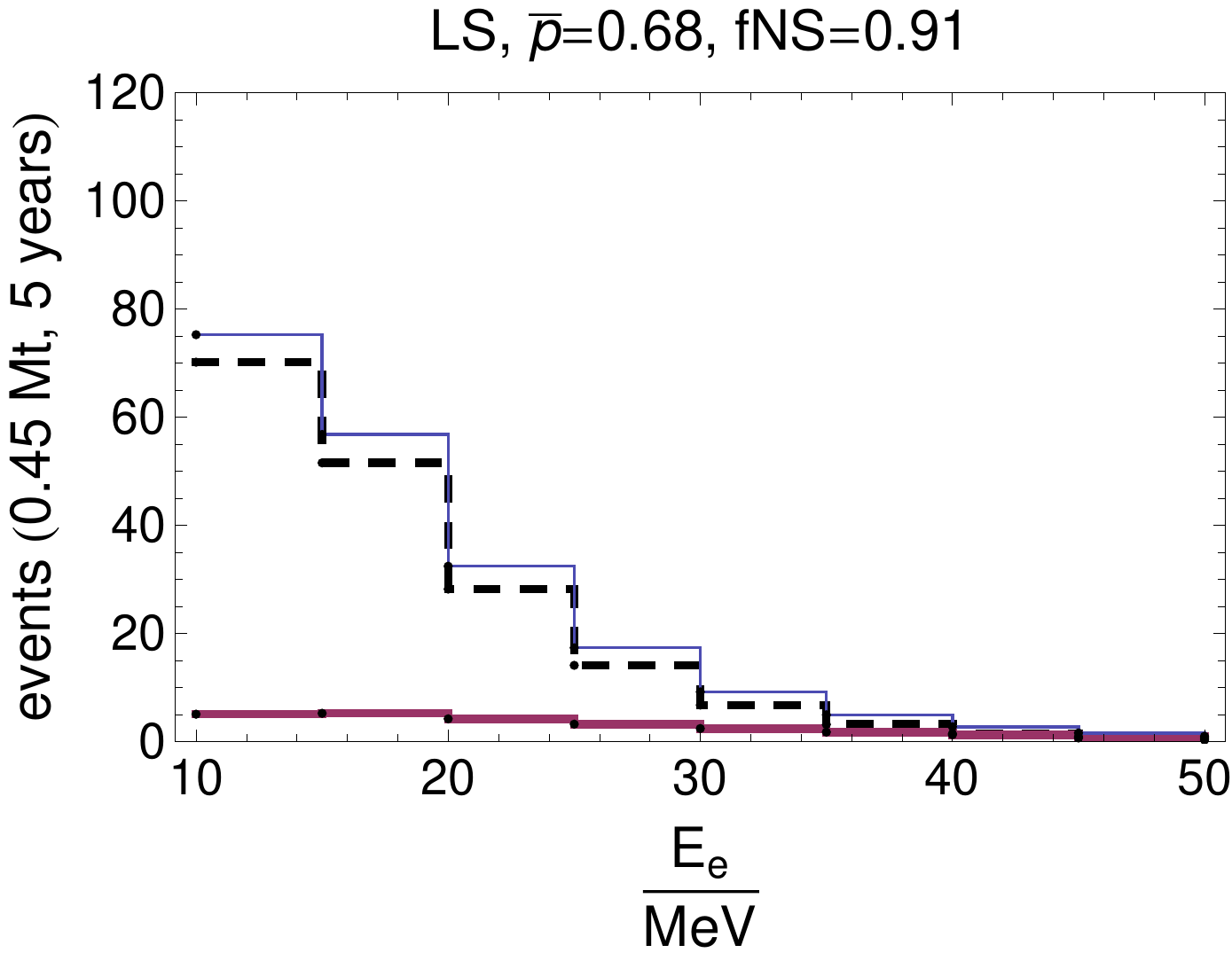}
 \includegraphics[width=0.39\textwidth]{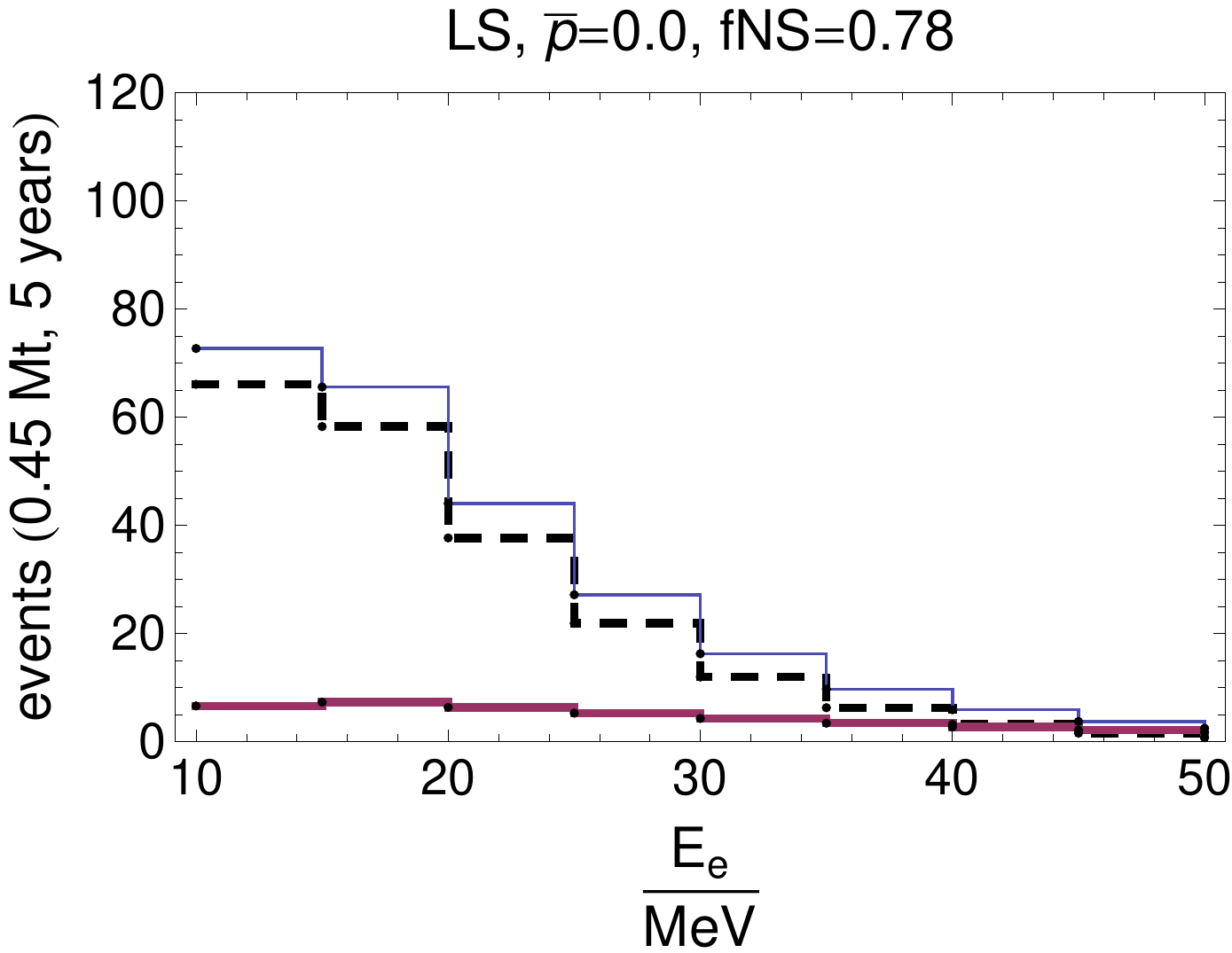}
 \includegraphics[width=0.39\textwidth]{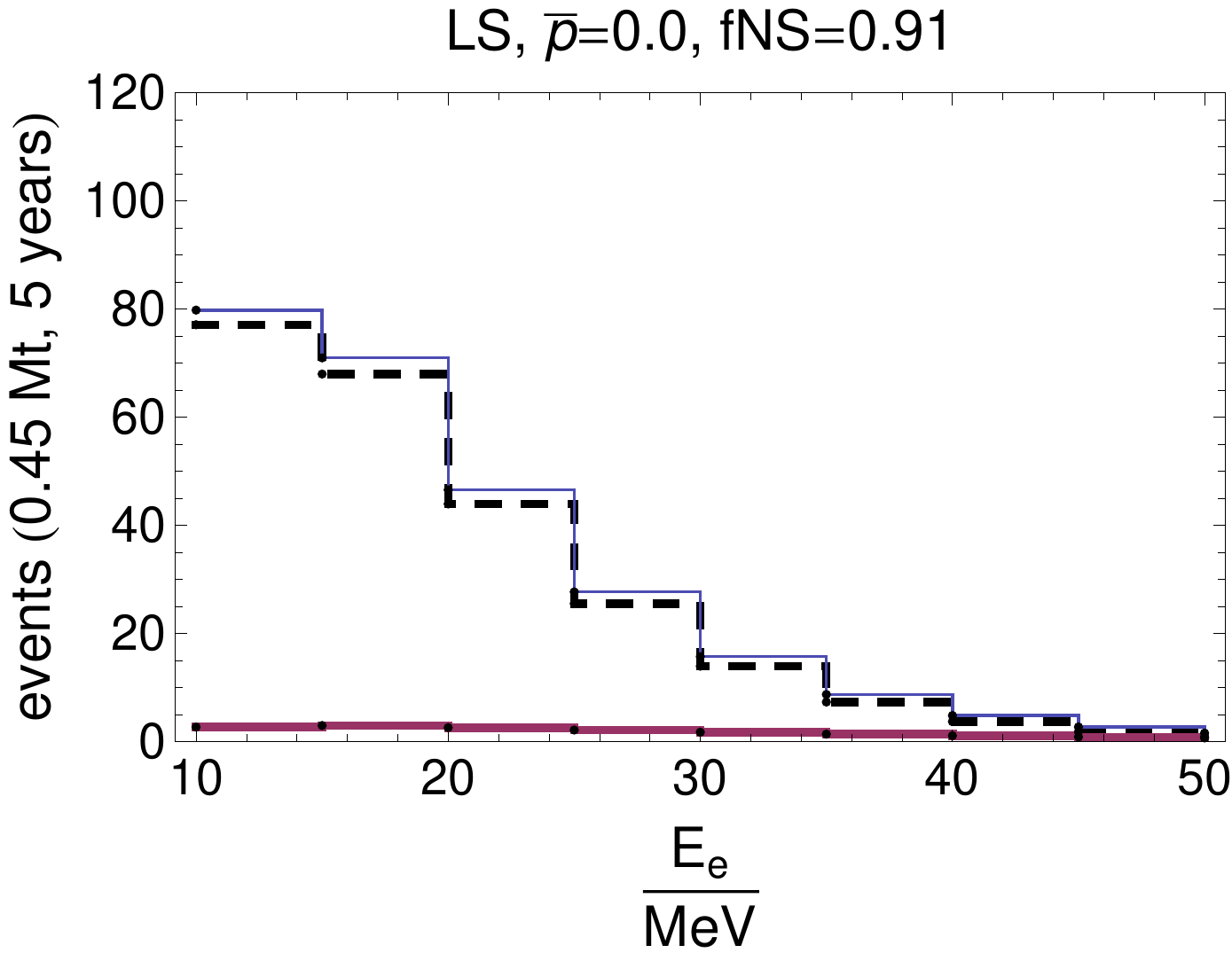}
 \caption{The same as fig. \ref{nuebareventplotS}, for the LS EoS. }
\label{nuebareventplotLS}
\end{figure}

Figs. \ref{nuebareventplotS} and \ref{nuebareventplotLS} show the spectrum of events in 5 MeV bins of positron energy due to \bh\ collapses, \nts\ collapses, and the total of the two.  As expected from the $\barnue$ fluxes (fig. \ref{DiffNueBarFlux}), the S EoS gives the largest event rate, reaching $\sim 40$ events per bin.   The events due to failed \sne\ could exceed those from \nsf\ in the most fortunate case  (maximum $\barnue$ survival and larger $f_{NS}$).  In general, for the  S EoS the contribution of failed \sne\ is at least $\sim 20\%$ (10\%) of the total in the $25-30$ (20-25) MeV bin, and could easily be at the level of  50\% or so depending on the parameters.
For the LS EoS the effect of failed \sne\ is more modest, but still reaches $\sim 50\%$ in the $25-30$ MeV bin, sufficient to conclude that \ns\ from \bhf\ can not be neglected, in general. 

Note that, thanks to their more energetic spectrum, the spectrum of events  from \bh\ collapses peaks in the 15-20 MeV bin, while events from \nts\ collapses  have their maximum around or below 10 MeV. Therefore, even a modest lowering of the energy threshold would allow to capture most of the events due to \bhf. The expected new threshold for \sk,  $E_e \simeq 16$ MeV \cite{smytalk,iidathesis}, would already be sufficient.

\begin{table}[htbp]
\begin{center}
\begin{tabular}{| c || c |c |c || c| c |c || c |c |}
\hline
\hline
 & \multicolumn{3 }{| c |}{  $\bar p=0.68$ } & \multicolumn{3 }{| c |}{ $\bar p=0$  }  &  &  \\
\cline{2-7}
  & NSFCs  & DBHFCs & Total  & NSFCs  & DBHFCs & Total   & atm.  & inv. muon  \\
\hline
 $18 < E_e/{\rm MeV} <28$     &  47.3   & 65.6   &   112.9   &  73.1   & 30.4   & 103.5    &   76 & 633 \\
   &  (55.2)  & (26.8)  &  (82.0)   &  (85.3)  & (12.4)  &  (97.7)    &    &   \\
 \hline
$18 <  E_e/{\rm MeV} <38$     &  58.86    &  105.34  &   164.20   &  96.62  & 53.16  &  149.78   &   268  & 1675   \\
   &  (68.68)  & (43.10)  &  (111.78)   &  (112.72)  & (21.75)  &  (134.47)    &    &   \\
 \hline
 $10 < E_e /{\rm MeV} <38$    &  148.33   &  164.67  &  313.00    &  200.17   &  76.97  & 277.14   &  283 &  1893  \\
   &  (173.05)  & (67.36)  &  (240.41)   &  (233.53)  & (31.49)  &  (265.02)    &    &   \\
\hline
\hline
\end{tabular}
\caption{The number of signal and background events (atmospheric \ns\  and invisible muons) in a water detector of exposure 2.25 Mt$\cdot$yr,  in three energy windows of interest (given in terms of the positron energy, $E_e$), for the S EoS.   The numbers of signal events are given for $\bar p=0,0.68$, $f_{NS}=0.78$ and $f_{NS}=0.91$ (the latter in parentheses in the table).  } 
\label{ratesmodelWa}
\end{center}
\end{table}

\begin{table}[htbp]
\begin{center}
\begin{tabular}{| c || c |c |c || c| c |c || c |c |}
\hline
\hline
 & \multicolumn{3 }{| c |}{  $\bar p=0.68$ } & \multicolumn{3 }{| c |}{ $\bar p=0$  }  &  &  \\
\cline{2-7}
  & NSFCs  & DBHFCs & Total  & NSFCs  & DBHFCs & Total   & atm.  & inv. muon  \\
\hline
  $18 < {\rm E_e/MeV} <28$     &  47.3   &  20.2  &  67.5    &   73.1  & 12.5   &  85.6   &  76  & 633 \\
   &  (55.2)  & (8.3)  &  (63.5)   &  (85.3)  & (5.1)  &   (90.4)   &    &   \\
 \hline
$18 < {\rm E_e/MeV} <38$     &  58.86    &  31.87  &   90.73   &   96.62  &  20.93 &  117.55   &   268  & 1675   \\
   &  (68.68)  & (13.04)  &  (81.72)   &  (112.72)  & (8.56)  &  (121.28)    &    &   \\
 \hline
 $10 < {\rm E_e/MeV} <38$    & 148.33    &  52.26  &   200.59   &  200.17   &  32.01  &  232.18  &  283 &  1893  \\
   &  (173.05)  & (21.38)  &  (194.43)   &  (233.53)  & (13.10)  &  (246.63)    &    &   \\
\hline
\hline
\end{tabular}
\caption{The same as Table \ref{ratesmodelWa}, for the LS EoS. } 
\label{ratesmodelWaLS}
\end{center}
\end{table}

Tables  \ref{ratesmodelWa} and \ref{ratesmodelWaLS}  give the numbers of background and signal events in water, for the S EoS and the LS EoS respectively, in different intervals of positron energy of experimental interest \footnote{For generality, all the background event rates refer to pure water, and do not include the expected reduction of the invisible muon events allowed by water+Gadolinium, which depends on the specific detector design and therefore is uncertain. As discussed in sec. \ref{detection}, a reduction of a factor of $\sim 5$ is expected. }. 
The total event rates exhibit the features already observed for the fluxes and for the positron energy distributions, mainly the enhancement of the signal by a factor of between $\sim 10\%$ and more than 2, compared to \nsf\ only. 

From the Tables one can estimate the statistical significance of the signal over the background. 
Following \cite{Fogli:2004ff}, we calculate the statistical error using the signal and background rates: 
\be
\sigma = \sqrt{N_{sig} + N_{bckg}}~. 
\label{stateq}
\ee
and compare it with the number of events from core collapse.   Note that $\sigma$ is dominated by the high number of invisible muon events, and therefore is much larger than the statistical error due to the signal only \footnote{The systematic uncertainties on the background rates are probably minor compared to the purely statistical errors.  Indeed, the atmospheric \n\ flux  has about 20\%-40\% systematic uncertainty \cite{Battistoni:2005pd}, which translates into a $\sim $10-20\% effect on $\sigma$.  Moreover, the normalization of the invisible muon event rate is known from the analysis of the SuperKamiokande data, and therefore is constrained within tens of per cent uncertainty \cite{Malek:2002ns}.  }.  
 For the 18-28 MeV energy window, we find that the signal has $\simeq 2.3-3.9 \sigma$ significance, resulting in the possibility to claim detection (in absence of systematic errors) for part of the parameter space.  For \nsf\ only, the significance would be $\simeq 1.7-3\sigma$; the comparison is indicative of how the contribution of \bhf\ improves the chances of detection of the \df.   
 
  If the flux from \nsf\ was known precisely, one could imagine analyzing the data to establish the flux from \bhf\  as a signal of its own. The number of events from failed \sne\ would not be a statistically significant excess for the exposure considered here ($\sim 2.2 \sigma$ at most), but would reach $3\sigma$ with about a double exposure, for the most optimistic parameters.
  
 Using a larger window, e.g. 18-38 MeV, decreases the signal to background ratio, thus decreasing the statistical significance of the \df\ data slightly.  

In the energy window $E \simeq 10-38$ MeV -- which is viable if spallation is subtracted --  the signal-to-background ratio is larger, resulting in a higher signal significance.  The events due to core collapse represent a 4 - 6.3$\sigma$ excess ($3\sigma$ or higher for \nts\ collapses only) over the 2176 background events in water.  A significance as high as 7.5$\sigma$ is reached for the optimized window of 10-28 MeV. 

With the  reduction of the invisible muon background by a factor of $\sim 5$, expected with Gd addition, the highest expected event rate due to \bhf\ would be significant by $\sim 5$ $\sigma$ above the total due to the background and the flux from \nsf, if the latter is assumed as known. However, for other flux parameters the contribution of direct \bh\ collapses would be below $3\sigma$ and require up to three times the exposure to become significant.
Our estimates on this are conservative, because they do not consider the potential of a full statistical, bin-by-bin, data analysis. In any case, 
every conclusion about significance is only indicative, due to the large uncertainty on the normalization of the \df\ relative to the backgrounds.   

\section{Neutrinos: flux and event rates in liquid argon}
\label{nu}

\subsection{$\nue$ flux}

\begin{figure}[htbp]
  \centering
 \includegraphics[width=0.39\textwidth]{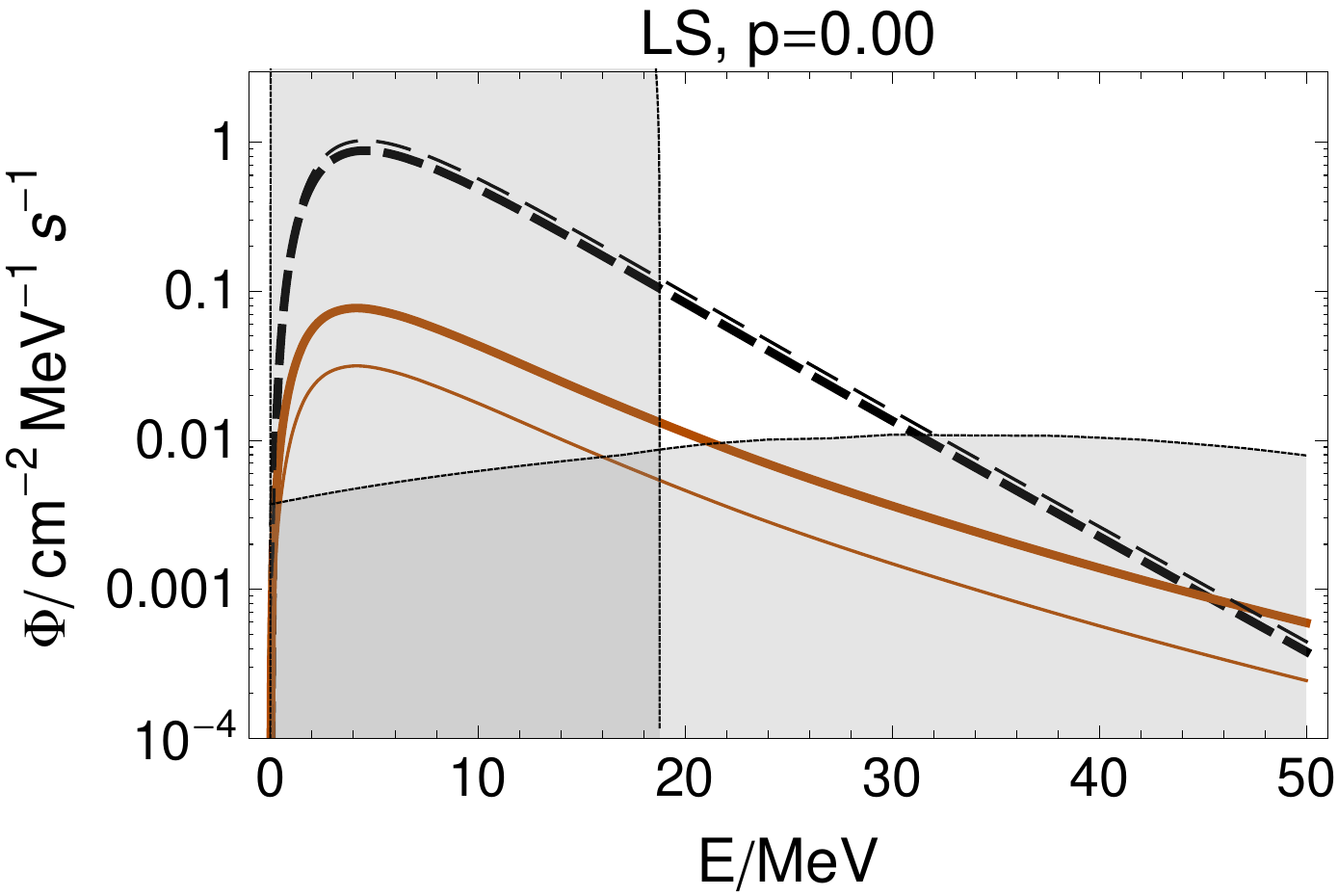}
   \vspace{0.3 truecm}
 \includegraphics[width=0.25\textwidth,height=2 truecm]{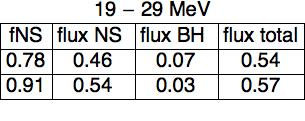}
  \vspace{0.3 truecm}
 \includegraphics[width=0.39\textwidth]{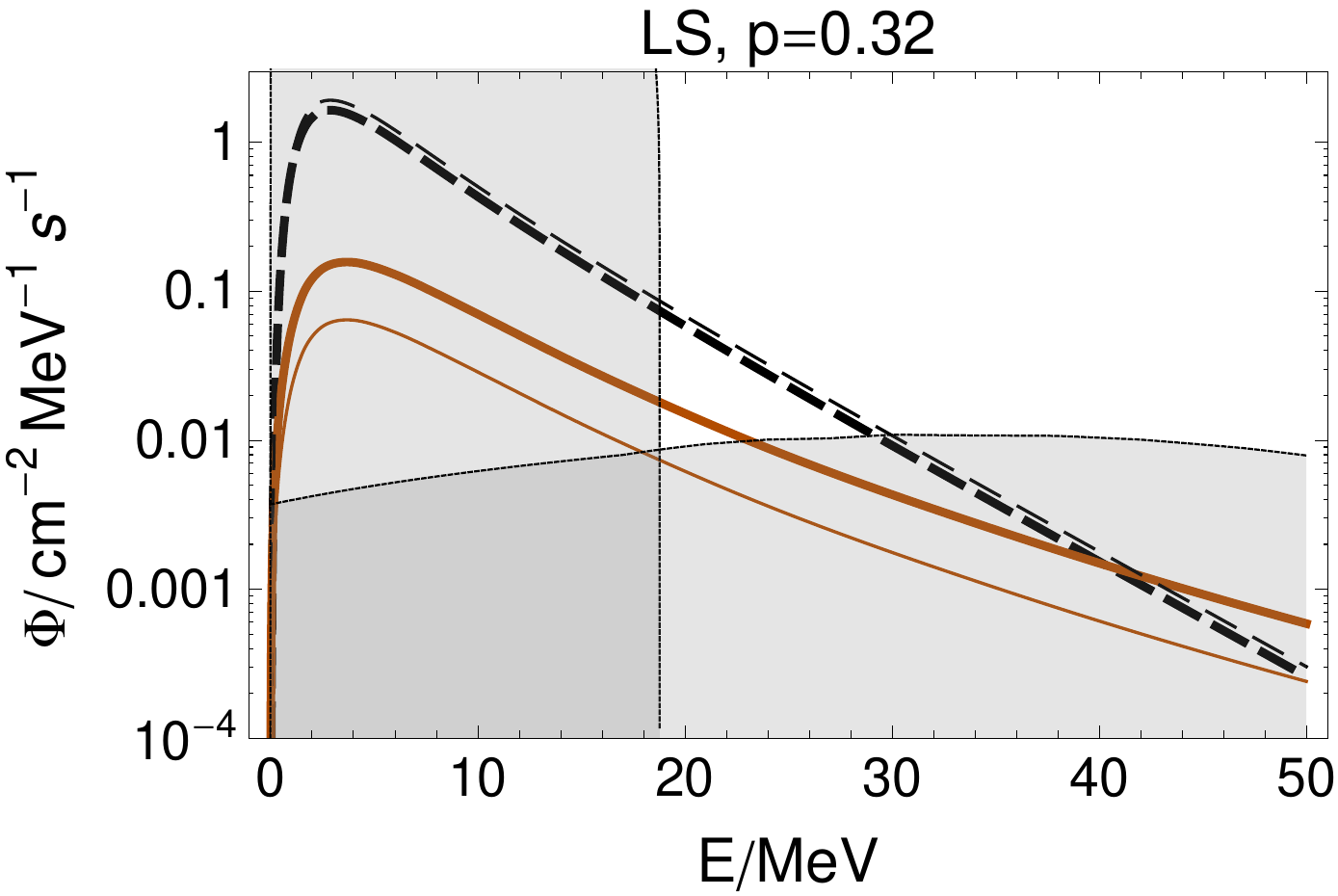}
   \vspace{0.3 truecm}
 \includegraphics[width=0.25\textwidth,height=2 truecm]{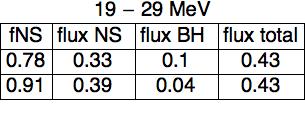}
  \vspace{0.3 truecm}
 \includegraphics[width=0.39\textwidth]{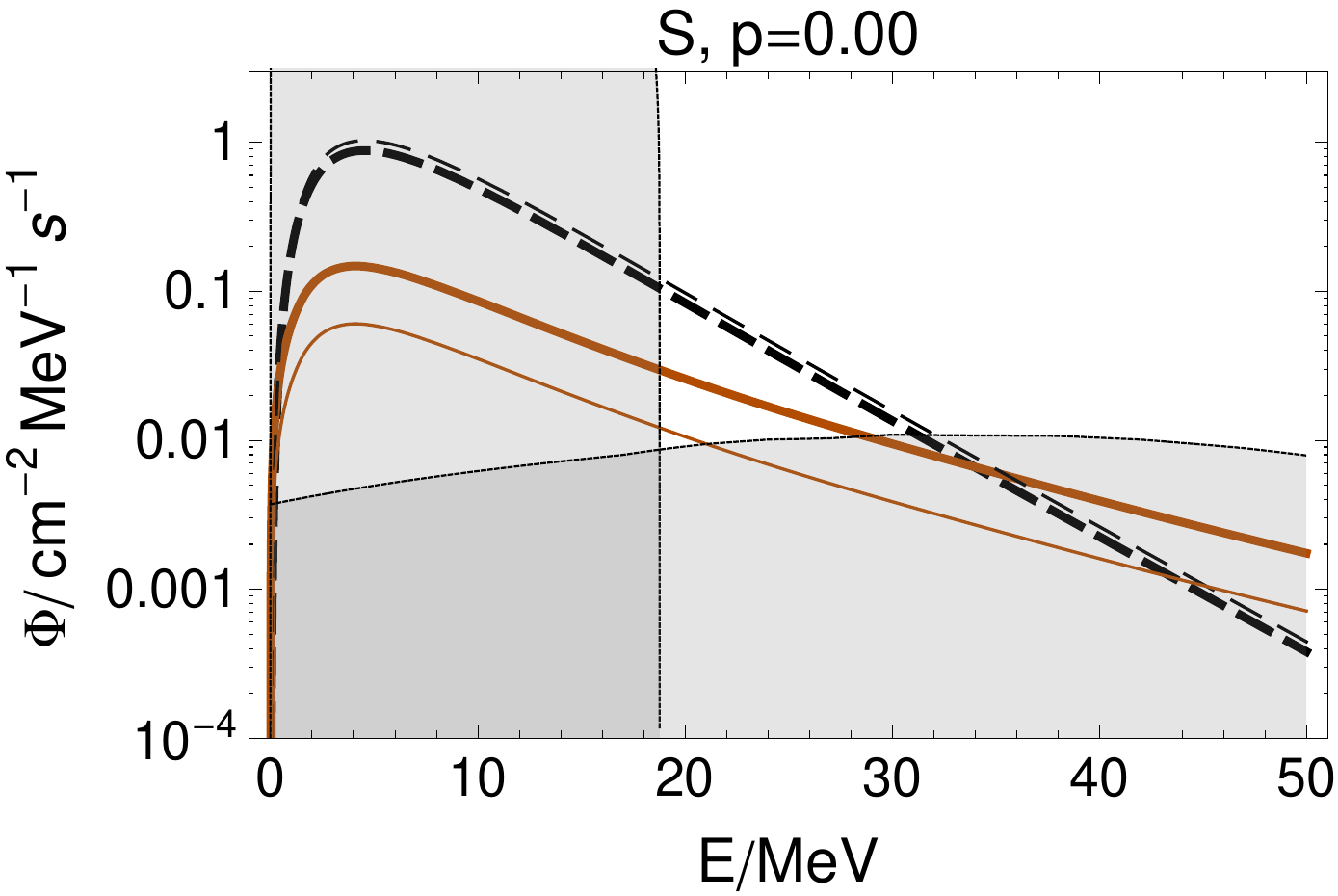}
    \vspace{0.3 truecm}
 \includegraphics[width=0.25\textwidth,height=2 truecm]{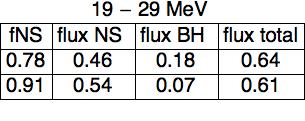}
  \vspace{0.3 truecm}
 \includegraphics[width=0.39\textwidth]{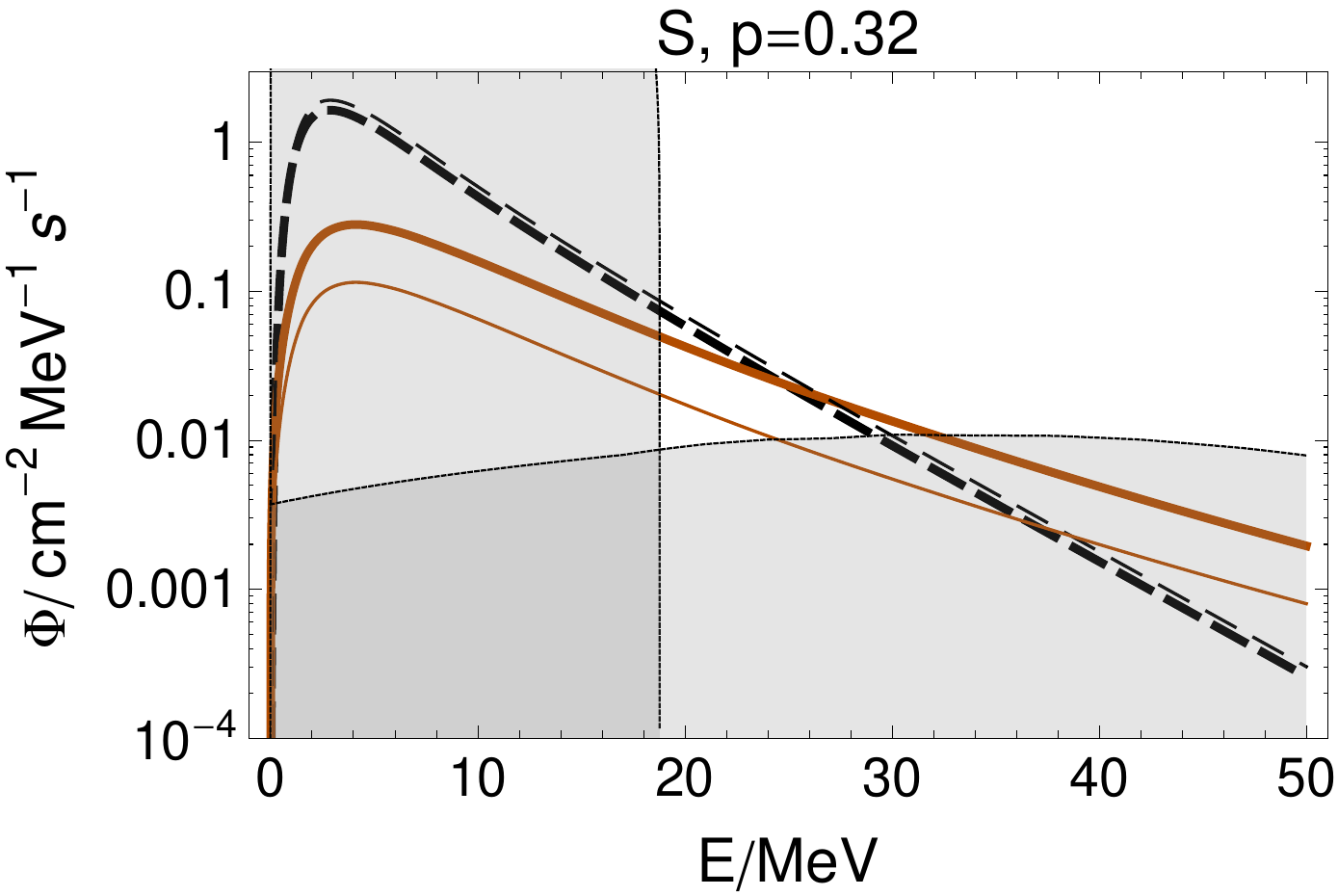}
    \vspace{0.3 truecm}
 \includegraphics[width=0.25\textwidth,height=2 truecm]{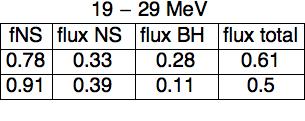}
 \caption{ The diffuse flux of $\nue$ from neutron star-forming collapses (dashed lines) and from failed supernovae (direct black-hole forming , solid lines).    We use two different equations of state (the S EoS and the LS EoS), the extreme values for the survival probability ($ p=0, 0.32$) and two values of the fraction of neutron star-forming collapses: $f_{NS}=0.78$ (thick lines) and $f_{NS}=0.91$ (thin).  For each case, we give the integrated flux in the energy interval of interest, in units of ${\rm cm^{-2} s^{-1}}$. }
\label{DiffNueFlux}
\end{figure}

Let us now discuss the $\nue$ components of the diffuse fluxes from \bhf\ and \nsf\ ($\Phi^{BH}_e$ and $\Phi^{NS}_e$). 
These are shown in fig. \ref{DiffNueFlux}, for several sets of parameters, together with the solar and atmospheric $\nue$ fluxes for comparison.   

Overall, the $\nue$ fluxes are similar to the $\barnue$ ones.
In particular,  $\Phi_e = \Phi_{\bar e}$  if there is complete flavor permutation in both channels  ($p=\bar p=0$), due to the equality of the original non-electron neutrino and antineutrino fluxes for both \nsf\ and \bhf.   If $p=\bar p \neq 0$, the $\nue$ and $\barnue$ components of the \bhf\ flux are still nearly identical, reflecting the strong similarity of the $\nue$ and  $\barnue$ fluxes at the production point  (see fig. \ref{spectra}).   
In general, however, the amounts of permutation in the $\nue$ and $\barnue$ channels are expected to be different, resulting in differences between the corresponding diffuse fluxes.  
As in the $\barnue$ channel, the largest $\Phi^{BH}_e$  -- in the energy window -- is realized for the maximum $p$, the S EoS and $f_{NS}=0.78$.  
Due to the stronger limit on the $\nue$ survival probability, $p \lta 0.32$,  we have that $\Phi^{BH}_e \lta \Phi^{BH}_{\bar e}$.  Still, for the most optimistic parameters $\Phi^{BH}_e$ dominates the total flux above $\sim 26$ MeV, where the signal is above the atmospheric background.  For all other cases, $\Phi^{BH}_e < \Phi^{NS}_e$ in the energy window. 

Looking at the fluxes integrated in the 19-29 MeV interval (fig. \ref{DiffNueFlux}, right), we see that the flux from failed \sne\ varies between $\sim 7\%$ and $\sim 50\%$ of the total flux. So,  the $\nue$ component of the \df\ could be doubled by the contribution of \bhf\ -- reaching the value of $\sim 0.64~{\rm s^{-1 } cm^{-2}}$ -- or be only moderately affected by it, depending on the parameters.

\subsection{Events in liquid argon}

\begin{figure}[htbp]
  \centering
 \includegraphics[width=0.39\textwidth]{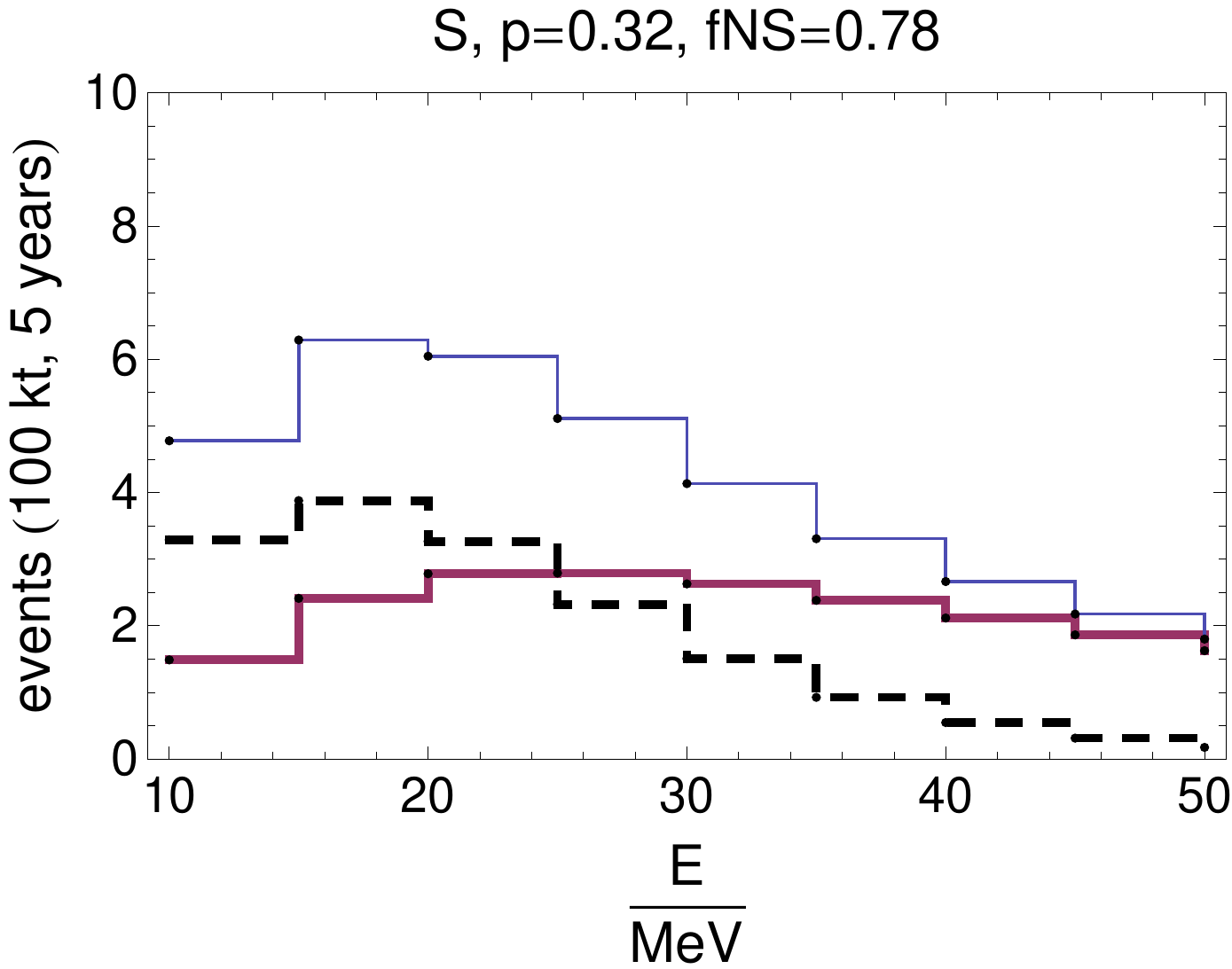}
 \includegraphics[width=0.39\textwidth]{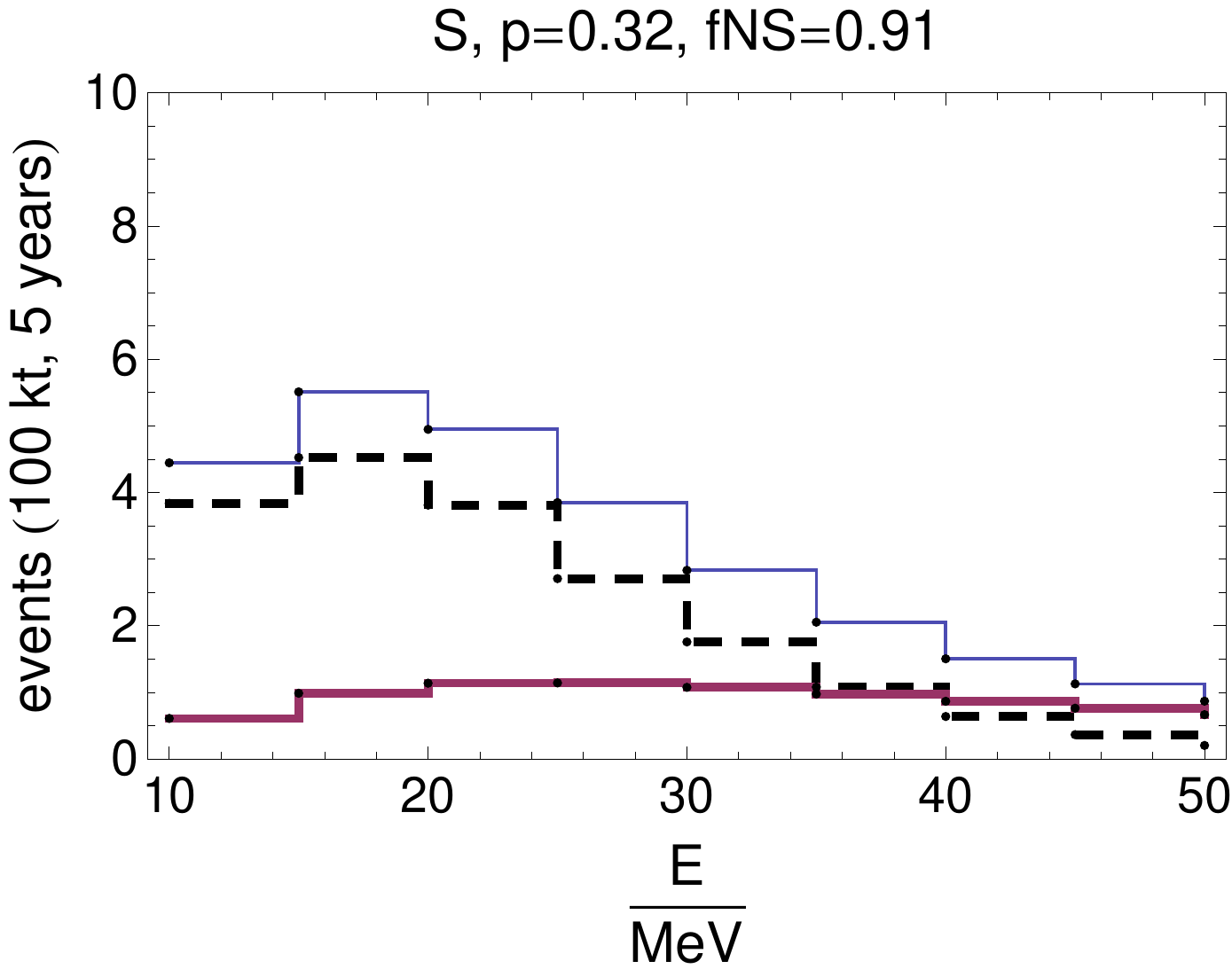}
 \includegraphics[width=0.39\textwidth]{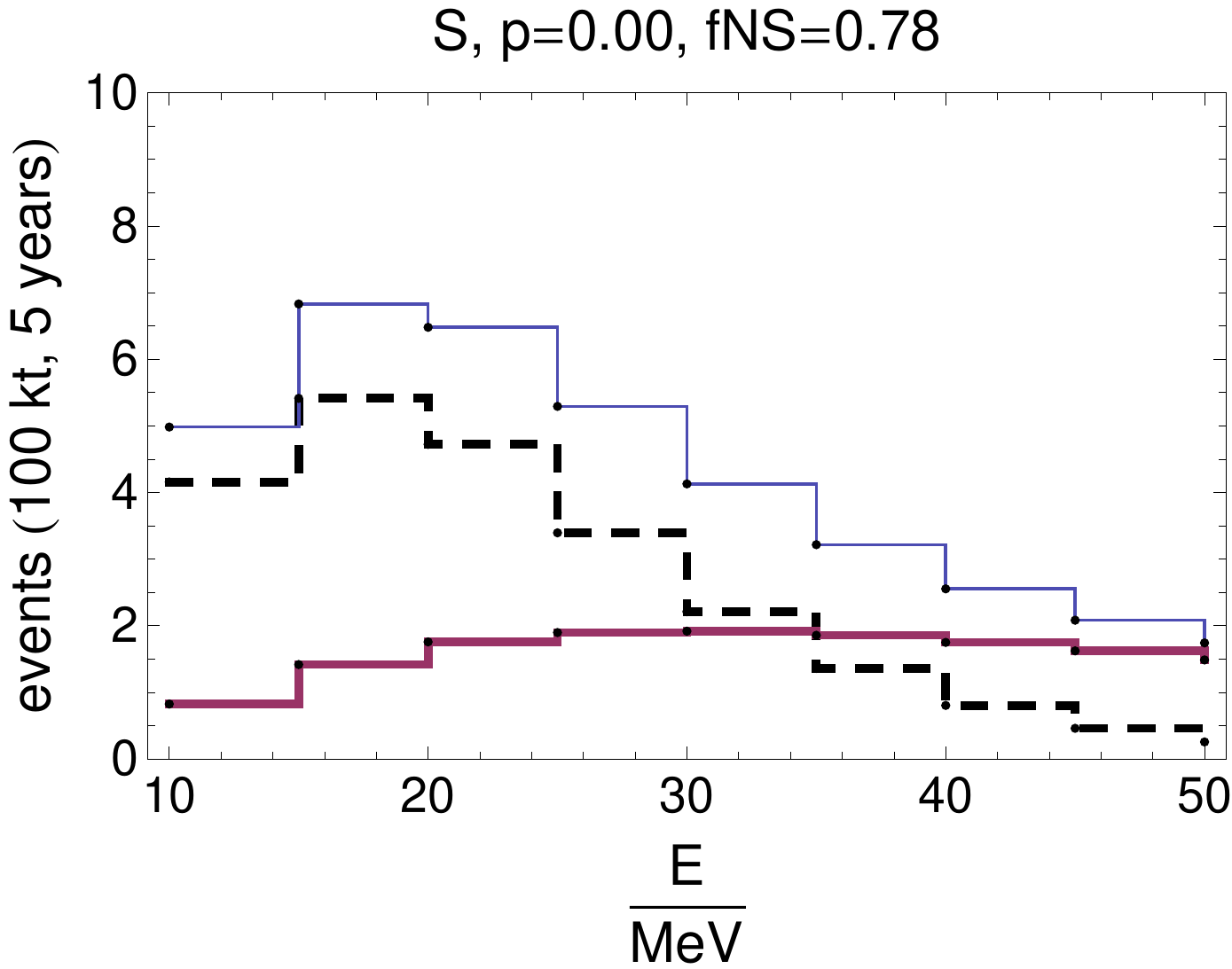}
 \includegraphics[width=0.39\textwidth]{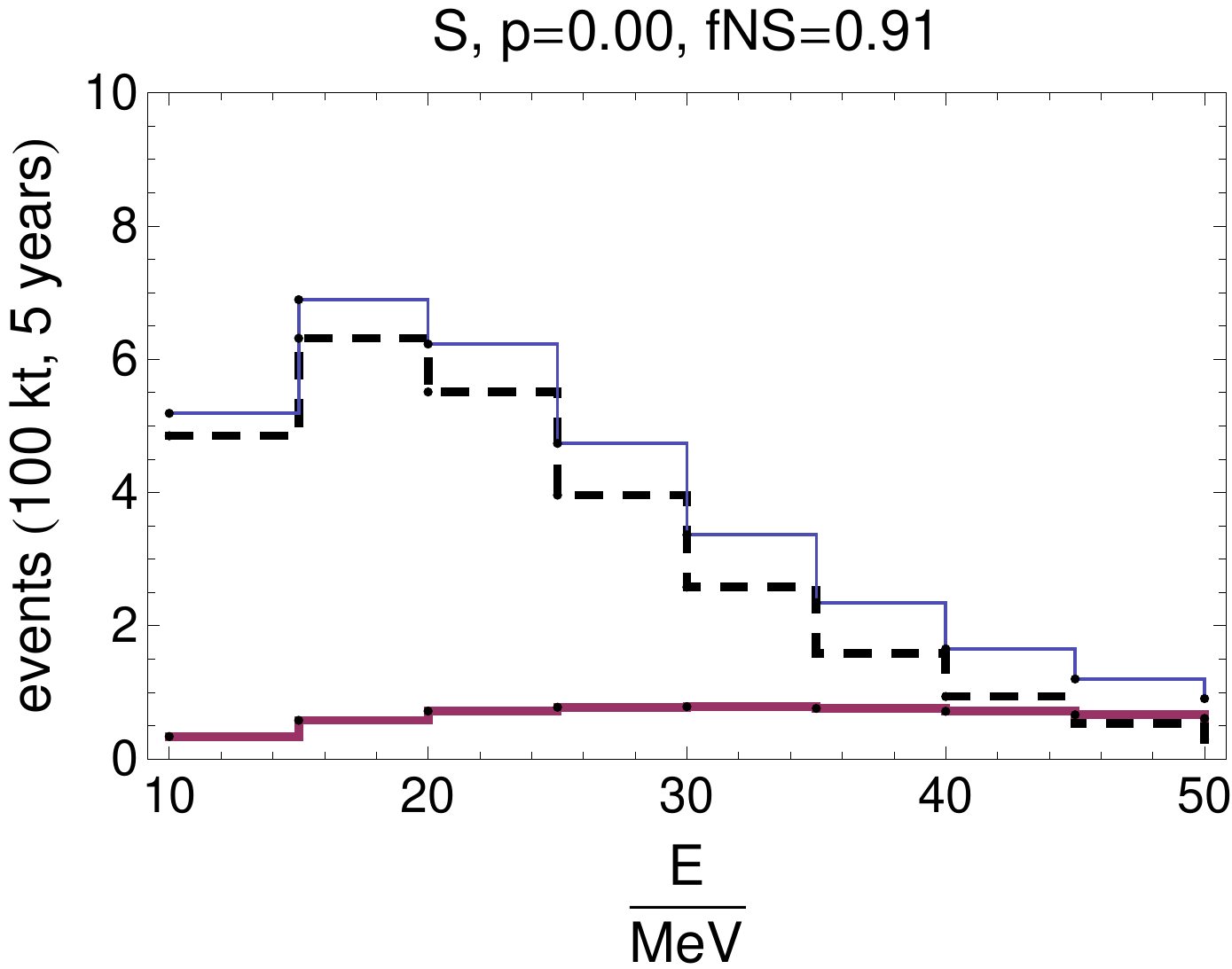}
 \caption{The number of $\nue$ events in a liquid argon  detector with exposure 0.5 Mt$\cdot$yr, for different sets of parameters. Histograms are shown,using the S EoS, for  direct \bh\ collapses only  (solid thick, purple), \nts\  collapses only (dashed, black), and total (solid, thin, blue). }
\label{nueeventplotS}
\end{figure}

\begin{figure}[htbp]
  \centering
 \includegraphics[width=0.39\textwidth]{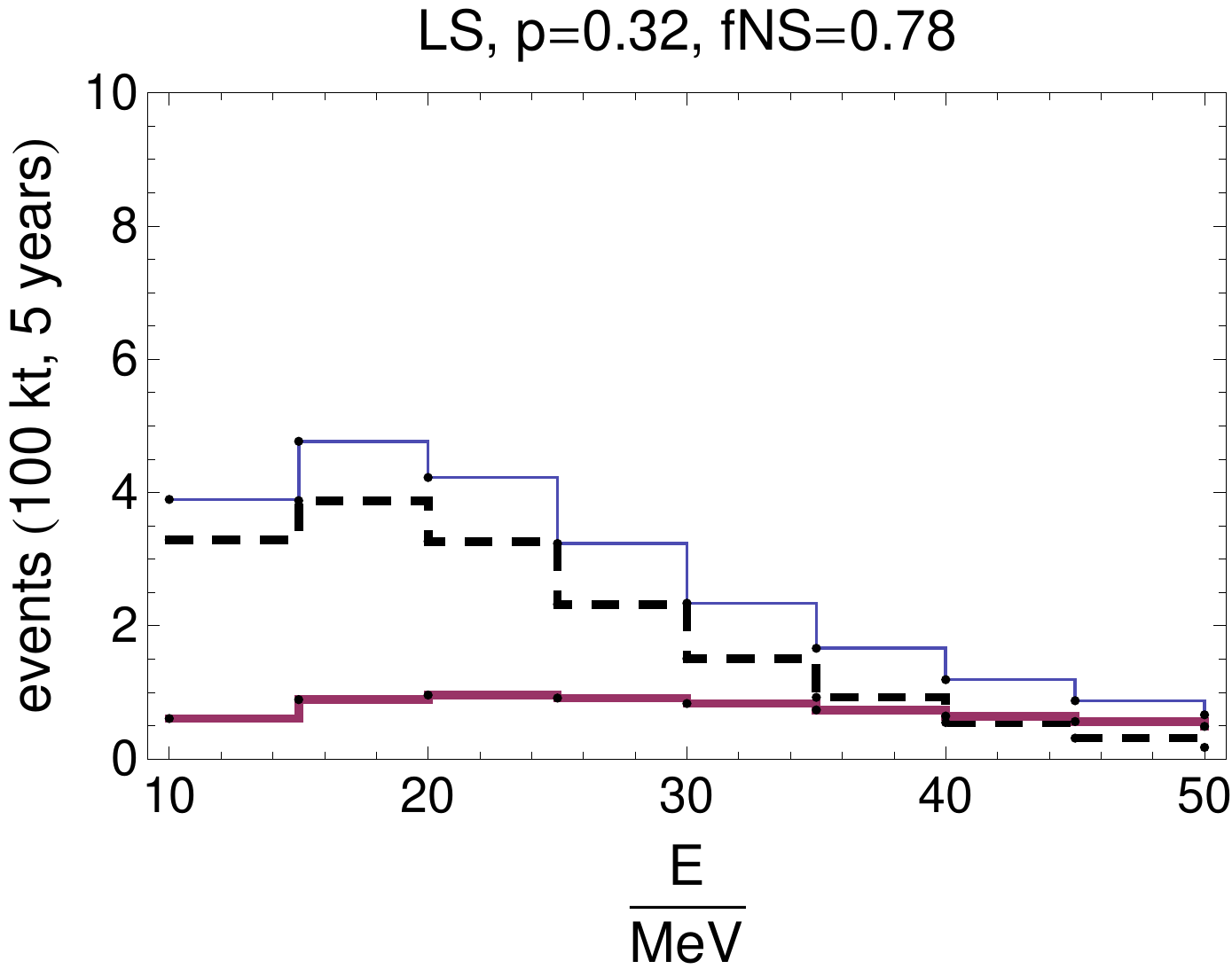}
 \includegraphics[width=0.39\textwidth]{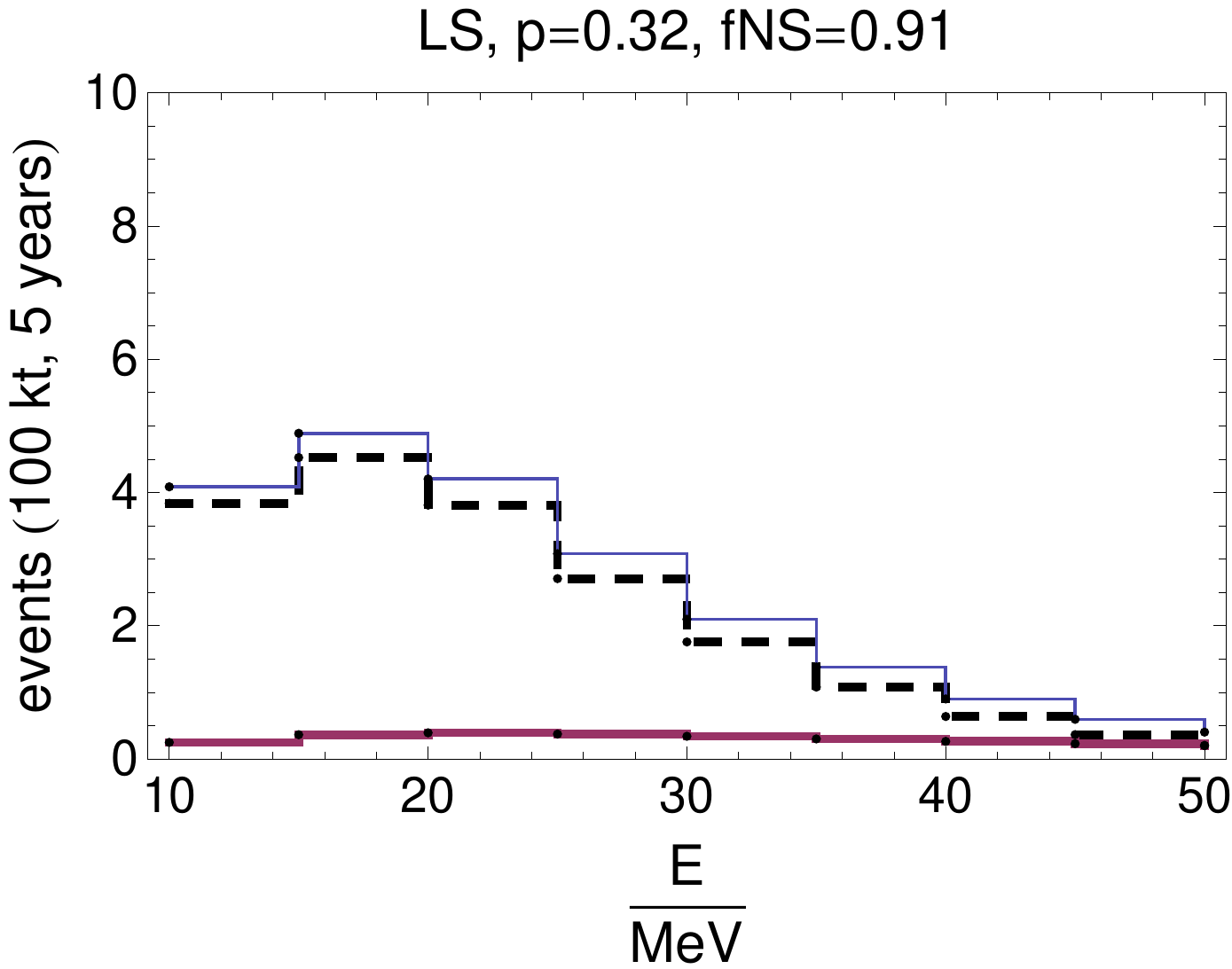}
 \includegraphics[width=0.39\textwidth]{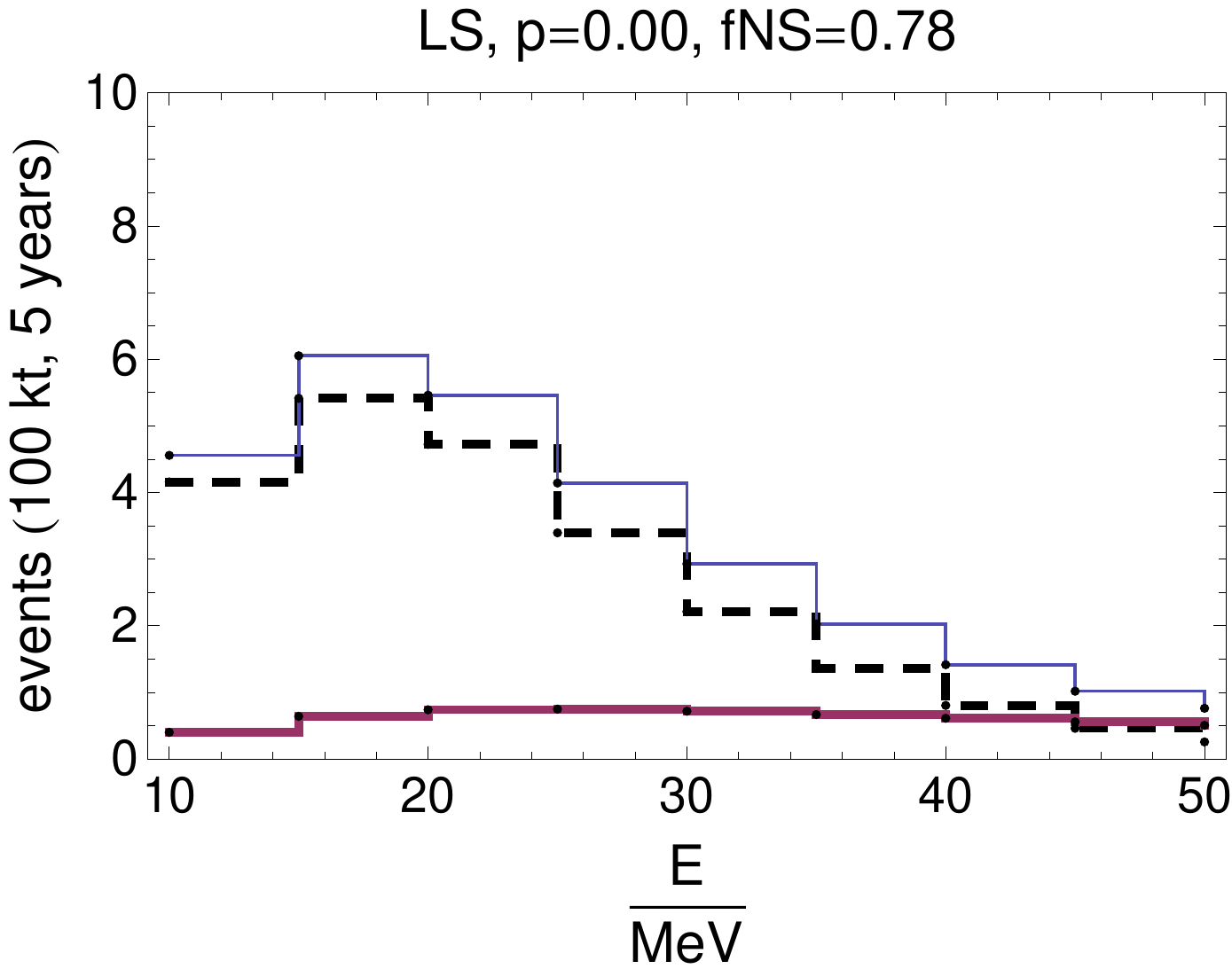}
 \includegraphics[width=0.39\textwidth]{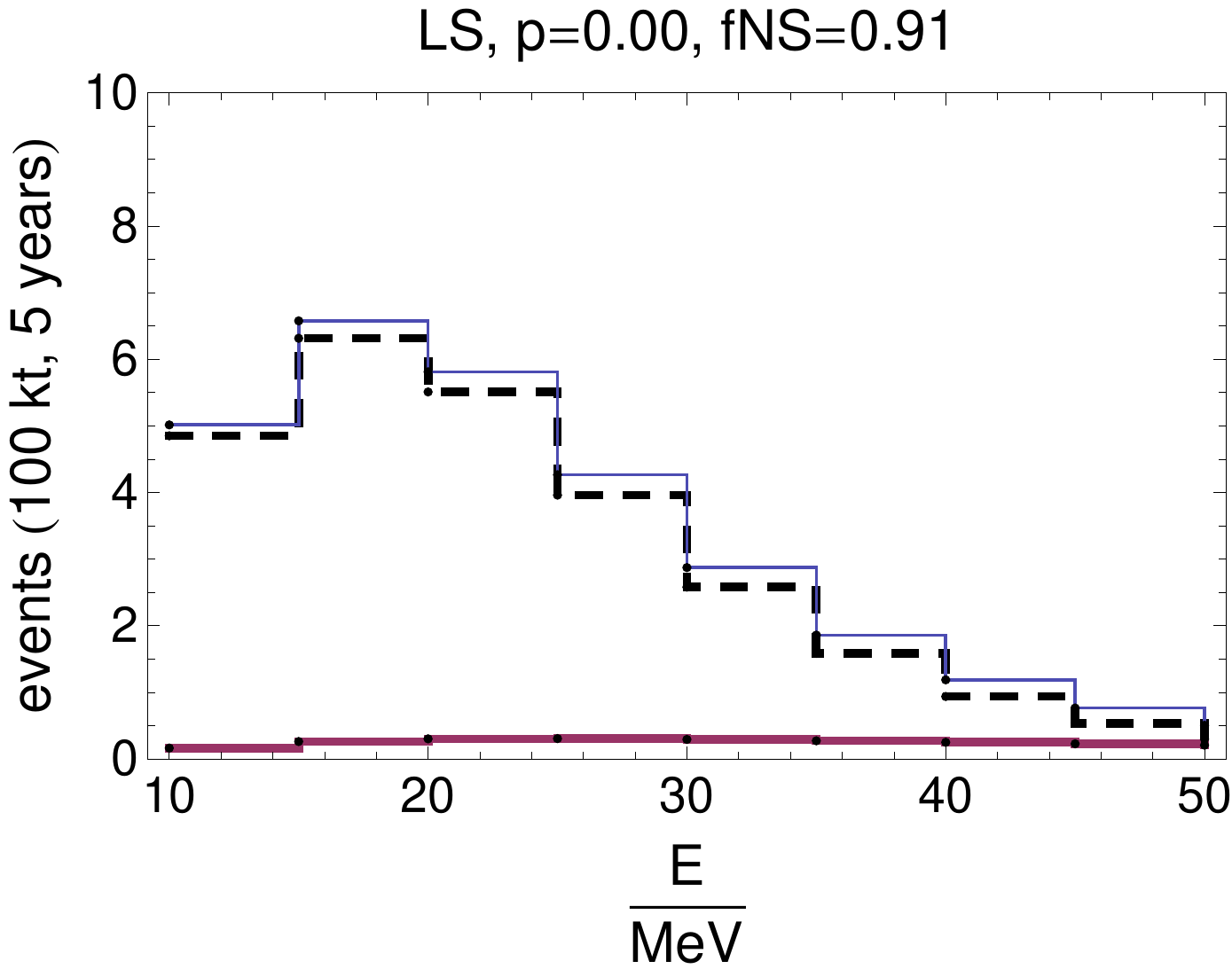}
 \caption{The same as fig. \ref{nueeventplotS} for the LS EoS.}
\label{nueeventplotLS}
\end{figure}

In figs. \ref{nueeventplotS} and \ref{nueeventplotLS} we show the expected event distributions (in neutrino energy) for a \lar\ experiment  with $0.5~{\rm Mt\cdot yr}$ exposure, for several sets of parameters.  %

 In contrast with a water detector, here the events from \bhf\ peak inside the detector energy window (or even slightly beyond, for the S EoS with $p=0$), $E \sim 19-30$ MeV, and not below it, thanks to the  faster increase of the detection cross section with energy.    Instead, the peak of the events from \nsf\ is below 19 MeV and therefore it is obscured by solar \ns.   Thus, the \lar\ technology, while needing threshold improvement to probe the bulk of \nsf\ events, is suitable as it is to study failed \sne.  For these, the most important improvement will probably be at the high end of the energy window, where  the atmospheric background is the limiting factor.

For the numbers of events, figs. \ref{nueeventplotS} and \ref{nueeventplotLS} confirm what was already observed for the  fluxes:
the contribution of failed \sne\ to the signal ranges from modest to dominant.
  Specifically, the \bhf\ contribute to each energy bin by at least $\sim 7\%$, and by more than $\sim 25\%$ for most of the parameter space.  Considering the low statistics, these excesses would probably be marginally significant at best, once theoretical uncertainties  -- especially on normalizations -- are included.  Still, the inclusion of failed \sne\ in the modeling of the signal would be important to have a reliable prediction as benchmark, and to estimate the theoretical error correctly. 
  
\begin{table}[htbp]
\begin{center}
\begin{tabular}{| c || c |c |c || c| c |c || c |}
\hline
\hline
 & \multicolumn{3 }{| c |}{  $ p=0.32$ } & \multicolumn{3 }{| c |}{ $ p=0$  }  &  \\
\cline{2-7}
  & \nsf\  & \bhf\ & Total  & \nsf\  & \bhf\  & Total   & atm.  \\
\hline
 
\hline
$19 <  E/{\rm MeV} <29$  &  5.9     & 5.5    &  11.5      &  8.6     & 3.6    &  12.2    &  2.25  \\
  												&  (6.9)   & (2.3)  & (9.2  )    &  (10.0)  & (1.5)  &  (11.5)  &    \\
\hline
$19 < E/{\rm MeV} <39$  &  8.6     & 10.6   &  19.3      &  12.5    & 7.4    &  19.9    &  8.6  \\
   												&  (10.0)  & (4.4)  &  (14.4 )   &  (14.6)  & (3.0)  &  (17.6)  &    \\
\hline
$10 < E/{\rm MeV} <39$  &  15.0    & 14.0   &   29.0     &  21.0    & 9.3    & 30.3     &   9.0 \\
  												&  (17.5)  & (5.7)  &  (23.3)    &  (24.5)  & (3.8)  &  (28.4)  &    \\
 \hline
\hline
\end{tabular}
\caption{The number of  $\nue$ interactions on $^{40}{\rm Ar}$ from the \df\ and from atmospheric neutrinos, in three energy windows of interest (given in terms of the neutrino energy, $E$) at a liquid argon detector of 0.5 Mt$\cdot$yr exposure. The rates for \bhf\ refer to the S EoS, $f_{NS}=0.78$ and $f_{NS}=0.91$ (the latter in parentheses).  } 
\label{ratesmodelAr}
\end{center}
\end{table}

\begin{table}[htbp]
\begin{center}
\begin{tabular}{| c || c |c |c || c| c |c || c |}
\hline
\hline
 & \multicolumn{3 }{| c |}{  $ p=0.32$ } & \multicolumn{3 }{| c |}{ $ p=0$  }  &  \\
\cline{2-7}
  & \nsf\  & \bhf\ & Total  & \nsf\  & \bhf\  & Total   & atm.  \\ \hline
 
\hline
$19 < E/{\rm MeV} <29$  &  5.9     &  1.9   &   7.8    &  8.6     &  1.5   &  10.1    &  2.25  \\
  												&  (6.9)   &  (0.8)   &   (7.7)    &  (10.0)  &  (0.6)   &  (10.6)    &    \\
\hline
$19 < E/{\rm MeV} <39$  &  8.6     &  3.5  &    12.1    &  12.5    &  2.9   &  15.4    &  8.6  \\
   												&  (10.0)  &  (1.4)  &   (11.5)     &  (14.6)  &  (1.2)   &   (15.8)   &    \\
\hline
$10 < E/{\rm MeV} <39$  &  15.0    &  4.8  &   19.8      &  21.0    &   9.3   &   24.8    &   9.0 \\
  												&  (17.5)  &  (2.0)  &   (19.5)      &  (24.5)  & (1.5)     &   (26.1)    &    \\
 \hline
\hline
\end{tabular}
\caption{The same as Table \ref{ratesmodelAr}  for the LS EoS. } 
\label{ratesmodelArLS}
\end{center}
\end{table}


Tables \ref{ratesmodelAr} and \ref{ratesmodelArLS} give the  numbers of events expected for different sets of parameters and different energy intervals.  We see that, for the S EoS and the interval $19-39$ MeV, failed \sne\ increase the event rate by $\sim 30-100\%$; specifically, the number of events goes from $\sim 8 - 12$ to $\sim 14-20$.  With $\sim 8-9$ background events expected, the significance of the signal changes from $\sim 2.3-3\sigma$ to $\sim3-3.7 \sigma $.  In the assumption of known $\Phi^{NS}_{e}$, the number of events due to $\Phi^{BH}_{e}$ would be too  small to be significant as a signal of its own, giving an excess of less than $1.7\sigma$.  For the largest $\Phi^{BH}_{e}$, a $3\sigma$ excess would be realized for a triple exposure, 1.5 Mt$\cdot$yr. 

For the smaller window $19-29$ MeV, the signal to background ratio increases, but the lower statistics compensate for this advantage, so that the significance of the signal due to the total \df\  becomes slightly worse, reaching $3\sigma$ in the most fortunate case.

For the LS EoS we expect  $\sim 12$ events from the \df\ in the window $19-39$ MeV; of these $1-3$ events would be from failed \sne.  These are not significant as a signal, but contribute to enhancing the statistical significance of the total number of events. This reaches $\sim 2.7\sigma$ at most, so that a significant excess could be established above the background with a moderately larger exposure. The significance is lower if the energy window is restricted, as observed previously. 

The enhancement of the signal due to failed \sne\ implies that a smaller exposure will be necessary (compared to \nsf\ only) for detection, therefore allowing a smaller mass and/or running time.  For example, for the most favourable set of parameters, a 10 kt detector running for 5 years and searching in the $19-39$ MeV window could see about 2 events from core collapses, compared to less than 1 from background and less than 1 expected from \nsf\ only. This is encouraging, in principle, although  errors would be insufficient to make firm conclusions in this case. 

\section{Directions of further study}
\label{directions}


Our results are limited by the still incomplete investigation of \n\ emission from failed \sne\ and from all collapses in general.  As these progress, a number of aspects will be included in the calculation of the diffuse flux. Here we discuss them briefly.

\begin{itemize}

\item Dependence on the EoS.  Little exists beyond the results with the LS and S EoS that we have used here for failed \sne.  However, initial studies evidence some trends.  Stiffer equations of state correspond to longer \n\ emission \cite{Sumiyoshi:2007pp,Hempel:2011mk}, and therefore to a more luminous time-integrated flux and diffuse flux.  The effect of the EoS is stronger for \bhf\ than for \nsf, where differences are mostly in the \n\ average energies and at the level of $\sim 10\%$ or so \cite{Hempel:2011mk}. It has to be stressed, though, that most results for \nsf\ refer to the first second post-bounce, while a full, $\sim 10$ s simulation is required to model the diffuse flux.  Progress in this direction exists \cite{Fischer:2009af,Huedepohl:2009wh}, but is still without systematic exploration of the EoS dependence.  Due to this lack of information, here we have neglected the EoS dependence of the \nsf\ diffuse flux.

Equations of state involving quarks, hyperons and/or pions have been considered for failed \sne\ \cite{Nakazato:2010ue,Nakazato:2010qy,Nakazato:2011vd}: they tend to shorten the \n\ emission and thus to decrease the diffuse flux. Pions, however, tend to increase the luminosity and average energy of the \n\ spectrum. Effects are of the order of tens of per cent,  so they may be difficult to distinguish in the diffuse flux.  

\item Effects of fallback black hole forming collapses (FBHFCs).  Ultimately, a robust prediction of the \df\ will require modeling the whole continuum between neutron star formation and direct black hole formation, with the inclusion of the intermediate case of black hole formation by fallback, which  is expected for at least some of the progenitors in the mass interval
$\sim 25-40 ~{\rm M_{\odot}}$. For FBHFCs, the initially formed neutron star collapses into a black hole as a result of accretion, after an explosion.  The explosion becomes weaker with the increase of the progenitor mass, until FBHFCs become indistinguishable from \bhf. 
For FBHFCs with a robust explosion, the  \n\ emission   resembles that of a neutron-star forming collapse for the first 10-20 s, and then exhibits a characteristic increase of the \n\ luminosity at later times as an effect of fallback \cite{Fryer:2007cf,Fryer:2009zs}.  This phenomenon was studied in a set of numerical simulations \cite{Fryer:2007cf}, and was  found to contribute by about $\sim 10\%$ to the total (time-integrated) \n\ flux.  However, the same simulations are still inconclusive about whether a black hole eventually forms, and systematic studies of heavy fallback for very massive progenitors are still needed to reach firm conclusions.

\item Effects of the diverse stellar population.  It is fascinating that the \df\ receives contributions from an enormous variety of stars, which may differ in many parameters like progenitor mass, magnetic fields, metallicity, rotation, etc.  The dependence of failed \sne\ on at least some of these parameters has been studied recently.   Studies with different models of progenitor stars in the 40-50 $M_\odot$ interval \cite{Nakazato:2008vj}  have shown that different stellar density profiles could result in very different oscillation effects, within the intervals in eq. (\ref{rangesprob}), while differences in the produced (pre-oscillation) \n\ fluxes are likely to be minor compared to the several uncertainties in the problem \cite{Sumiyoshi:2008zw}. A detailed study of direct black hole collapse with and without rotation \cite{O'Connor:2010tk} has shown that rotation tends to prevent or delay the black hole formation, with overall lower \n\ luminosities and average energies.   The same study predicts that up to 15\% of low metallicity stars (metallicity less than $10^{-4}$ times that of our Sun) can undergo direct black hole formation, compared to the maximum of $\sim 7\%$ for stars with solar metallicity.  Therefore, one might expect an enhanced contribution to the \df\ from low metallicity stars. However, this enhancement is probably overcompensated by the relative rarity of such stars at low redshift.  

Finally, we recall that some failed \sne\ may generate collapsars, the hosts of the Gamma Ray Bursts (and their accompanying jets of $\sim$ TeV \ns).  These collapsars continue to emit ${\mathcal O}(10) $ MeV  \ns\ after the black hole formation, due to the presence of an accretion disk around the black hole itself \cite{Nagataki:2002bn,McLaughlin:2006yy}. While very luminous, this \n\ flux should contribute to the \df\ at the level of  $\sim 10\%$ or less, due to the rarity of collapsars, see e.g., \cite{Nagataki:2002bn}.

\end{itemize}


\section{Discussion and conclusions}
\label{disc}

Let us summarize our results. 

\begin{itemize}

\item  The diffuse flux from \bhf\ reflects the features of the original \n\ flux from a failed \sn: it is more luminous and more energetic than the flux from \nts\ collapses, with the most energetic spectra being realized for the stiffer, Shen et al. equation of state. In energy windows relevant for detection, the $\barnue$  component of this flux  is at a maximum for the largest $\barnue$ survival probability, due to the especially large flux of $\barnue$s originally produced in the star. An analogous result holds for the $\nue$ component as well.  This contrasts with the case of \nts\ collapses, where the luminosity is roughly equipartitioned among the \n\ species.  

\item Because of its more energetic spectrum, the flux from \bhf\ has a cosmological component -- from stars with $z\gta 1$ -- as large as $\sim 40\%$ above 20 MeV. This is
interestingly larger than the $\sim 10\%$ or less expected in the same interval for \nsf, 
 for which the cosmological component largely accumulates below the experimental energy threshold.  This could result in new possibilities  to use \ns\ to test the rate of collapses at cosmological distances.
 
 \item The harder spectrum of the \bhf\ flux can  result in a wider energy window of detection  for the \df\ (defined as the energy interval where the core collapse flux exceeds the background fluxes of \ns\ of other origin). The window can be up to roughly 7 MeV wider than for \nsf\ only, depending on the  magnitude of the atmospheric background relative to the signal (fig. \ref{backgroundsites}).   
 
 \item The diffuse flux of \ns\ from failed \sne\ could be substantial, up to $\phi^{BH}_{\bar e}=0.38~{\rm  ~s^{-1}cm^{-2}}$ ($\phi^{BH}_{ e}=0.28 ~ {\rm ~ s^{-1}cm^{-2}}$) for $\barnue$ ($\nue$) in the interval $19.3 - 29.3$ MeV, normalized to a local rate of core collapses of $R_{cc}(0)=10^{-4}~{\rm yr^{-1}Mpc^{-3}}$.  This is only a factor of $\sim 4$ lower  than the current sensitivity of \sk, indicating the possibility of detection in the near future. 

\item 
  Depending on the parameters (the oscillation probabilities, the fraction of \bh\ collapses and the EoS), the flux from failed \sne\  ranges from 6-10\% to a dominant fraction of the total \df, for energies of experimental interest. 
The total flux is enhanced  -- compared to \nts\ collapses only -- by up to a factor $\sim 2.3$, reaching $\phi_{\bar e} \simeq 0.67~{\rm  ~s^{-1}cm^{-2}}$ in the  $19.3 - 29.3$ MeV window, and   $\phi^{BH}_{\bar e}=0.89~{\rm  ~s^{-1}cm^{-2}}$ in the open interval $E>19.3 $ MeV. The latter estimate is only a factor of $\sim 2$ lower than of the current \sk\ limit, Eq. (\ref{sklim}), and therefore it is very promising for the next phase of experimental searches.

\item The  \sk\ limit  constrains the multi-dimensional region of the parameter space.  This loose constraint can be expressed in terms of conditional limits on the individual parameters: for example, the rate of core collapses is constrained to  $R_{cc}(0) < 2.1 \cdot 10^{-4}~{\rm yr^{-1}Mpc^{-3}}$ when all the other parameters are fixed to maximize $\Phi^{BH}_{\bar e}$.  Similarly, one gets a limit on the fraction of failed \sne, $f_{BH}=1-f_{NS} \lta 0.7  $, for the same set of parameters and $R_{cc}(0) =  10^{-4}~{\rm yr^{-1}Mpc^{-3}}$. 

\item in a detector, the most immediate effect of the \n\ flux from \bhf\ is an enhancement of the event rate, which reflects the enhancement of the flux compared to \nts\ collapses only.  In a water \ck\ detector with a 2.25 Mt$\cdot$yr exposure (e.g., 0.45 Mt for 5 years) we expect $\sim 5 - 65$ events from failed \sne\ in the window 18-28 MeV of positron energy, out of a total of 63-113 events from all collapses. These represent an excess of $2.3 - 3.9\sigma$, after background rates have been included to calculate errors. 
  For the extended window 10-38 MeV, relevant to water plus Gadolinium, we get 13-165 events from failed \sne\ and a total of $\sim 190-310$ events from all collapses, corresponding to an excess of 4 - 6$\sigma$ above background.  
 
  \item in liquid argon the spectrum of events from \bhf\ peaks above $\sim 19$ MeV, where the solar \n\ flux terminates. This is a distinctive feature of liquid argon, and is  due to the fast increase of the cross section with the \n\ energy. 
  
  \item   For a liquid argon detector with exposure of $0.5$ Mt$\cdot$yr, the larger energy window of 19-39 MeV is overall  convenient to increase statistics at the price of a slightly worse signal-to-background ratio.  For this window we predict 1-11 events from \bhf, and a total of 12-20 signal events, with 9 events from background.  Statistical significance of $3\sigma$ is realized for the S EoS, in the absence of background systematics. 

\end{itemize}

Our results show that, with an improvement by a factor of 2 (in flux) of its sensitivity, \sk\ can start to probe the parameter space of \ns\ from failed \sne\ at the basic level and that next generation detectors should cover a substantial portion of this space. 

Due to uncertainties in the normalizations, the most robust signature of failed \sne\ in the diffuse flux would be the harder spectrum, and possibly a deviation from the characteristic exponential shape of the spectrum of the \df\ \cite{Lunardini:2006pd} where the two contributions, from \nts\ and \bh\ collapses, are comparable. Such spectral distortion could be visible with the extended energy window of a water+Gd detector or with a liquid scintillator detector \cite{Wurm:2007cy}, which both have the advantage of a better energy resolution.

To establish the presence of a flux from failed \sne\ would already be a fundamental result, being the first detection of  new type of \n\ source.  Beyond the discovery phase, with a high statistics signal it might be possible to  distinguish between different models of \bh\ collapse, at the level of favouring one EoS over another, although a model-independent discrimination might not be possible due to the large errors.  The cases with the largest $\Phi^{BH}$ -- maximum survival of the electron flavors and smallest $f_{NS}$  -- might be established or ruled out relatively easily, while other scenarios might be more difficult to probe because their lower flux is more shadowed by the  atmospheric background. 

 For a given model of \n\ spectra from \bh\ and \nts\ collapses, the position of the spectral distortion (with respect to an exponential spectrum) might be used to probe the relative frequency of the two types, in other words $f_{NS}$, and in turn the minimum progenitor mass required to produce a direct \bh\ collapse (sec. \ref{general}). 
   It is likely that in the space of a few years the rate of \nts\ collapses will be known with good precision from astronomy \cite{snap,snls}, and this will allow translation of the information on $f_{NS}$ obtained from \ns\  into an absolute (as opposed to relative) rate of failed \sne. 
   
Data on \ns\ from failed \sne\ would also constitute a new ground to test \n\ oscillations, and therefore \n\ masses and mixings. Realistically, only average survival probabilities could be extracted from a fit to high statistics data.  It would be especially interesting to look for differences in the oscillation patterns for \bhf\ and \nsf, as these could give insight on the different physics at play in the two types of collapses (e.g., different matter density profiles influencing the MSW resonances). 
   
To conclude, the detection of a diffuse \n\ flux from failed \sne\ is a realistic possibility. It would have profound implications on the study of these invisible objects, on which we have no data so far. The flux is uncertain by more than one order of magnitude,  and therefore it remains to be established whether it dominates the total flux or just modifies it at the 10\% level. In the first case, a change of perspective in the field will be needed. In the latter, failed \sne\ would be an ingredient of precision modeling of the \df\ and their parameter space would be constrained experimentally. 

\subsection*{Acknowledgments}
We are grateful to G. Mangano and M. Vagins for useful exchanges. 
C.L. acknowledges the support of
the NSF under Grant No. PHY-0854827. J.G.K. acknowledges the resources available to him at Arizona State University during his undergraduate studies. 

\bibliography{draft}

\begin{thebibliography}{70}
\expandafter\ifx\csname natexlab\endcsname\relax\def\natexlab#1{#1}\fi
\expandafter\ifx\csname bibnamefont\endcsname\relax
  \def\bibnamefont#1{#1}\fi
\expandafter\ifx\csname bibfnamefont\endcsname\relax
  \def\bibfnamefont#1{#1}\fi
\expandafter\ifx\csname citenamefont\endcsname\relax
  \def\citenamefont#1{#1}\fi
\expandafter\ifx\csname url\endcsname\relax
  \def\url#1{\texttt{#1}}\fi
\expandafter\ifx\csname urlprefix\endcsname\relax\def\urlprefix{URL }\fi
\providecommand{\bibinfo}[2]{#2}
\providecommand{\eprint}[2][]{\url{#2}}

\bibitem[{\citenamefont{Hirata et~al.}(1987)}]{Hirata:1987hu}
\bibinfo{author}{\bibfnamefont{K.}~\bibnamefont{Hirata}} \bibnamefont{et~al.}
  (\bibinfo{collaboration}{KAMIOKANDE-II}), \bibinfo{journal}{Phys. Rev. Lett.}
  \textbf{\bibinfo{volume}{58}}, \bibinfo{pages}{1490} (\bibinfo{year}{1987}).

\bibitem[{\citenamefont{Bionta et~al.}(1987)}]{Bionta:1987qt}
\bibinfo{author}{\bibfnamefont{R.~M.} \bibnamefont{Bionta}}
  \bibnamefont{et~al.}, \bibinfo{journal}{Phys. Rev. Lett.}
  \textbf{\bibinfo{volume}{58}}, \bibinfo{pages}{1494} (\bibinfo{year}{1987}).

\bibitem[{\citenamefont{Malek et~al.}(2003)}]{Malek:2002ns}
\bibinfo{author}{\bibfnamefont{M.}~\bibnamefont{Malek}} \bibnamefont{et~al.}
  (\bibinfo{collaboration}{Super-Kamiokande}), \bibinfo{journal}{Phys. Rev.
  Lett.} \textbf{\bibinfo{volume}{90}}, \bibinfo{pages}{061101}
  (\bibinfo{year}{2003}), \eprint{hep-ex/0209028}.

\bibitem[{\citenamefont{Eguchi et~al.}(2004)}]{Eguchi:2003gg}
\bibinfo{author}{\bibfnamefont{K.}~\bibnamefont{Eguchi}} \bibnamefont{et~al.}
  (\bibinfo{collaboration}{KamLAND}), \bibinfo{journal}{Phys. Rev. Lett.}
  \textbf{\bibinfo{volume}{92}}, \bibinfo{pages}{071301}
  (\bibinfo{year}{2004}), \eprint{hep-ex/0310047}.

\bibitem[{\citenamefont{Aharmim et~al.}(2006)}]{Aharmim:2006wq}
\bibinfo{author}{\bibfnamefont{B.}~\bibnamefont{Aharmim}} \bibnamefont{et~al.}
  (\bibinfo{collaboration}{SNO Collaboration}), \bibinfo{journal}{Astrophys.J.}
  \textbf{\bibinfo{volume}{653}}, \bibinfo{pages}{1545} (\bibinfo{year}{2006}),
  \eprint{hep-ex/0607010}.

\bibitem[{\citenamefont{Lunardini and Peres}(2008)}]{Lunardini:2008xd}
\bibinfo{author}{\bibfnamefont{C.}~\bibnamefont{Lunardini}} \bibnamefont{and}
  \bibinfo{author}{\bibfnamefont{O.~L.} \bibnamefont{Peres}},
  \bibinfo{journal}{JCAP} \textbf{\bibinfo{volume}{0808}}, \bibinfo{pages}{033}
  (\bibinfo{year}{2008}), \eprint{0805.4225}.

\bibitem[{\citenamefont{Sumiyoshi et~al.}(2006)\citenamefont{Sumiyoshi, Yamada,
  Suzuki, and Chiba}}]{Sumiyoshi:2006id}
\bibinfo{author}{\bibfnamefont{K.}~\bibnamefont{Sumiyoshi}},
  \bibinfo{author}{\bibfnamefont{S.}~\bibnamefont{Yamada}},
  \bibinfo{author}{\bibfnamefont{H.}~\bibnamefont{Suzuki}}, \bibnamefont{and}
  \bibinfo{author}{\bibfnamefont{S.}~\bibnamefont{Chiba}},
  \bibinfo{journal}{Phys. Rev. Lett.} \textbf{\bibinfo{volume}{97}},
  \bibinfo{pages}{091101} (\bibinfo{year}{2006}), \eprint{astro-ph/0608509}.

\bibitem[{\citenamefont{Lunardini}(2009)}]{Lunardini:2009ya}
\bibinfo{author}{\bibfnamefont{C.}~\bibnamefont{Lunardini}},
  \bibinfo{journal}{Phys. Rev. Lett.} \textbf{\bibinfo{volume}{102}},
  \bibinfo{pages}{231101} (\bibinfo{year}{2009}), \eprint{0901.0568}.

\bibitem[{\citenamefont{Liebendoerfer et~al.}(2004)}]{Liebendoerfer:2002xn}
\bibinfo{author}{\bibfnamefont{M.}~\bibnamefont{Liebendoerfer}}
  \bibnamefont{et~al.}, \bibinfo{journal}{Astrophys. J. Suppl.}
  \textbf{\bibinfo{volume}{150}}, \bibinfo{pages}{263} (\bibinfo{year}{2004}),
  \eprint{astro-ph/0207036}.

\bibitem[{\citenamefont{Sumiyoshi et~al.}(2007)\citenamefont{Sumiyoshi, Yamada,
  and Suzuki}}]{Sumiyoshi:2007pp}
\bibinfo{author}{\bibfnamefont{K.}~\bibnamefont{Sumiyoshi}},
  \bibinfo{author}{\bibfnamefont{S.}~\bibnamefont{Yamada}}, \bibnamefont{and}
  \bibinfo{author}{\bibfnamefont{H.}~\bibnamefont{Suzuki}},
  \bibinfo{journal}{Astrophys.J.} \textbf{\bibinfo{volume}{667}},
  \bibinfo{pages}{382} (\bibinfo{year}{2007}), \eprint{0706.3762}.

\bibitem[{\citenamefont{Fischer et~al.}(2009)\citenamefont{Fischer, Whitehouse,
  Mezzacappa, Thielemann, and Liebendorfer}}]{Fischer:2008rh}
\bibinfo{author}{\bibfnamefont{T.}~\bibnamefont{Fischer}},
  \bibinfo{author}{\bibfnamefont{S.~C.} \bibnamefont{Whitehouse}},
  \bibinfo{author}{\bibfnamefont{A.}~\bibnamefont{Mezzacappa}},
  \bibinfo{author}{\bibfnamefont{F.~K.} \bibnamefont{Thielemann}},
  \bibnamefont{and}
  \bibinfo{author}{\bibfnamefont{M.}~\bibnamefont{Liebendorfer}},
  \bibinfo{journal}{Astron. Astrophys.} \textbf{\bibinfo{volume}{499}},
  \bibinfo{pages}{1} (\bibinfo{year}{2009}), \eprint{0809.5129}.

\bibitem[{\citenamefont{Sumiyoshi et~al.}(2008)\citenamefont{Sumiyoshi, Yamada,
  and Suzuki}}]{Sumiyoshi:2008zw}
\bibinfo{author}{\bibfnamefont{K.}~\bibnamefont{Sumiyoshi}},
  \bibinfo{author}{\bibfnamefont{S.}~\bibnamefont{Yamada}}, \bibnamefont{and}
  \bibinfo{author}{\bibfnamefont{H.}~\bibnamefont{Suzuki}},
  \bibinfo{journal}{Astrophys. J.} \textbf{\bibinfo{volume}{688}},
  \bibinfo{pages}{1176} (\bibinfo{year}{2008}), \eprint{0808.0384}.

\bibitem[{\citenamefont{Nakazato et~al.}(2008)\citenamefont{Nakazato,
  Sumiyoshi, Suzuki, and Yamada}}]{Nakazato:2008vj}
\bibinfo{author}{\bibfnamefont{K.}~\bibnamefont{Nakazato}},
  \bibinfo{author}{\bibfnamefont{K.}~\bibnamefont{Sumiyoshi}},
  \bibinfo{author}{\bibfnamefont{H.}~\bibnamefont{Suzuki}}, \bibnamefont{and}
  \bibinfo{author}{\bibfnamefont{S.}~\bibnamefont{Yamada}},
  \bibinfo{journal}{Phys. Rev.} \textbf{\bibinfo{volume}{D78}},
  \bibinfo{pages}{083014} (\bibinfo{year}{2008}), \eprint{0810.3734}.

\bibitem[{\citenamefont{Nakazato
  et~al.}(2010{\natexlab{a}})\citenamefont{Nakazato, Sumiyoshi, and
  Yamada}}]{Nakazato:2010ue}
\bibinfo{author}{\bibfnamefont{K.}~\bibnamefont{Nakazato}},
  \bibinfo{author}{\bibfnamefont{K.}~\bibnamefont{Sumiyoshi}},
  \bibnamefont{and} \bibinfo{author}{\bibfnamefont{S.}~\bibnamefont{Yamada}},
  \bibinfo{journal}{Astrophys. J.} \textbf{\bibinfo{volume}{721}},
  \bibinfo{pages}{1284} (\bibinfo{year}{2010}{\natexlab{a}}),
  \eprint{1001.5084}.

\bibitem[{\citenamefont{Beacom and Vagins}(2004)}]{Beacom:2003nk}
\bibinfo{author}{\bibfnamefont{J.~F.} \bibnamefont{Beacom}} \bibnamefont{and}
  \bibinfo{author}{\bibfnamefont{M.~R.} \bibnamefont{Vagins}},
  \bibinfo{journal}{Phys. Rev. Lett.} \textbf{\bibinfo{volume}{93}},
  \bibinfo{pages}{171101} (\bibinfo{year}{2004}), \eprint{hep-ph/0309300}.

\bibitem[{\citenamefont{Lien et~al.}(2010)\citenamefont{Lien, Fields, and
  Beacom}}]{Lien:2010yb}
\bibinfo{author}{\bibfnamefont{A.}~\bibnamefont{Lien}},
  \bibinfo{author}{\bibfnamefont{B.~D.} \bibnamefont{Fields}},
  \bibnamefont{and} \bibinfo{author}{\bibfnamefont{J.~F.}
  \bibnamefont{Beacom}}, \bibinfo{journal}{Phys. Rev.}
  \textbf{\bibinfo{volume}{D81}}, \bibinfo{pages}{083001}
  (\bibinfo{year}{2010}), \eprint{1001.3678}.

\bibitem[{\citenamefont{Iida}(2005)}]{iida}
\bibinfo{author}{\bibfnamefont{T.}~\bibnamefont{Iida}}, \bibinfo{journal}{J.
  Phys.: Conf. Ser.} \textbf{\bibinfo{volume}{136}}, \bibinfo{pages}{042075}
  (\bibinfo{year}{2005}).

\bibitem[{\citenamefont{Iida}()}]{iidathesis}
\bibinfo{author}{\bibfnamefont{T.}~\bibnamefont{Iida}}, \bibinfo{note}{phD
  thesis, U. of Tokyo, 2010. Available at
  http://www-sk.icrr.u-tokyo.ac.jp/sk/pub/}.

\bibitem[{\citenamefont{Lazauskas et~al.}(2009)\citenamefont{Lazauskas,
  Lunardini, and Volpe}}]{Lazauskas:2009yh}
\bibinfo{author}{\bibfnamefont{R.}~\bibnamefont{Lazauskas}},
  \bibinfo{author}{\bibfnamefont{C.}~\bibnamefont{Lunardini}},
  \bibnamefont{and} \bibinfo{author}{\bibfnamefont{C.}~\bibnamefont{Volpe}},
  \bibinfo{journal}{JCAP} \textbf{\bibinfo{volume}{0904}}, \bibinfo{pages}{029}
  (\bibinfo{year}{2009}), \eprint{0901.0581}.

\bibitem[{\citenamefont{Boyd et~al.}(2010)\citenamefont{Boyd, Kajino, and
  Onaka}}]{Boyd:2010ak}
\bibinfo{author}{\bibfnamefont{R.~N.} \bibnamefont{Boyd}},
  \bibinfo{author}{\bibfnamefont{T.}~\bibnamefont{Kajino}}, \bibnamefont{and}
  \bibinfo{author}{\bibfnamefont{T.}~\bibnamefont{Onaka}},
  \bibinfo{journal}{Astrobiology} \textbf{\bibinfo{volume}{10}},
  \bibinfo{pages}{561} (\bibinfo{year}{2010}), \eprint{1001.3849}.

\bibitem[{\citenamefont{Hopkins and Beacom}(2006)}]{Hopkins:2006bw}
\bibinfo{author}{\bibfnamefont{A.~M.} \bibnamefont{Hopkins}} \bibnamefont{and}
  \bibinfo{author}{\bibfnamefont{J.~F.} \bibnamefont{Beacom}},
  \bibinfo{journal}{Astrophys. J.} \textbf{\bibinfo{volume}{651}},
  \bibinfo{pages}{142} (\bibinfo{year}{2006}), \eprint{astro-ph/0601463}.

\bibitem[{\citenamefont{Woosley et~al.}(2002)\citenamefont{Woosley, Heger, and
  Weaver}}]{Woosley:2002zz}
\bibinfo{author}{\bibfnamefont{S.~E.} \bibnamefont{Woosley}},
  \bibinfo{author}{\bibfnamefont{A.}~\bibnamefont{Heger}}, \bibnamefont{and}
  \bibinfo{author}{\bibfnamefont{T.~A.} \bibnamefont{Weaver}},
  \bibinfo{journal}{Rev. Mod. Phys.} \textbf{\bibinfo{volume}{74}},
  \bibinfo{pages}{1015} (\bibinfo{year}{2002}).

\bibitem[{\citenamefont{Salpeter}(1955)}]{Salpeter:1955it}
\bibinfo{author}{\bibfnamefont{E.~E.} \bibnamefont{Salpeter}},
  \bibinfo{journal}{Astrophys. J.} \textbf{\bibinfo{volume}{121}},
  \bibinfo{pages}{161} (\bibinfo{year}{1955}).

\bibitem[{\citenamefont{Keil et~al.}(2003)\citenamefont{Keil, Raffelt, and
  Janka}}]{Keil:2002in}
\bibinfo{author}{\bibfnamefont{M.~T.} \bibnamefont{Keil}},
  \bibinfo{author}{\bibfnamefont{G.~G.} \bibnamefont{Raffelt}},
  \bibnamefont{and} \bibinfo{author}{\bibfnamefont{H.-T.} \bibnamefont{Janka}},
  \bibinfo{journal}{Astrophys. J.} \textbf{\bibinfo{volume}{590}},
  \bibinfo{pages}{971} (\bibinfo{year}{2003}), \eprint{astro-ph/0208035}.

\bibitem[{\citenamefont{Heger et~al.}(2003)\citenamefont{Heger, Fryer, Woosley,
  Langer, and Hartmann}}]{Heger:2002by}
\bibinfo{author}{\bibfnamefont{A.}~\bibnamefont{Heger}},
  \bibinfo{author}{\bibfnamefont{C.~L.} \bibnamefont{Fryer}},
  \bibinfo{author}{\bibfnamefont{S.~E.} \bibnamefont{Woosley}},
  \bibinfo{author}{\bibfnamefont{N.}~\bibnamefont{Langer}}, \bibnamefont{and}
  \bibinfo{author}{\bibfnamefont{D.~H.} \bibnamefont{Hartmann}},
  \bibinfo{journal}{Astrophys. J.} \textbf{\bibinfo{volume}{591}},
  \bibinfo{pages}{288} (\bibinfo{year}{2003}), \eprint{astro-ph/0212469}.

\bibitem[{woo()}]{woosley}
\bibinfo{note}{S. E. Woosley and T. Weaver, Astrophys. J. Suppl. 101, 181
  (1995)}.

\bibitem[{\citenamefont{Shen et~al.}(1998)\citenamefont{Shen, Toki, Oyamatsu,
  and Sumiyoshi}}]{shenetal}
\bibinfo{author}{\bibfnamefont{H.}~\bibnamefont{Shen}},
  \bibinfo{author}{\bibfnamefont{H.}~\bibnamefont{Toki}},
  \bibinfo{author}{\bibfnamefont{K.}~\bibnamefont{Oyamatsu}}, \bibnamefont{and}
  \bibinfo{author}{\bibfnamefont{K.}~\bibnamefont{Sumiyoshi}},
  \bibinfo{journal}{Nucl. Phys. A} \textbf{\bibinfo{volume}{637}},
  \bibinfo{pages}{435} (\bibinfo{year}{1998}).

\bibitem[{\citenamefont{Lattimer and Swesty}(1991)}]{lattimer91generalized}
\bibinfo{author}{\bibfnamefont{J.~M.} \bibnamefont{Lattimer}} \bibnamefont{and}
  \bibinfo{author}{\bibfnamefont{F.~D.} \bibnamefont{Swesty}},
  \bibinfo{journal}{Nucl. Phys. A} \textbf{\bibinfo{volume}{535}},
  \bibinfo{pages}{331} (\bibinfo{year}{1991}).

\bibitem[{\citenamefont{Dighe and Smirnov}(2000)}]{Dighe:1999bi}
\bibinfo{author}{\bibfnamefont{A.~S.} \bibnamefont{Dighe}} \bibnamefont{and}
  \bibinfo{author}{\bibfnamefont{A.~Y.} \bibnamefont{Smirnov}},
  \bibinfo{journal}{Phys. Rev.} \textbf{\bibinfo{volume}{D62}},
  \bibinfo{pages}{033007} (\bibinfo{year}{2000}), \eprint{hep-ph/9907423}.

\bibitem[{\citenamefont{Dasgupta and Dighe}(2008)}]{Dasgupta:2007ws}
\bibinfo{author}{\bibfnamefont{B.}~\bibnamefont{Dasgupta}} \bibnamefont{and}
  \bibinfo{author}{\bibfnamefont{A.}~\bibnamefont{Dighe}},
  \bibinfo{journal}{Phys. Rev.} \textbf{\bibinfo{volume}{D77}},
  \bibinfo{pages}{113002} (\bibinfo{year}{2008}), \eprint{0712.3798}.

\bibitem[{\citenamefont{Lunardini and Smirnov}(2003)}]{Lunardini:2003eh}
\bibinfo{author}{\bibfnamefont{C.}~\bibnamefont{Lunardini}} \bibnamefont{and}
  \bibinfo{author}{\bibfnamefont{A.~Y.} \bibnamefont{Smirnov}},
  \bibinfo{journal}{JCAP} \textbf{\bibinfo{volume}{0306}}, \bibinfo{pages}{009}
  (\bibinfo{year}{2003}), \eprint{hep-ph/0302033}.

\bibitem[{\citenamefont{Duan et~al.}(2006)\citenamefont{Duan, Fuller, and
  Qian}}]{Duan:2005cp}
\bibinfo{author}{\bibfnamefont{H.}~\bibnamefont{Duan}},
  \bibinfo{author}{\bibfnamefont{G.~M.} \bibnamefont{Fuller}},
  \bibnamefont{and} \bibinfo{author}{\bibfnamefont{Y.-Z.} \bibnamefont{Qian}},
  \bibinfo{journal}{Phys. Rev.} \textbf{\bibinfo{volume}{D74}},
  \bibinfo{pages}{123004} (\bibinfo{year}{2006}), \eprint{astro-ph/0511275}.

\bibitem[{\citenamefont{Hannestad et~al.}(2006)\citenamefont{Hannestad,
  Raffelt, Sigl, and Wong}}]{Hannestad:2006nj}
\bibinfo{author}{\bibfnamefont{S.}~\bibnamefont{Hannestad}},
  \bibinfo{author}{\bibfnamefont{G.~G.} \bibnamefont{Raffelt}},
  \bibinfo{author}{\bibfnamefont{G.}~\bibnamefont{Sigl}}, \bibnamefont{and}
  \bibinfo{author}{\bibfnamefont{Y.~Y.~Y.} \bibnamefont{Wong}},
  \bibinfo{journal}{Phys. Rev.} \textbf{\bibinfo{volume}{D74}},
  \bibinfo{pages}{105010} (\bibinfo{year}{2006}), \eprint{astro-ph/0608695}.

\bibitem[{\citenamefont{Raffelt and Smirnov}(2007)}]{Raffelt:2007cb}
\bibinfo{author}{\bibfnamefont{G.~G.} \bibnamefont{Raffelt}} \bibnamefont{and}
  \bibinfo{author}{\bibfnamefont{A.~Y.} \bibnamefont{Smirnov}},
  \bibinfo{journal}{Phys.Rev.} \textbf{\bibinfo{volume}{D76}},
  \bibinfo{pages}{081301} (\bibinfo{year}{2007}), \eprint{0705.1830}.

\bibitem[{\citenamefont{Esteban-Pretel
  et~al.}(2007)\citenamefont{Esteban-Pretel, Pastor, Tomas, Raffelt, and
  Sigl}}]{EstebanPretel:2007ec}
\bibinfo{author}{\bibfnamefont{A.}~\bibnamefont{Esteban-Pretel}},
  \bibinfo{author}{\bibfnamefont{S.}~\bibnamefont{Pastor}},
  \bibinfo{author}{\bibfnamefont{R.}~\bibnamefont{Tomas}},
  \bibinfo{author}{\bibfnamefont{G.~G.} \bibnamefont{Raffelt}},
  \bibnamefont{and} \bibinfo{author}{\bibfnamefont{G.}~\bibnamefont{Sigl}},
  \bibinfo{journal}{Phys.Rev.} \textbf{\bibinfo{volume}{D76}},
  \bibinfo{pages}{125018} (\bibinfo{year}{2007}), \eprint{0706.2498}.

\bibitem[{\citenamefont{Fogli et~al.}(2007)\citenamefont{Fogli, Lisi, Marrone,
  and Mirizzi}}]{Fogli:2007bk}
\bibinfo{author}{\bibfnamefont{G.~L.} \bibnamefont{Fogli}},
  \bibinfo{author}{\bibfnamefont{E.}~\bibnamefont{Lisi}},
  \bibinfo{author}{\bibfnamefont{A.}~\bibnamefont{Marrone}}, \bibnamefont{and}
  \bibinfo{author}{\bibfnamefont{A.}~\bibnamefont{Mirizzi}},
  \bibinfo{journal}{JCAP} \textbf{\bibinfo{volume}{0712}}, \bibinfo{pages}{010}
  (\bibinfo{year}{2007}), \eprint{0707.1998}.

\bibitem[{\citenamefont{Duan and Kneller}(2009)}]{Duan:2009cd}
\bibinfo{author}{\bibfnamefont{H.}~\bibnamefont{Duan}} \bibnamefont{and}
  \bibinfo{author}{\bibfnamefont{J.~P.} \bibnamefont{Kneller}},
  \bibinfo{journal}{J.Phys.G} \textbf{\bibinfo{volume}{G36}},
  \bibinfo{pages}{113201} (\bibinfo{year}{2009}), \eprint{arXiv:0904.0974}.

\bibitem[{\citenamefont{Duan et~al.}(2007)\citenamefont{Duan, Fuller, Carlson,
  and Zhong}}]{Duan:2007bt}
\bibinfo{author}{\bibfnamefont{H.}~\bibnamefont{Duan}},
  \bibinfo{author}{\bibfnamefont{G.~M.} \bibnamefont{Fuller}},
  \bibinfo{author}{\bibfnamefont{J.}~\bibnamefont{Carlson}}, \bibnamefont{and}
  \bibinfo{author}{\bibfnamefont{Y.-Q.} \bibnamefont{Zhong}},
  \bibinfo{journal}{Phys. Rev. Lett.} \textbf{\bibinfo{volume}{99}},
  \bibinfo{pages}{241802} (\bibinfo{year}{2007}), \eprint{0707.0290}.

\bibitem[{\citenamefont{Dasgupta et~al.}(2009)\citenamefont{Dasgupta, Dighe,
  Raffelt, and Smirnov}}]{Dasgupta:2009mg}
\bibinfo{author}{\bibfnamefont{B.}~\bibnamefont{Dasgupta}},
  \bibinfo{author}{\bibfnamefont{A.}~\bibnamefont{Dighe}},
  \bibinfo{author}{\bibfnamefont{G.~G.} \bibnamefont{Raffelt}},
  \bibnamefont{and} \bibinfo{author}{\bibfnamefont{A.~Y.}
  \bibnamefont{Smirnov}}, \bibinfo{journal}{Phys.Rev.Lett.}
  \textbf{\bibinfo{volume}{103}}, \bibinfo{pages}{051105}
  (\bibinfo{year}{2009}), \eprint{arXiv:0904.3542}.

\bibitem[{\citenamefont{Fogli et~al.}(2009)\citenamefont{Fogli, Lisi, Marrone,
  and Tamborra}}]{Fogli:2009rd}
\bibinfo{author}{\bibfnamefont{G.}~\bibnamefont{Fogli}},
  \bibinfo{author}{\bibfnamefont{E.}~\bibnamefont{Lisi}},
  \bibinfo{author}{\bibfnamefont{A.}~\bibnamefont{Marrone}}, \bibnamefont{and}
  \bibinfo{author}{\bibfnamefont{I.}~\bibnamefont{Tamborra}},
  \bibinfo{journal}{JCAP} \textbf{\bibinfo{volume}{0910}}, \bibinfo{pages}{002}
  (\bibinfo{year}{2009}), \eprint{arXiv:0907.5115}.

\bibitem[{\citenamefont{Friedland}(2010)}]{Friedland:2010sc}
\bibinfo{author}{\bibfnamefont{A.}~\bibnamefont{Friedland}},
  \bibinfo{journal}{Phys.Rev.Lett.} \textbf{\bibinfo{volume}{104}},
  \bibinfo{pages}{191102} (\bibinfo{year}{2010}), \eprint{arXiv:1001.0996}.

\bibitem[{\citenamefont{Nakazato
  et~al.}(2010{\natexlab{b}})\citenamefont{Nakazato, Sumiyoshi, Suzuki, and
  Yamada}}]{Nakazato:2010qy}
\bibinfo{author}{\bibfnamefont{K.}~\bibnamefont{Nakazato}},
  \bibinfo{author}{\bibfnamefont{K.}~\bibnamefont{Sumiyoshi}},
  \bibinfo{author}{\bibfnamefont{H.}~\bibnamefont{Suzuki}}, \bibnamefont{and}
  \bibinfo{author}{\bibfnamefont{S.}~\bibnamefont{Yamada}},
  \bibinfo{journal}{Phys.Rev.} \textbf{\bibinfo{volume}{D81}},
  \bibinfo{pages}{083009} (\bibinfo{year}{2010}{\natexlab{b}}),
  \eprint{arXiv:1004.0291}.

\bibitem[{\citenamefont{Ando and Sato}(2004)}]{Ando:2004hc}
\bibinfo{author}{\bibfnamefont{S.}~\bibnamefont{Ando}} \bibnamefont{and}
  \bibinfo{author}{\bibfnamefont{K.}~\bibnamefont{Sato}}, \bibinfo{journal}{New
  J. Phys.} \textbf{\bibinfo{volume}{6}}, \bibinfo{pages}{170}
  (\bibinfo{year}{2004}), \eprint{astro-ph/0410061}.

\bibitem[{\citenamefont{Smartt et~al.}(2009)\citenamefont{Smartt, Eldridge,
  Crockett, and Maund}}]{Smartt:2008zd}
\bibinfo{author}{\bibfnamefont{S.~J.} \bibnamefont{Smartt}},
  \bibinfo{author}{\bibfnamefont{J.~J.} \bibnamefont{Eldridge}},
  \bibinfo{author}{\bibfnamefont{R.~M.} \bibnamefont{Crockett}},
  \bibnamefont{and} \bibinfo{author}{\bibfnamefont{J.~R.} \bibnamefont{Maund}},
  \bibinfo{journal}{Mon. Not. Roy. Astron. Soc.}
  \textbf{\bibinfo{volume}{395}}, \bibinfo{pages}{1409} (\bibinfo{year}{2009}),
  \eprint{0809.0403}.

\bibitem[{\citenamefont{O'Connor and Ott}(2011)}]{O'Connor:2010tk}
\bibinfo{author}{\bibfnamefont{E.}~\bibnamefont{O'Connor}} \bibnamefont{and}
  \bibinfo{author}{\bibfnamefont{C.~D.} \bibnamefont{Ott}},
  \bibinfo{journal}{Astrophys. J.} \textbf{\bibinfo{volume}{730}},
  \bibinfo{pages}{70} (\bibinfo{year}{2011}), \eprint{1010.5550}.

\bibitem[{\citenamefont{Battistoni et~al.}(2005)\citenamefont{Battistoni,
  Ferrari, Montaruli, and Sala}}]{Battistoni:2005pd}
\bibinfo{author}{\bibfnamefont{G.}~\bibnamefont{Battistoni}},
  \bibinfo{author}{\bibfnamefont{A.}~\bibnamefont{Ferrari}},
  \bibinfo{author}{\bibfnamefont{T.}~\bibnamefont{Montaruli}},
  \bibnamefont{and} \bibinfo{author}{\bibfnamefont{P.~R.} \bibnamefont{Sala}},
  \bibinfo{journal}{Astropart. Phys.} \textbf{\bibinfo{volume}{23}},
  \bibinfo{pages}{526} (\bibinfo{year}{2005}).

\bibitem[{\citenamefont{Wurm et~al.}(2007)}]{Wurm:2007cy}
\bibinfo{author}{\bibfnamefont{M.}~\bibnamefont{Wurm}} \bibnamefont{et~al.},
  \bibinfo{journal}{Phys. Rev.} \textbf{\bibinfo{volume}{D75}},
  \bibinfo{pages}{023007} (\bibinfo{year}{2007}), \eprint{astro-ph/0701305}.

\bibitem[{dus()}]{duselwhite}
\bibinfo{note}{DUSEL white paper, in preparation. Available at \\
  http://citeseerx.ist.psu.edu/viewdoc/summary?doi=10.1.1.156.7734}.

\bibitem[{vag()}]{vaginsprivate}
\bibinfo{note}{M. Vagins, private communication}.

\bibitem[{\citenamefont{Hochmuth et~al.}(2007)}]{Hochmuth:2005nh}
\bibinfo{author}{\bibfnamefont{K.~A.} \bibnamefont{Hochmuth}}
  \bibnamefont{et~al.}, \bibinfo{journal}{Astropart. Phys.}
  \textbf{\bibinfo{volume}{27}}, \bibinfo{pages}{21} (\bibinfo{year}{2007}),
  \eprint{hep-ph/0509136}.

\bibitem[{\citenamefont{Terashima et~al.}(2008)}]{Terashima:2008zz}
\bibinfo{author}{\bibfnamefont{A.}~\bibnamefont{Terashima}}
  \bibnamefont{et~al.} (\bibinfo{collaboration}{KamLAND}), \bibinfo{journal}{J.
  Phys. Conf. Ser.} \textbf{\bibinfo{volume}{120}}, \bibinfo{pages}{052029}
  (\bibinfo{year}{2008}).

\bibitem[{\citenamefont{Bahcall et~al.}(2005)\citenamefont{Bahcall, Serenelli,
  and Basu}}]{Bahcall:2004pz}
\bibinfo{author}{\bibfnamefont{J.~N.} \bibnamefont{Bahcall}},
  \bibinfo{author}{\bibfnamefont{A.~M.} \bibnamefont{Serenelli}},
  \bibnamefont{and} \bibinfo{author}{\bibfnamefont{S.}~\bibnamefont{Basu}},
  \bibinfo{journal}{Astrophys. J.} \textbf{\bibinfo{volume}{621}},
  \bibinfo{pages}{L85} (\bibinfo{year}{2005}), \eprint{astro-ph/0412440}.

\bibitem[{\citenamefont{Lunardini}(2010)}]{Lunardini:2010ab}
\bibinfo{author}{\bibfnamefont{C.}~\bibnamefont{Lunardini}}
  (\bibinfo{year}{2010}), \eprint{arXiv:1007.3252}.

\bibitem[{\citenamefont{Strumia and Vissani}(2003)}]{Strumia:2003zx}
\bibinfo{author}{\bibfnamefont{A.}~\bibnamefont{Strumia}} \bibnamefont{and}
  \bibinfo{author}{\bibfnamefont{F.}~\bibnamefont{Vissani}},
  \bibinfo{journal}{Phys. Lett.} \textbf{\bibinfo{volume}{B564}},
  \bibinfo{pages}{42} (\bibinfo{year}{2003}), \eprint{astro-ph/0302055}.

\bibitem[{\citenamefont{Fogli et~al.}(2005)\citenamefont{Fogli, Lisi, Mirizzi,
  and Montanino}}]{Fogli:2004ff}
\bibinfo{author}{\bibfnamefont{G.~L.} \bibnamefont{Fogli}},
  \bibinfo{author}{\bibfnamefont{E.}~\bibnamefont{Lisi}},
  \bibinfo{author}{\bibfnamefont{A.}~\bibnamefont{Mirizzi}}, \bibnamefont{and}
  \bibinfo{author}{\bibfnamefont{D.}~\bibnamefont{Montanino}},
  \bibinfo{journal}{JCAP} \textbf{\bibinfo{volume}{0504}}, \bibinfo{pages}{002}
  (\bibinfo{year}{2005}), \eprint{hep-ph/0412046}.

\bibitem[{\citenamefont{Cocco et~al.}(2004)\citenamefont{Cocco, Ereditato,
  Fiorillo, Mangano, and Pettorino}}]{Cocco:2004ac}
\bibinfo{author}{\bibfnamefont{A.~G.} \bibnamefont{Cocco}},
  \bibinfo{author}{\bibfnamefont{A.}~\bibnamefont{Ereditato}},
  \bibinfo{author}{\bibfnamefont{G.}~\bibnamefont{Fiorillo}},
  \bibinfo{author}{\bibfnamefont{G.}~\bibnamefont{Mangano}}, \bibnamefont{and}
  \bibinfo{author}{\bibfnamefont{V.}~\bibnamefont{Pettorino}},
  \bibinfo{journal}{JCAP} \textbf{\bibinfo{volume}{0412}}, \bibinfo{pages}{002}
  (\bibinfo{year}{2004}), \eprint{hep-ph/0408031}.

\bibitem[{\citenamefont{Kolbe et~al.}(2003)\citenamefont{Kolbe, Langanke,
  Martinez-Pinedo, and Vogel}}]{Kolbe:2003ys}
\bibinfo{author}{\bibfnamefont{E.}~\bibnamefont{Kolbe}},
  \bibinfo{author}{\bibfnamefont{K.}~\bibnamefont{Langanke}},
  \bibinfo{author}{\bibfnamefont{G.}~\bibnamefont{Martinez-Pinedo}},
  \bibnamefont{and} \bibinfo{author}{\bibfnamefont{P.}~\bibnamefont{Vogel}},
  \bibinfo{journal}{J. Phys.} \textbf{\bibinfo{volume}{G29}},
  \bibinfo{pages}{2569} (\bibinfo{year}{2003}), \eprint{nucl-th/0311022}.

\bibitem[{sna(2004)}]{snap}
\bibinfo{journal}{SNAP letter of intent, at http://snap.lbl.gov/}
  (\bibinfo{year}{2004}).

\bibitem[{snl(2010)}]{snls}
\bibinfo{journal}{SNLS collaboration, http://www.cfht.hawaii.edu/SNLS/}
  (\bibinfo{year}{2010}).

\bibitem[{\citenamefont{Smy}(2009)}]{smytalk}
\bibinfo{author}{\bibfnamefont{M.}~\bibnamefont{Smy}} (\bibinfo{year}{2009}),
  \bibinfo{note}{talk at the workshop {\it Supernova Physics and DUSEL}, Los
  Angeles}.

\bibitem[{\citenamefont{Hempel et~al.}(2011)\citenamefont{Hempel, Fischer,
  Schaffner-Bielich, and Liebendorfer}}]{Hempel:2011mk}
\bibinfo{author}{\bibfnamefont{M.}~\bibnamefont{Hempel}},
  \bibinfo{author}{\bibfnamefont{T.}~\bibnamefont{Fischer}},
  \bibinfo{author}{\bibfnamefont{J.}~\bibnamefont{Schaffner-Bielich}},
  \bibnamefont{and}
  \bibinfo{author}{\bibfnamefont{M.}~\bibnamefont{Liebendorfer}}
  (\bibinfo{year}{2011}), \eprint{1108.0848}.

\bibitem[{\citenamefont{Fischer et~al.}(2010)\citenamefont{Fischer, Whitehouse,
  Mezzacappa, Thielemann, and Liebendorfer}}]{Fischer:2009af}
\bibinfo{author}{\bibfnamefont{T.}~\bibnamefont{Fischer}},
  \bibinfo{author}{\bibfnamefont{S.~C.} \bibnamefont{Whitehouse}},
  \bibinfo{author}{\bibfnamefont{A.}~\bibnamefont{Mezzacappa}},
  \bibinfo{author}{\bibfnamefont{F.~K.} \bibnamefont{Thielemann}},
  \bibnamefont{and}
  \bibinfo{author}{\bibfnamefont{M.}~\bibnamefont{Liebendorfer}},
  \bibinfo{journal}{Astron. Astrophys.} \textbf{\bibinfo{volume}{517}},
  \bibinfo{pages}{A80} (\bibinfo{year}{2010}), \eprint{0908.1871}.

\bibitem[{\citenamefont{Hudepohl et~al.}(2010)\citenamefont{Hudepohl, Muller,
  Janka, Marek, and Raffelt}}]{Huedepohl:2009wh}
\bibinfo{author}{\bibfnamefont{L.}~\bibnamefont{Hudepohl}},
  \bibinfo{author}{\bibfnamefont{B.}~\bibnamefont{Muller}},
  \bibinfo{author}{\bibfnamefont{H.~T.} \bibnamefont{Janka}},
  \bibinfo{author}{\bibfnamefont{A.}~\bibnamefont{Marek}}, \bibnamefont{and}
  \bibinfo{author}{\bibfnamefont{G.~G.} \bibnamefont{Raffelt}},
  \bibinfo{journal}{Phys. Rev. Lett.} \textbf{\bibinfo{volume}{104}},
  \bibinfo{pages}{251101} (\bibinfo{year}{2010}), \eprint{0912.0260}.

\bibitem[{\citenamefont{Nakazato et~al.}(2011)\citenamefont{Nakazato, Furusawa,
  Sumiyoshi, Ohnishi, Yamada et~al.}}]{Nakazato:2011vd}
\bibinfo{author}{\bibfnamefont{K.}~\bibnamefont{Nakazato}},
  \bibinfo{author}{\bibfnamefont{S.}~\bibnamefont{Furusawa}},
  \bibinfo{author}{\bibfnamefont{K.}~\bibnamefont{Sumiyoshi}},
  \bibinfo{author}{\bibfnamefont{A.}~\bibnamefont{Ohnishi}},
  \bibinfo{author}{\bibfnamefont{S.}~\bibnamefont{Yamada}},
  \bibnamefont{et~al.} (\bibinfo{year}{2011}), \eprint{1111.2900}.

\bibitem[{\citenamefont{Fryer}(2009)}]{Fryer:2007cf}
\bibinfo{author}{\bibfnamefont{C.~L.} \bibnamefont{Fryer}},
  \bibinfo{journal}{Astrophys. J.} \textbf{\bibinfo{volume}{699}},
  \bibinfo{pages}{409} (\bibinfo{year}{2009}), \eprint{0711.0551}.

\bibitem[{\citenamefont{Fryer et~al.}(2009)}]{Fryer:2009zs}
\bibinfo{author}{\bibfnamefont{C.~L.} \bibnamefont{Fryer}}
  \bibnamefont{et~al.}, \bibinfo{journal}{Astrophys. J.}
  \textbf{\bibinfo{volume}{707}}, \bibinfo{pages}{193} (\bibinfo{year}{2009}),
  \eprint{0908.0701}.

\bibitem[{\citenamefont{Nagataki et~al.}(2003)\citenamefont{Nagataki, Kohri,
  Ando, and Sato}}]{Nagataki:2002bn}
\bibinfo{author}{\bibfnamefont{S.}~\bibnamefont{Nagataki}},
  \bibinfo{author}{\bibfnamefont{K.}~\bibnamefont{Kohri}},
  \bibinfo{author}{\bibfnamefont{S.}~\bibnamefont{Ando}}, \bibnamefont{and}
  \bibinfo{author}{\bibfnamefont{K.}~\bibnamefont{Sato}},
  \bibinfo{journal}{Astropart. Phys.} \textbf{\bibinfo{volume}{18}},
  \bibinfo{pages}{551} (\bibinfo{year}{2003}), \eprint{astro-ph/0203481}.

\bibitem[{\citenamefont{McLaughlin and Surman}(2007)}]{McLaughlin:2006yy}
\bibinfo{author}{\bibfnamefont{G.~C.} \bibnamefont{McLaughlin}}
  \bibnamefont{and} \bibinfo{author}{\bibfnamefont{R.}~\bibnamefont{Surman}},
  \bibinfo{journal}{Phys. Rev.} \textbf{\bibinfo{volume}{D75}},
  \bibinfo{pages}{023005} (\bibinfo{year}{2007}), \eprint{astro-ph/0605281}.

\bibitem[{\citenamefont{Lunardini}(2007)}]{Lunardini:2006pd}
\bibinfo{author}{\bibfnamefont{C.}~\bibnamefont{Lunardini}},
  \bibinfo{journal}{Phys. Rev.} \textbf{\bibinfo{volume}{D75}},
  \bibinfo{pages}{073022} (\bibinfo{year}{2007}), \eprint{astro-ph/0612701}.

\bibitem[{\citenamefont{Kochanek et~al.}(2008)}]{Kochanek:2008mp}
\bibinfo{author}{\bibfnamefont{C.~S.} \bibnamefont{Kochanek}}
  \bibnamefont{et~al.}, \bibinfo{journal}{Astrophys. J.}
  \textbf{\bibinfo{volume}{684}}, \bibinfo{pages}{1336} (\bibinfo{year}{2008}),
  \eprint{0802.0456}.

\end{thebibliography}

\end{document}